\documentclass{aastex701}
\usepackage{amsmath}
\usepackage{tocloft}
\makeatletter
\@ifundefined{if@restonecol}{\newif\if@restonecol}{}
\makeatother
\usepackage{subcaption}
\usepackage{graphicx}
\usepackage{tabularx}
\usepackage{refcount}
\usepackage{footmisc}
\usepackage{url}
\usepackage[utf8]{inputenc}
\usepackage[polish,english]{babel}
\usepackage{tikz}
\usetikzlibrary{arrows.meta, decorations.markings}
\usetikzlibrary{calc}

\providecommand{\contentsname}{Contents}

\newcommand{\mapimg}[1]{%
\includegraphics[
width=\textwidth,
trim={0.8cm 0cm 0.8cm 1.1cm},
clip
]{#1}%
}

\hypersetup{
  colorlinks=true,
  linkcolor=black,
  citecolor=black,
  urlcolor=black
}

\makeatletter
\newcommand{\capratoc}{%
  \par\vspace{-0.8em}%
  \begingroup
    \small
    \color{black}%
    \phantomsection
    {\centering\section*{\contentsname}\par}%
    \@starttoc{toc}%
  \endgroup
  \vspace{-0.8em}%
}
\makeatother

\usepackage{etoolbox}   

\AtEndEnvironment{abstract}{%
  \par\vspace{-0.8em}%
  {%
    \let\clearpage\relax
    \begingroup
    \small
    \hypersetup{linkcolor=black}%
    \color{black}%
    \tableofcontents
    \endgroup
  }%
  \vspace{-0.8em}%
}

\begin{document}

\title{\textsc{Capra}: Scalable HEALPix-Native Intensity Reconstruction for High-Resolution IMAP Analyses}

\author[orcid=0000-0003-1129-5310]{Nikola Bukowiecka}
\altaffiliation{Present affiliation. Parts of the work performed while author was affiliated with Space Research Centre.}
\affiliation{Department of Physics, University of Rhode Island, Kingston, RI 02881, USA}
\email[show]{nikola.bukowiecka@uri.edu}

\author[0000-0003-1874-9450]{Daniel B. Reisenfeld}
\affiliation{Los Alamos National Laboratory, Los Alamos, NM 87545, USA}
\email{dreisenfeld@lanl.gov}

\author[0000-0003-3957-2359]{Maciej Bzowski}
\affiliation{Space Research Centre PAS (CBK PAN), Warsaw, Poland}
\email{bzowski@cbk.waw.pl}

\begin{abstract}

We present \textsc{Capra}, a HEALPix-native pipeline for 
reconstructing all-sky energetic neutral atom (ENA) intensity maps from the NASA \textit{Interstellar Boundary Explorer} (IBEX) and  \textit{Interstellar Mapping and Acceleration} (IMAP) missions' event-counting data. \textsc{Capra} produces two products: (1) a baseline boresight-assigned map, and (2) an optional smoothed map on the HEALPix grid. This produces smooth, physically interpretable maps with consistent resolution changes and reference-frame transformations. 

The smoothed Capra maps recover the large-scale ENA morphology while reducing sampling-driven mottling and strip-like artifacts.
Relative comparison with the \textsc{Theseus} reconstruction shows broadly consistent large-scale structure, with the discrepancies reflecting differences in reconstruction methods, rather than major morphological discrepancies.

We verify invariance (preservation) of count numbers, rates, and intensities across tessellations under coordinate transformations, and demonstrate numerical convergence up to HEALPix resolution  $N_{\mathrm{side}}=64$ for a representative dataset. 

We show that thread-based parallelism performance exhibits approximately linear strong scaling, with bounded memory usage, enabling efficient high-resolution reconstruction and making \textsc{Capra} well-suited for high-resolution IMAP reconstruction.

\end{abstract}
\vspace{-0.8em}



\section{\textbf{Introduction}}

Spherical imaging from event-counting sky surveys continues to pose a challenge in various astrophysical missions. When the instrument repeatedly scans the sky along similar paths, the resulting coverage may exhibit stripes and gaps that arise from the sampling pattern rather than the underlying physical signal.

Energetic neutral atom (ENA) intensity sky maps produced by the ENA imagers on the \textit{Interstellar 
Boundary Explorer} (IBEX) \citep{mccomas_etal:09a} and the newly launched \textit{Interstellar 
Mapping and Acceleration Probe} IMAP \citep{mccomas_etal:25a} provide unique global probes of the heliosphere. While IBEX faces the above-mentioned problems of incomplete spatial coverage, it is too early to tell what challenges IMAP will face. We expect the new probe to be able to collect data nearly continuously due to its placement at L1 (rather than Earth's orbit, like IBEX), however some gaps may still occur during prolonged solar storms and spacecraft maneuvers. The goal is to produce sky images of ENA intensities that are physically interpretable and quantitatively faithful to the data, while remaining robust, transformable, and scalable across coordinate systems and pixel resolutions.

The existing IBEX map-making approaches -- including both the standard approach taken by the IBEX Science Operation Center (ISOC) \citep{Schwadron2009} and the 
\textsc{Theseus} method described in \citet{Osthus02042024} -- treat each detected atom as originating from the imager's instantaneous boresight direction. The ISOC map rendering methodology then applies a smoothing function that decreases linearly with the angle in both longitude and latitude from a given pixel's center in an attempt to fill in gaps in the maps, which also spatially smooths the data. By contrast, \textsc{THESEUS} applies a sharpening algorithm, where individual counts are reassigned to locations based on information in adjacent cells. This method statistically models the data, which removes much of the aforementioned gaps and holes, while also increasing the resolution of resulting maps by deconvolution of the instrument response; however this process is complex and computationally expensive.

In this paper, we introduce \textsc{Capra}, a HEALPix-native (Appendix \ref{sec:hp}, \citet{gorski_etal:05a}) pipeline for reconstructing all-sky ENA maps from IBEX and IMAP event-counting data \citep{Bukowiecka_Capra}. It is mathematically simple and computationally inexpensive; moreover, it is parallelized. There are two products, one directly supported by the data and one optionally offered for better interpretability: \textsc{Capra} produces a baseline map that follows the sky survey sampling (binning) as closely as possible, and an optional smoothed map that applies a smoothing kernel to reduce sampling artifacts. Given the incomplete spatial coverage and low statistics of our baseline IBEX dataset, in this paper we primarily make use of the Gaussian kernel to present the results and showcase \textsc{Capra} capabilities.

The baseline product follows the native sampling geometry as closely as possible and produces a minimally processed map that exposes the potential incomplete spatial data coverage, like  gaps and low-exposure regions. The optional smoothed product applies an explicit, user-controlled kernel on the HEALPix grid to reduce sampling-driven mottling and improve visual interpretability. This smoothing step is therefore a reconstruction choice rather than part of the instrument response.

The second aspect of \textsc{Capra} that we highlight in the paper is its numerical transparency. In addition to presentatioon of the map-production process maps, we discuss its validation and verification, and we show that \textsc{Capra} is stable under transformations. Due to the grid being HEALPix native, it is straightforward to test the validity and convergence of the method. We present a suite of preservation checks for the count numbers, counting rates, and intensity-like quantities.
We perform two scaling tests. First, we test the resolution scaling by increasing
$N_{\rm side}$ and comparing the resulting maps to assess convergence of the
reconstruction. Second, we test parallel strong scaling by holding the input
dataset and HEALPix resolution fixed while varying the number of parallel kernels. The reconstruction shows convergence with increasing HEALPix resolution and
approximately linear strong scaling over the tested range of parallel kernels.

Given its good qualitative agreement with the state-of-the-art, which we show in the following Sections validation and verification testing and its scalability, \textsc{Capra} is a reliable map reconstruction pipeline that provides a baseline product, an optional configurable smoothing layer, and a consistent, scalable performance.

In Section \ref{subsec:IBEX-data} we review the IBEX-Hi data structure used in the paper. In Section \ref{sec:data} we summarize the measurement geometry, binning, and exposure time considerations.
In Section \ref{sec:algorithm}, we present the map reconstruction pipeline. Section \ref{sec:postprocessing} describes the post-processing framework: normalization schemes, grid transformations, and area-weighted statistics. In Section \ref{subsec:results} we present the results: ENA intensities for all energy channels and all data products for a selected energy channel; comparison of \textsc{Capra} with baseline IBEX maps constructed directly from the same orbit-arc data; and a detailed comparison with \textsc{Theseus}. In Section \ref{sec:dat_manip}, we demonstrate the flexibility of HEALPix-based data manipulation: reference-frame changes, tessellation changes, and variable-resolution grids. In Section \ref{sec:conclusions} we summarize the main conclusions. Appendix \ref{sec:hp} provides additional information about HEALPix discretization scheme, and Appendix \ref{sec:conservation} contains validation tests, convergence studies, and performance benchmarks.

\section{\textbf{Application to IBEX and IMAP data}} \label{subsec:IBEX-data}
The IBEX and IMAP missions detect ENAs that arrive at the spacecraft's ENA imagers. The ENA detectors (\citep{funsten_etal:09a, funsten_etal:26}), are installed on spin-stabilized spacecraft with the rotation axis directed within a few degrees off the Sun's center. The direction of spin axis is regularly adjusted, so that the instrument scanning circles cover the entire sky during half of the year. The time intervals when the rotation axis is fixed are referred to as orbital arcs. 

On detection, ENAs first pass through a collimator, which only accepts atoms incoming within a certain finite angular field-of-view (FoV) with respect to the instantaneous boresight of the instrument. Due to the mechanical design of the collimator, the acceptance probability decreases with an increasing angle from the center of the FoV. This response function, referred to as the instrument point spread function (PSF), leads to an unavoidable defocusing of the incident ENA distribution during the measurement process. 
 We calculate the intensities through a non-spherical surface (the IBEX detector) and to aid visualization we represent them on a unit sphere.

It is important to note this response is distinct from any further smoothing which may be applied during map reconstruction. In \textsc{Capra}, optional smoothing kernels can be applied to the reconstructed maps, with an effect of a reduction of sampling artifacts and apparent noise in the map. This is not part of the instrument response and does not attempt to simulate any detector physics; rather, it is a choice made by the researcher to facilitate visual analysis. Note that any smoothing applied will degrade the angular resolution of the map. The users are offered a choice: using a built-in function for Gaussian smoothing, inserting their own smoothing function, or using nothing at all; the latter may be most appropriate when the data have nearly complete spatial coverage (as expected for IMAP).

The IBEX spacecraft has two instruments, IBEX-Hi \citep{funsten_etal:09a} and IBEX-Lo \citep{fuselier_etal:09b}, but from here on, we will only discuss IBEX-Hi. To generate the HEALPix maps in our pipeline, we used data from the IBEX-Hi instrument collected over the second half of 2018, corresponding to the orbit arc data that compose \textit{Map2018B} (orbits 412a-431b) binned at $1\arcdeg$ resolution, the \textsc{HP} dataset \citep{Bukowiecka_Capra}. We then compared our results against both the baseline IBEX maps\footnote{For validation purposes we map the $1\arcdeg$ binned IBEX-Hi data directly onto a rectangular grid, with no other treatment. The same data is used as input for our \textsc{Capra} pipeline.} and the \textsc{Theseus} products that were processed using the \textsc{Theseus} method by D.~Reisenfeld’s team, using the same underlying IBEX orbit arc data as input.

Our IBEX-Hi input data are grouped into files, each one corresponding to one orientation of the IBEX spin axis, referred to as orbit arcs.
For every orbit arc $n \in \{ 1,2,...,n_o \}$, where $n_o = 40$ - the collection of which make up a full-sky map -- each file contains the longitude $l_{n,i}$, latitude $\theta_{n,i}$, count number $c_{n,i}$, exposure time $e_{n,i}$, and background rate $b_{n,i}$, for each bin $i \in \{1,2,...,n_b \}$, where $n_b = 360$. The measurements are performed in $k$ energy channels (ESA) $E_k$, where $k \in \{1,2,...,6 \}$, so for each bin there are six sets of data $(c_{n,i,k}, e_{n,i,k}, b_{n,i,k})$. The procedure is performed identically for all energy steps. The IBEX-Hi collimator has a characteristic FWHM of $6.5 \arcdeg$; in the reconstruction we use this angular scale to define the selection-circle radius $R_c$ \citep{2009:funsten}. We also use the longitude and latitude of the spin axis direction for each orbital arc $r_n$.

Our pipeline can be used to make visualizations and perform efficient and flexible analysis of available IBEX data, as demonstrated in this paper, and can be easily extended for the upcoming IMAP data \citep{mccomas_etal:25a}. 

\section{\textbf{Data structure}} \label{sec:data}
\subsection{Measurement geometry}

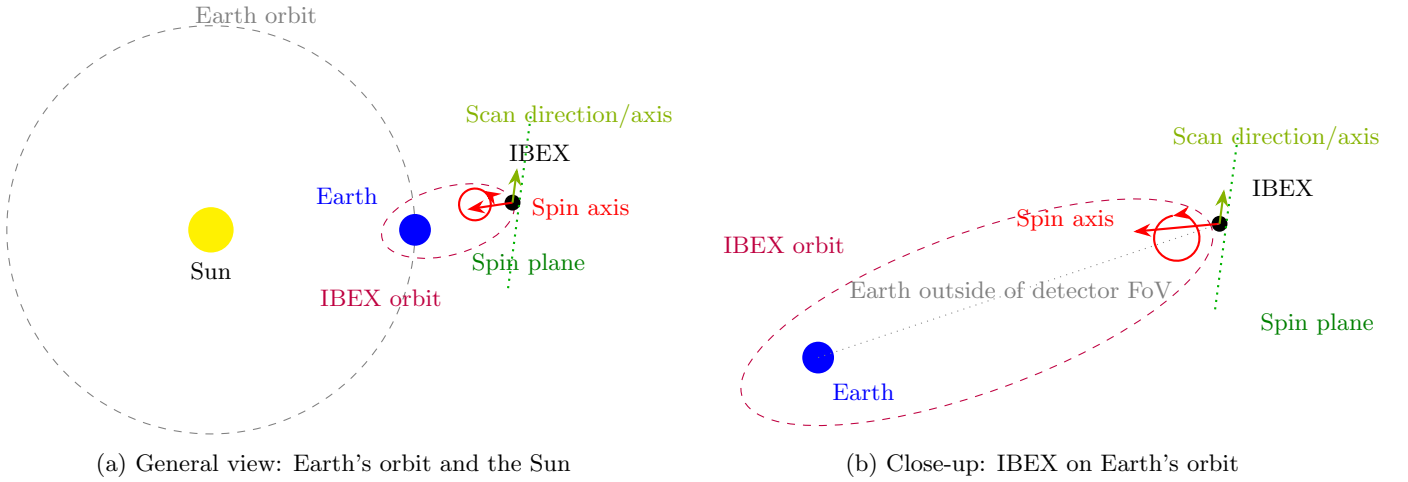
\begin{figure}[htbp]
\centering

\begin{subfigure}[b]{0.48\textwidth}
\centering
\begin{tikzpicture}[scale=3, >=Stealth]

\fill[yellow] (0,0) circle (0.1);
\node at (0,-0.18) {Sun};

\draw[gray, dashed] (0,0) circle (0.9);
\node[gray] at (0.2,0.95) {Earth orbit};

\fill[blue] (0.9,0) circle (0.07);
\node[text=blue] at (0.6,0.15) {Earth};

\draw[purple, dashed, rotate around={15:(0.9,0)}] (1.05,0) ellipse (0.3 and 0.15);
\node[purple] at (0.75,-0.3) {IBEX orbit};

\coordinate (IBEX) at (1.33,0.12);
\fill (IBEX) circle (0.035);
\node at (1.45,0.34) {IBEX};

\coordinate (SpinEnd) at ($(IBEX)+(-0.2,-0.03)$);
\draw[red, ->, thick] (IBEX) -- (SpinEnd);
\node[red] at ($(SpinEnd)+(0.5,0)$) {Spin axis};

\draw[green!70!black, dotted, thick] 
  ($(IBEX)+(0.08,0.38)$) -- ($(IBEX)+(-0.02,-0.38)$);
\node[green!50!black] at (1.4,-0.15) {Spin plane};

\draw[->, lime!70!black, thick]
  (IBEX) -- ($(IBEX)+0.2*(0.1,0.76)$);
\node[lime!70!black] at ($(IBEX)+(0.25,0.38)$) {Scan direction/axis};

\coordinate (SpinMid) at ($(IBEX)!0.8!(SpinEnd)$);
\draw[->, red, thick] 
  ($(SpinMid)+(0.06,0.04)$) arc[start angle=20, end angle=420, radius=0.07];

\end{tikzpicture}
\caption{General view: Earth's orbit and the Sun}
\end{subfigure}
\hfill
\begin{subfigure}[b]{0.48\textwidth}
\centering
\begin{tikzpicture}[scale=3, >=Stealth]

\fill[blue] (-1.2,-0.2) circle (0.07);
\node[text=blue] at (-1,-0.35) {Earth};

\draw[purple, dashed, rotate around={20:(-0.5,0)}] (-0.5,0) ellipse (1.1 and 0.35);
\node[purple] at (-1.35,0.3) {IBEX orbit};

\coordinate (IBEX) at (0.57,0.39);
\fill (IBEX) circle (0.035);
\node at (0.85,0.55) {IBEX};

\coordinate (SpinEnd) at ($(IBEX)+(-0.38,-0.035)$);
\draw[red, ->, thick] (IBEX) -- (SpinEnd);
\node[red] at ($(SpinEnd)+(-0.3,0.05)$) {Spin axis};

\draw[green!70!black, dotted, thick] 
  ($(IBEX)+(0.08,0.38)$) -- ($(IBEX)+(-0.02,-0.38)$);
\node[green!50!black] at (1.0,-0.05) {Spin plane};

\draw[->, lime!70!black, thick]
  (IBEX) -- ($(IBEX)+0.2*(0.1,0.76)$);
\node[lime!70!black] at ($(IBEX)+(0.25,0.38)$) {Scan direction/axis};

\draw[gray, dotted] (-1.2,-0.2) -- (IBEX);
\node[gray] at (-0.35,0.1) {Earth outside of detector FoV};

\coordinate (SpinMid) at ($(IBEX)!0.7!(SpinEnd)$);
\draw[->, red, thick] 
  ($(SpinMid)+(0.06,0.06)$) arc[start angle=100, end angle=460, radius=0.1];

\end{tikzpicture}
\caption{Close-up: IBEX on Earth's orbit}
\end{subfigure}

\caption{The geometry of the IBEX mission, showing IBEX's orbit around the Earth and the orientation of the spin axis and the scanning (``spin'') plane. The spin axis is pointed in the direction of the Sun's center, and drifts up to $\sim 4\arcdeg.5$ away from it as the Earth carries IBEX around the Sun, until it is re-pointed at the Sun every 4.5 days \citep{mccomas_etal:11a}. The ENA imagers (IBEX-Hi and IBEX-Lo) are mounted on the spacecraft such that their boresights are perpendicular to the spin axis. The boresight direction sweeps out a great circle in the sky as the spacecraft spins.}
\label{fig:IBEX_geometry}
\end{figure}

In Figure \ref{fig:IBEX_geometry}, we present the orbital and viewing geometry of the IBEX mission to help the readers better understand the data structure and the algorithm. For more information, see  \citep{mccomas:09a} for general insight on the IBEX mission, and \citep{funsten_etal:09a} for details on the IBEX-Hi instrument. Here, we briefly recapitulate the most important highlights:
\begin{itemize}
    \item IBEX is moving in a highly elliptical orbit around Earth with an apogee of $\sim 50 R_E$ and an orbital period of $\sim 9.1$ days.
    \item ENA observations are performed when the spacecraft is safely outside the radiation belts, or above  $15 R_E$.
    \item IBEX is a spinning spacecraft ($\sim 4$ rpm), with a spin axis (red arrow in Figure \ref{fig:IBEX_geometry}) fixed with respect to the stars (i.e., in an inertial reference frame). During one spin IBEX-Hi views a great circle of the sky. For $\sim 4.5$ days this axis remains stationary, resulting in thousands of repeated observations of the same strip of the sky.  After 4.5 days\footnote{Currently, IBEX repoints twice per orbit (every 4.5 days), once at perigee and once at apogee.  Originally, IBEX had a 7.5-day orbit and only repointed at perigee, but soon after launch it was discovered that this orbit was unstable due to perturbations from the Moon, and so in 2011 June the IBEX orbit was changed to a 9.1 day orbit.}, IBEX repoints back to the Sun, tracking the Sun's apparent motion (which moves roughly $\sim 1\arcdeg$ per day with respect to the background stars).  IBEX then begins collecting data from a new strip, shifted by $4\arcdeg.5$ in ecliptic longitude from the last one.  In this manner, it takes 6 months for IBEX to view the whole sky.
    \item The IBEX-Hi and IBEX-Lo entrance apertures are aligned perpendicularly to the spin axis. For a given instrument, the vector normal to the entrance aperture, or boresight, represents the mean instantaneous viewing direction of the instrument (green arrow).  Thanks to the spacecraft's rotation, each instrument's boresight scans a great circle in the sky ( the ``scan circle'') in the plane perpendicular to the spin axis, called the spin plane (dotted line).\footnote{ 
    (i) The \textit{spin plane} is the entire path that the boresight travels during one rotation of the spacecraft. It incorporates all possible directions the instrument points during one orbit. (ii) The \textit{Boresight direction} is a \textbf{fixed} vector with respect to the spacecraft. (iii) The \textit{Scan direction} is the \textbf{instantaneous} viewing vector of the instrument in the spacecraft-inertial frame. It is the direction that the boresight points at at a given moment during the rotation. Physically, the boresight and the scan direction are the same vector; the distinction is that one is intrinsic to the instrument and doesn't change, and the other is the instantaneous orientation of that vector in inertial space.
    }
\end{itemize}



\subsection{Orbit-arc and scan-angle binning} \label{subsec:binning}
In this work we reconstruct sky maps from binned event data rather than from individual event times. The input data are histograms of detected counts, organized by spacecraft spin-axis orientation and scan angle in Ecliptic J2000 coordinates. Each scan-angle bin corresponds to an interval in the spacecraft spin phase and is associated with a mean boresight direction on the sky. 
The data used here are binned at $1^\circ$ resolution, matching the input used for the \textsc{THESEUS} maps; the standard ISOC products are commonly represented using $6^\circ$ bins.

IBEX repoints its spin axis twice per orbit, so each orbit is divided into two observing arcs, hereafter the \textit{orbit arcs}. We label the ascending arc, from the perigee to apogee, with ``a'', and the descending arc, from the apogee to perigee, with ``b''. Therefore, a label 238a would denote one spin-axis orientation and its corresponding set of scan-angle bins. In this notation, the orbit-arc index is denoted by $n$, and the scan-angle bin within that arc by $i$.

The conversion from time-domain measurements to sky coordinates is set by the spacecraft rotation. 
As IBEX spins, the instrument boresight sweeps out a scan circle in the plane perpendicular to the spin axis (see Figure \ref{fig:IBEX_geometry} or Figure \ref{fig:hp_coll_two}). Dividing this scan circle into equally spaced angular bins assigns each observation to a scan-angle interval and therefore to a viewing direction. For each orbit arc and bin, the data product contains the total number of counts $c_{n,i}$, the corresponding exposure time $e_{n,i}$, and the background rate $b_{n,i}$. Bin size has to be carefully chosen -- it has to be small enough with respect to the collimator resolution, so that while binning the counts one does not introduce unnecessary smoothing. 

\subsection{Exposure times and background} \label{subsec:goodtimes}
The exposure time $e_{n,i}$ is the total \textbf{valid} observing time accumulated in orbit arc $n$ and scan-angle bin $i$. It measures how long the instrument boresight sampled the corresponding angular bin after the data have been filtered for valid observing conditions.

The filtering is performed using the \textit{goodtimes} list, which specifies intervals in time and spin angle for which the measurements are accepted. This semi-automated culling removes data affected by local backgrounds or other invalid observing conditions. Because the filtering is applied to specific time and spin-angle intervals, it does not remove entire orbit arcs uniformly - different scan-angle bins within the same orbit arc can have different exposures. If no valid measurements remain in a bin, both the counts and exposure time for that bin are set to zero.

The same processing also estimates the isotropic background contribution, primarily from penetrating radiation such as cosmic rays.
For each orbit arc and scan-angle bin, the background is represented by an average rate $b_{n,i}$, which is later multiplied by the exposure time to obtain the expected number of background counts in that bin. 
The background-subtracted signal used in the reconstruction is therefore given by
\begin{equation}
    s_{n,i} = c_{n,i} - b_{n,i} e_{n,i}.
\end{equation} \label{signal_eq}

\begin{figure} [hbt!]
    \centering
	\includegraphics[width=0.45\textwidth]{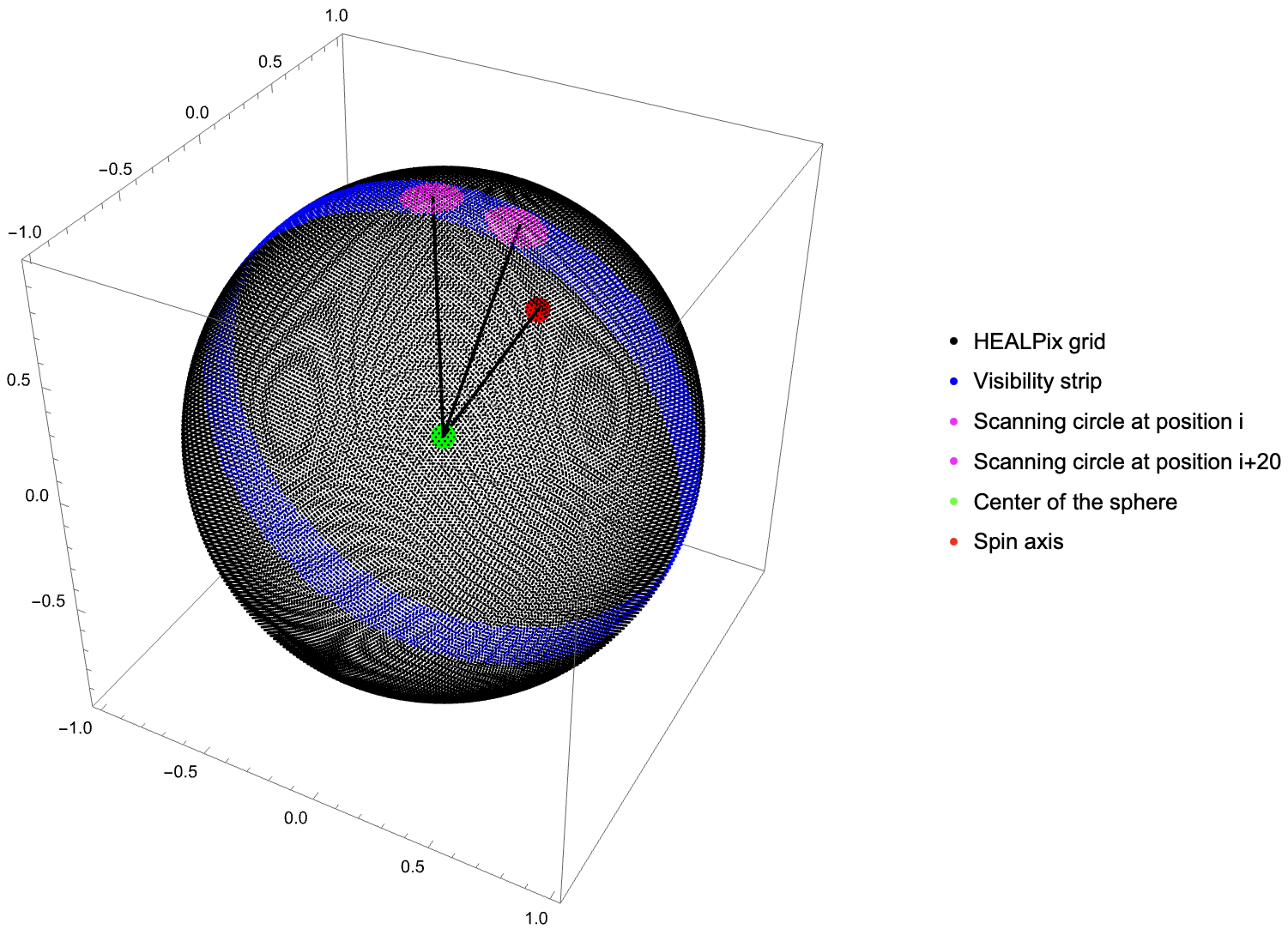}\includegraphics[width=0.5\textwidth]{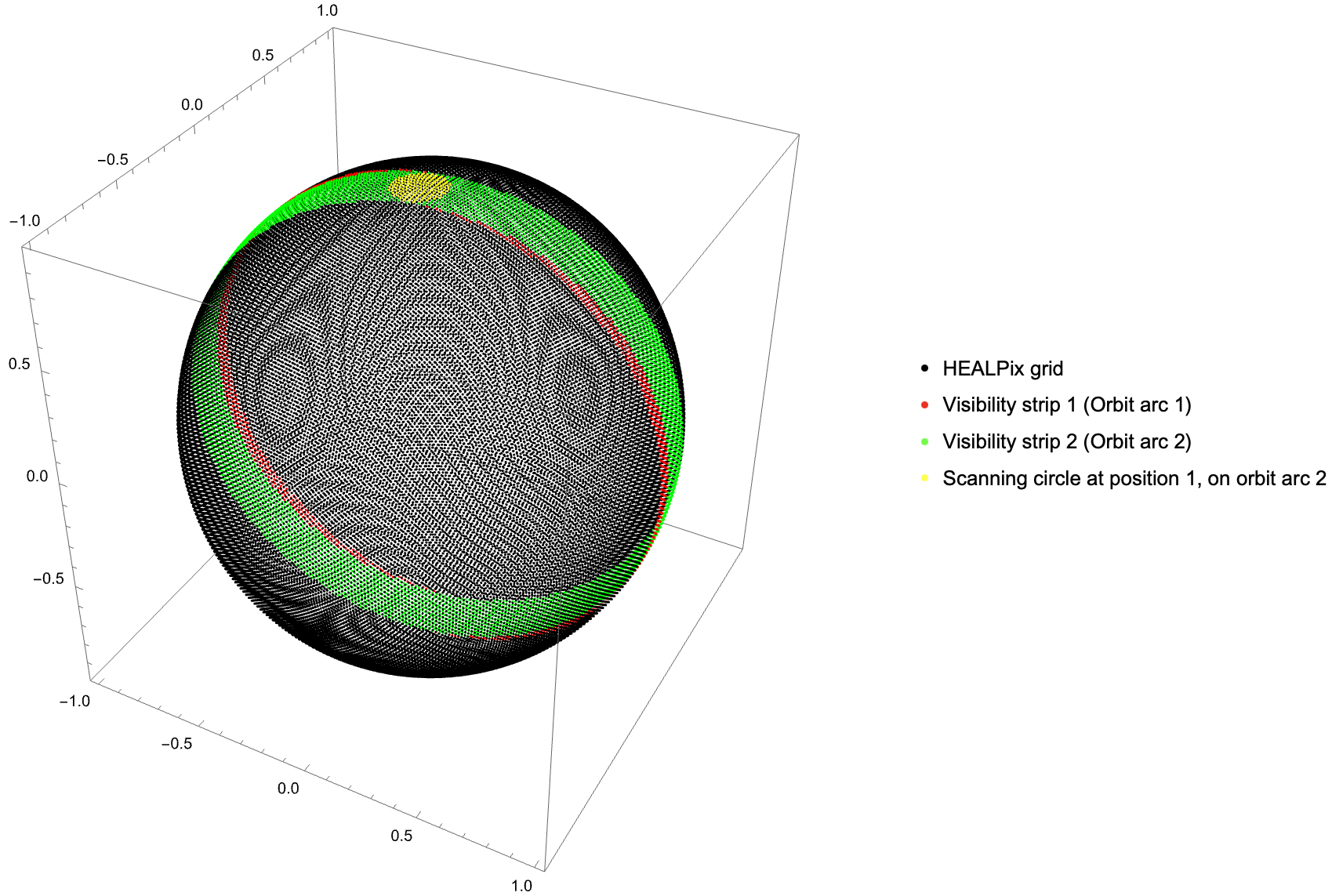}
    \caption{Left: An example sphere discretized with the HEALPix grid ($N_{\mathrm{side}}$ = 64), with the spacecrafts rotation axis for an example orbit arc, and the two fields of view of the selection circle centered at two selected event location positions. Right: two neighboring orbit arcs.}
    \label{fig:hp_coll_two}
\end{figure}

Because successive orbit arcs correspond to different spin-axis pointings, their scan circles only partially overlap in the sky, see Figure \ref{fig:hp_coll_two}. As a result, the accumulated exposure is not uniform over the final all-sky map. 
Regions near the scan-circle intersections, especially near the ecliptic poles, receive more repeated coverage (which is why the exposure times map are brighter in the polar regions), while other regions receive less. 
The exposure map is, therefore, a central part of the reconstruction: it is needed to convert counts or background-subtracted signal into the respective rates and intensities, and it provides an important diagnostic of the reconstructed sky map.

\section{\textbf{Algorithm description}} \label{sec:algorithm}

In this section, we present a system for reconstruction of all-sky ENA intensity maps. The pipeline provides multiple outputs; however, the most physically relevant one, and the most used for postprocessing, is the exposure-corrected, geometrically normalized intensity. The distributions of counts, exposure times, signals and signal rates are used for validation and convergence testing. Throughout the paper, all vectors are represented in a Cartesian coordinate basis, unless explicitly specified otherwise. 

Let $I_-$ denote the set of all HEALPix pixel centers (Section~\ref{sec:hp}), with the elements $\overrightarrow{I_j} \in \mathbb{R}^3$, where $j \in \{1,2,\dots, N_{\mathrm{pix}}\}$ is the HEALPix pixel index, and $N_{\mathrm{pix}} = 12\,N_{\mathrm{side}}^2$. Similarly, let $R_{n,-,-}$, where $R_{n,-,-} \subseteq I_-$, denote the set of pixel centers that fall within the $n$-th orbit arc and $i$-th bin, with elements $\overrightarrow{R_{n,i,k}} \in \mathbb{R}^3$, where $i \in \{1,2,\dots, n_{\mathrm{b}}\}$ indexes the bins in that orbit arc, $k \in \{1,2,\dots, n_{\mathrm{bpix}}\}$ indexes the pixels in the respective bins, and
$n^{(n,i)}_{\mathrm{bpix}} = |R_{n,i,-}|$.
There exists a unique map from $\overrightarrow{R_{n,i,k}}$ vectors to $\overrightarrow{I_j}$ HEALPix vectors via a transformation function $f$ given in \cite{gorski_etal:05a}\footnote{In \citep{gorski_etal:05a}, Equations [17-18] give the relations for the ring and longitude indices, and Equations [4,5,8-9] give transformations of those indices into $(z,\phi)$ coordinates, which can be further transformed into Cartesian coordinates.}:

\begin{align*}
f: R_{n,-,-} &\rightarrow I_-, \\
\overrightarrow{R_{n,i,k}}  &\rightarrow \overrightarrow{I_j}.
\end{align*}
For each orbit arc $n$ (each iteration), the algorithm constructs partial all-sky maps of $c_{n,j}$, $e_{n,j}$, $s_{n,j}$, $r_{n,j}$ and $f_{n,j}$, which are subsequently superposed to obtain the final maps $c'_{j}$, $e'_{j}$, $s'_{j}$, $r'_{j}$ and $f'_{j}$.

The algorithm can be divided into steps, described in detail in the following sections.

\subsection{Data initialization and filtering}
We initialize to zero five lists, representing: the number of total counts (both signal and background) $c'_{j}$, the associated exposure times $e'_{j}$, the signal $s'_{j}$, the signal rate $r'_{j}$, and the energetic neutral atom (ENA) intensity $f'_{j}$ where the $j \in \{1,2,...,N_{\mathrm{pix}}\}$ represents the HEALPix pixel index. Those represent the all-sky maps and are filled iteratively orbit arc by orbit arc, through map-superposition, i.e., element-wise summation. 
From the input data, we select the exposure times $e_{n,i}$ and their corresponding filtered count values $c_{n,i}$, background values $b_{n,i}$, longitudes $l_{n,i}$ and latitudes $\theta_{n,i}$, where the orbit arc number $n \in \{1,2,...,n_o\}$, and the bin number $i \in \{1,2,...,n_b\}$. For a given orbit arc, there is an associated spin axis vector $\overrightarrow{r_n}$.

\subsection{The visibility strip selection}
\noindent \underline{Per orbit arc:}

We begin the main algorithm by creating the \textit{visibility strip}, corresponding to the scanning circle (shown in blue in Figure \ref{fig:hp_coll_two}). For each orbit arc $n$, we calculate the dot product between the spacecraft spin-axis vector $\overrightarrow{r_n}$  and all HEALPix pixel vectors $\overrightarrow{I_j}$, and select the subset of pixels $R_{n,-,-}$ that fall within a band bounded by the collimator radius around the scanning circle:
\begin{equation}
\forall n \in \{ 1,2,...,n_o \}, \qquad R_{n,-,-} = \{ \overrightarrow{v} \in I_{-} | \mathrm{vis_{min}} \leq \overrightarrow{r_n} \cdot \overrightarrow{v} \leq \mathrm{vis_{max}} \},
\end{equation}
where the boundaries are calculated as:
\begin{equation}
\mathrm{vis_{min/max}} = \cos{\left( \frac{\pi}{2} \pm R_c \right)} \mp dM,
\end{equation}
the margin \textit{dM} is dependent on the tessellation and should be at least the size of the HEALPix pixel ($\Omega=\frac{\pi}{3\cdot N_{\mathrm{side}}^2}$ in steradians). 

We choose the width of our visibility strip, selection circle, and smoothing function (all related) to be comparable with the collimator PSF (approximated well enough by the Gaussian distribution, however not actually Gaussian itself). It is a good choice due to the measurement geometry, but there is nothing fundamental with regards to the collimator itself that dictates this choice.

\subsection{Selection circle and pixel assignment (no smoothing)}

To specify the region in the sky corresponding to a given viewing direction, we introduce a notion of the \textit{selection circle}, defined to be a region whose radius is set to be the same as that of the collimator (and is the radius of the smoothing function, when one is applied), i.e., $r = R_c$. %
The algorithm assigns the measured counts from each bin to the HEALPix pixels within the corresponding \textit{selection circle}. We first describe the simplest implementation, in which no smoothing is applied. In this case, the bin is assigned to the pixel whose center lies closest to the measurement direction. This provides the conceptual baseline for the algorithm and explains the role of the optional smoothing function introduced later.

\noindent \underline{Per bin, within the orbit arc:}

\begin{figure} [hbt!]
    \centering
    \includegraphics[scale=0.3]{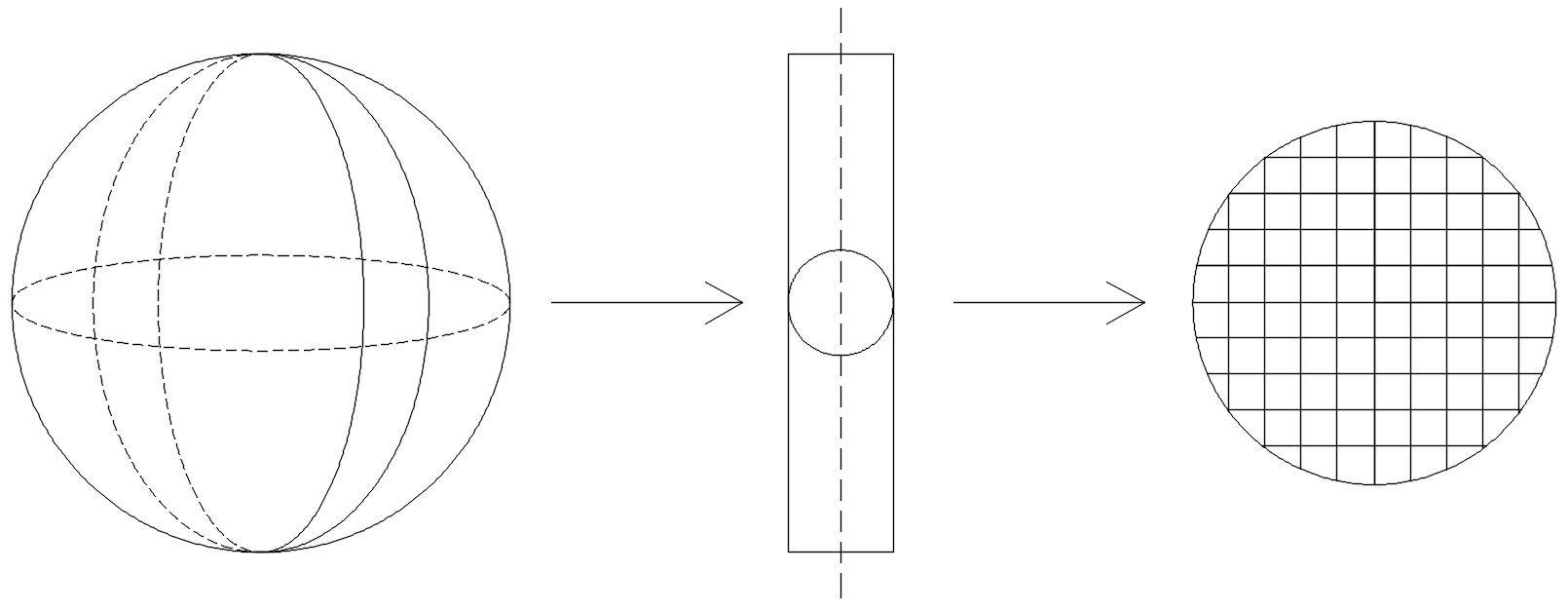}
    \caption{A schematic placing the \textit{selection circle} on a discretized sphere and obtaining the position of an individual point, as needed to calculate the \textit{selection circle} weights. Figure made by Iwona Nowierska.}
    \label{fig:coll3}
\end{figure}

The subset $R_{n,i,-}$ corresponds to the \textit{selection circle} (shown in pink in Figure \ref{fig:hp_coll_two}) and the vectors outside that shape are assigned the value 0 (like in Figure \ref{fig:coll3}).

To now identify the pixels within the \textit{selection circle}, we calculate the \textit{selection circle} center vector along the orbit arc, using the measurements coordinates $\left( l_{n,i}, \theta_{n,i} \right)$ (i.e., the scan direction):
\begin{equation}
\forall n \in \{ 1,2,...,n_o \}, \forall i \in \{ 1,2,...,n_b \} \; \overrightarrow{m_{n,i}} = \left( \cos{\left( l_{n,i} \right)}\cos{\left( \theta_{n,i} \right)}, \sin{\left( l_{n,i} \right)}\cos{\left( \theta_{n,i} \right)}, \sin{\left( \theta_{n,i} \right)} \right),
\end{equation}
and use it to perform a dot product with all the vectors within the visibility strip, restricted to those satisfying the collimator-radius condition:
\begin{equation}
\forall n \in \{ 1,2,...,n_o \}, \forall i \in \{ 1,2,...,n_b \} , \qquad R_{n,i,-} = \{ \overrightarrow{v} \in R_{n,-,-} | \overrightarrow{m_{n,i}} \cdot \overrightarrow{v} \geq \cos{\left( R_c \right)} \}.
\end{equation}
An example visibility strip $R_{n,-,-}$ with the example \textit{selection circle} $R_{n,i,-}$ is visualized in Figure \ref{fig:hp_coll_two}.

Having chosen the \textit{n}-th visibility strip $R_{n,-,-}$ and placed the selection-circle in the center of the \textit{i}-th bin, we find the center pixel by minimizing the angular separation between the measurement vector and pixel vectors:

\begin{equation}
\forall n \in \{ 1,2,...,n_o \}, \forall i \in \{ 1,2,...,n_b \} , \qquad
k_{\mathrm{center}} = \operatorname*{arg\,min}_{k \in \{1,2,\dots,n_{\mathrm{pix}}\}}
\arccos\!\left(
\overrightarrow{m_{n,i}} \cdot \overrightarrow{R_{n,i,k}}
\right).
\end{equation}
With this, we calculate the weights $t_{n,i,k}$ associated with each $\overrightarrow{R_{n,i,k}}$ vector (the HEALPix pixel $I_j$ whose center it represents) using a delta function: 1 for the center pixel (associated with the measurement) and 0 for the rest.
\begin{equation}
  t_{n,i,k} =
    \begin{cases}
      1 & \text{if $k$ = $k_{\mathrm{center}}$}\\
      0 & \text{otherwise}
    \end{cases}       
\end{equation}
To obtain the signal value $s_{n,i,k}$, we multiply the \textit{selection circle} weights $t_{n,i,k}$ by the counts $c_{n,i}$ assigned to bin $(n,i)$ and subtract the correspondingly weighted background contribution $t_{n,i,k} b_{n,i} e_{n,i}$, i.e., $s_{n,i,k} = t_{n,i,k} \; c_{n,i} - (t_{n,i,k} b_{n,i} e_{n,i})$ - multiply weights by Equation~\ref{signal_eq}. Notice that the first term is in the units of counts, whereas the background rate is in the units of counts per time, which needs to be multiplied by the exposure time to obtain the number of background counts before it can be subtracted from the number of counts:
\begin{equation}
\forall n \in \{ 1,2,...,n_o \}, \forall i \in \{ 1,2,...,n_b \}, \forall k \in \{ 1,2,...,n_{\mathrm{pix}} \} , \qquad
s_{n,i,k}
  = t_{n,i,k}\,\bigl(c_{n,i} - b_{n,i}\,e_{n,i}\bigr).
\end{equation}
Similarly, we multiply the exposure times by the \textit{selection circle} weights to obtain directional exposure values. 

As mentioned before, each vector $\overrightarrow{R_{n,i,k}}$ (with its associated value $v_{n,i,k}$) can be mapped to the HEALPix vector $\overrightarrow{I_j}$ (and conversely, each HEALPix vector $\overrightarrow{I_j}$ can be mapped to the $\overrightarrow{R_{n,i,k}}$ vector with the inverse function), as explained in the beginning of this Section. We perform that procedure to fill the event count list $c_{n,j}$, the exposure time list $e_{n,j}$, and the signal list $s_{n,j}$; one pass through this procedure produces the partial map corresponding to a single orbit arc.

An important detail is the treatment of the signal after the application of the smoothing kernel: pixels with no geometric coverage are stored as Null, while pixels that were observed but whose background-subtracted signal is not positive are assigned a value of zero. This preserves the distinction betwxeen ``unobserved'' and ``observed but background-dominated'' regions and ensures that physical quantities such as rates and intensities remain non-negative.

\subsection{Smoothing kernel} \label{sec:smoothing}
The baseline product is a ``boresight''-type map -- all detected counts are first binned into 1$\arcdeg$ wide bins chosen based on the look direction of the instrument boresight at the detection time of each count, and then each bin is assigned to the HEALPix pixel that contains the bin-averaged look direction. This approach preserves the native Poisson statistics of the data but also any sampling limitations, such as potential incomplete spatial coverage. The smoothed map is generated by redistribution of each count over neighboring pixels according to a user-specific kernel. This kernel-application step is an explicit smoothing choice, introduced to fill gaps and suppress small-scale mottling.

For generality, the core body of the pipeline looks the same regardless of whether the user decides to use a smoothing function or not. If the \textit{smoothing} flag has not been set as the input parameter, it will receive the \textsc{Null} value by default. Currently, the user can specify \textit{smoothing} $\rightarrow$ \textit{Gaussian}, to choose the Gaussian kernel truncated at an angular radius $\alpha \leq R_c$, but it is very straightforward to implement one's own kernel function. This flag will determine what kind of operation will be performed on the $R_{n,i,-}$ pixels. If the smoothing flag is set to \textit{Gaussian}, all pixels within the \textit{selection circle} will be assigned a weight according to the normal distribution and then this template would be multiplied by the measurement values.

Having chosen the \textit{n}-th visibility strip $R_{n,-,-}$ and centered the selection-circle on the measurement direction of the \textit{i}-th bin, we calculate the smoothing values associated with each $\overrightarrow{R_{n,i,k}}$ vector (the HEALPix pixel $I_j$ whose center it represents).

We first define the angular separation between the bin (measurement) vector
$\overrightarrow{m_{n,i}}$ and each pixel vector
$\overrightarrow{R_{n,i,k}}$:
\begin{equation}
\forall n \in \{ 1,2,...,n_o \}, \forall i \in \{ 1,2,...,n_b \}, \forall k \in \{ 1,2,...,n_\mathrm{pix} \} , \qquad
l_{n,i,k}
= \arccos\!\left(
    \overrightarrow{m_{n,i}} \cdot \overrightarrow{R_{n,i,k}}
  \right).
\end{equation}
The weight for pixel $k$ in bin $(n,i)$ is then
\begin{equation}
\forall n \in \{ 1,2,...,n_o \}, \forall i \in \{ 1,2,...,n_b \}, \forall k \in \{ 1,2,...,n_\mathrm{pix} \} , \qquad
t_{n,i,k} = \mathrm{coll} \!\left(l_{n,i,k}\right),
\end{equation}
where the function $\mathrm{coll}$ is dependent on the smoothing function chosen by the user. 


If the smoothing function is used, the raw angular kernel values are evaluated at
the HEALPix pixel centers inside the \textit{selection circle}. These raw values are then renormalized as discrete weights,
\[
w_{n,i,k} =
\frac{g(\alpha_{n,i,k})}
{\sum_{\ell\in R_{n,i,-}} g(\alpha_{n,i,\ell})},
\]
so that
\[
\sum_{k\in R_{n,i,-}} w_{n,i,k}=1.
\]
The weights are therefore dimensionless fractions of the bin contribution assigned
to each HEALPix pixel (not an intensity per steradian). The signal, counts, and
exposure associated with bin $(n,i)$ are distributed using the same weights,
\[
s_{n,i,k}=w_{n,i,k}\left(c_{n,i}-b_{n,i}e_{n,i}\right),
\qquad
e_{n,i,k}=w_{n,i,k}e_{n,i},
\]
and analogously for the counts. 

Consequently, if the smoothing function has been used, we need to divide the \textit{selection circle} weights $t_{n,i,k}$ by the appropriate normalization factor. To avoid introducing too many subscripts, particularly since one does not need to introduce the smoothing function, we will continue using $t_{n,i,k}$ to denote the \textit{selection circle} weights. 

To reiterate, if the \textit{smoothing} flag is not set, meaning that the user does not want to use a smoothing function, the special case of the $\mathrm{coll}$ function will be called on the \textit{selection circle} pixels -- the delta function. There is no need to perform normalization, as the weights of the \textit{selection circle} pixels sum to 1. After multiplying with the counts, backgrounds, and exposures, the only pixel with a non-zero signal will be the central one.

The subsequent calculation of the signal, exposure time, and derived quantities proceeds identically to the no-smoothing case, with the only difference being the use of the kernel weights $t_{n,i,k}$.

\subsection{Final map creation}
The procedure presented in Section \ref{sec:smoothing} is performed separately for every orbit arc, in effect producing \textit{n} all-sky  distribution maps of counts, exposure times, and signals, each holding values only for one orbit arc. Now, we perform the superposition of the previously constructed lists:
\begin{align*}
\forall j \in \{ 1,2,...,N_{\mathrm{pix}} \},  \qquad c'_j &= \sum_n c_{n,j}, \\
\forall j \in \{ 1,2,...,N_{\mathrm{pix}} \}, \qquad e'_j &= \sum_n e_{n,j}. \\
\forall j \in \{ 1,2,...,N_{\mathrm{pix}} \},  \qquad s'_j &= \sum_n s_{n,j}. \\
\end{align*}

Having done that, we can now calculate the signal rates and intensity maps (a quantity expressed in the units of $\# \;\text{cm}^{-2}\text{s}^{-1}\text{sr}^{-1}\mathrm{keV}^{-1}$):
\begin{align*}
\forall j \in \{ 1,2,...,N_{\mathrm{pix}} \} , \qquad r'_j &=  \frac{s'_{j}}{e'_{j}}, \\
\forall j \in \{ 1,2,...,N_{\mathrm{pix}} \} , \qquad f'_j &=  \frac{s'_{j}}{e'_{j}\cdot G C_e}, \\
\end{align*}
where $G$ is the geometric factor for the specific energy channel in $\mathrm{cm^2sr}$, and $C_e$ is the central energy value for that specific energy channel in $\mathrm{keV}$. In keeping with our convention, we are not using separate indices for energy channels. The values for $G$ and $E$ for IBEX-Hi energy channels can be found in Table \ref{tab:G_E_values}, where the energy channels are specific ranges of particle energies that an instrument, in our case IBEX-Hi, groups together to measure how many particles fall within that specific energy interval. 

\begin{table}[h!]
\centering
\caption{Values of the geometric factor $G$ and corresponding energy values $E$ for different energy channels.}
\begin{tabular}{ccc}
\hline
\hline
ESA & $G$ & $C_e$ (keV) \\
\hline
1 & 0.00013 & 0.45 \\
2 & 0.00037 & 0.71 \\
3 & 0.00073 & 1.10 \\
4 & 0.00140 & 1.74 \\
5 & 0.00250 & 2.73 \\
6 & 0.00420 & 4.29 \\
\hline
\end{tabular}
\label{tab:G_E_values}
\end{table}

\section{\textbf{Postprocessing}} \label{sec:postprocessing}

The reconstruction described in Section~\ref{sec:algorithm} produces the intensity, rate, signal, count, and exposure maps on the native HEALPix grid. The reconstructed background-subtracted signal rates and derived flux/intensity maps are divided by the detector-efficiency factor, accounting for the gradual decrease in the detector efficiency
during the mission. For the 2018 maps, the value of that factor is $\epsilon_{\rm eff}=0.96$. The postprocessing stage has two separate purposes. First, we define normalized intensity maps used for comparison and interpretation. Second, we transform maps between the native HEALPix grid and rectangular comparison grids when direct comparison with IBEX and \textsc{THESEUS} products is needed.

In the first stage, we can apply two complementary normalization procedures to the (ENA) intensity maps, to separate purely geometric effects from instrumental and physical ones. These schemes define the \textit{relative intensity} and the \textit{GDF-normalized} (globally distributed flux) intensity. Here we use the term GDF-normalized intensity, although the historical IBEX literature often refers to the distributed component as the flux. The GDF is the relatively smooth ENA emission distributed across the entire sky, primarily originating from the inner heliosheath \citep{mccomas_etal:09b, McComas_2020}. 
Each method is designed to answer a distinct question about the sky distribution and therefore uses a different weighting or scaling approach.  
Both were applied to: 1) the HEALPix maps produced by the \textsc{Capra} pipeline, 2) the maps obtained from the \textsc{Theseus} \citep{Osthus02042024} mapping algorithm, 3) the baseline IBEX-Hi maps used to create the HEALPix maps (for consistency checks). 

We use \textit{relative normalization} for relative algorithmic comparison, when the goal is to check whether spatial patterns (ribbon morphology, contrast, and gradients) agree between maps, independent of the absolute calibration. This normalization does not change the morphology of the map and should not be interpreted as an absolute calibration; it only places maps with different mean intensity levels on a common dimensionless scale.
We use \textit{GDF-normalization} for physical validation against each dataset’s own baseline, testing how well each reproduces its internal off-Ribbon levels and the Ribbon contrast. The Ribbon is a narrow, arc-shaped region of enhanced ENA emission superimposed on the GDF, believed to form outside the heliopause, where the interstellar magnetic field is perpendicular to the line of sight \citep{McComas_2020,Schwadron2009,funsten_etal:09b}. We treat the GDF as a baseline because it represents the large-scale emission, while the Ribbon is a localized signal enhancement; normalizing by the GDF isolates the Ribbon contrast independent of the absolute intensity, and therefore allows us to evaluate the physical contrast between the Ribbon and GDF emissions.


\subsection{Normalized intensity}

\subsubsection{Relative normalization} \label{subsec:relnorm}

\underline{Part 1}, or the \textit{relative normalization} routine computes the solid-angle-weighted all-sky mean intensity,
\begin{equation}
\label{eq:rel_mean}
\langle F \rangle_\Omega = 
\frac{\sum_{j\in\mathcal{V}} F_j \, \Omega_j}
     {\sum_{j\in\mathcal{V}} \Omega_j},
\end{equation}
where 
\[
\mathcal{V} = \{j \mid F_j \in \mathbb{R}\}.
\]
is the set of valid (observed) cells, not necessarily the full sky if some cells are Null. The mean $\langle F \rangle_\Omega$ has the same units as $F$, i.e., $[\# \; s^{-1}cm^{-2}sr^{-1}\mathrm{keV^{-1}}]$. Dividing the map by this mean,
\begin{equation}
F_\mathrm{rel}(j) = \frac{F_j}{\langle F \rangle_\Omega}
\end{equation}
gives us a dimensionless map whose area-weighted mean is unity.

In the case of \textsc{Theseus}, the IBEX data (or \textsc{Capra} -- often refereed to as\textsc{HP} throughout the paper for simplicity, data that have been transformed to a rectangular grid), pixels are not equi-areal, and therefore do not have equal weights. We can compute them using Equation \ref{weights}, because the field sizes will vary with latitude, but will remain constant with given latitude strips (see Section \ref{subsec:gridtransformation} and Figure \ref{fig:spher}).

\subsubsection{GDF-normalized intensity} \label{subsec:gdfnorm}

\underline{Part 2} or the \textit{GDF-normalized} routine determines a dataset-specific physical baseline (a reference level) for each intensity map using the corresponding exposure map. We use this normalization to compare each map relative to its own quiet off-Ribbon reference level. Each map is divided by characteristic low-intensity reference level (referred to as \textit{baseline}) estimated from a ``quiet'' off-Ribbon region.

We define $M_j$, as an off-Ribbon reference mask. It is chosen to select a quiet reference region rather than all non-Ribbon sky. We exclude pixels within $30^\circ$ of the fitted Ribbon locus, remove the nose and tail directions, where large-scale heliospheric structure can bias the baseline even when the pixels are geometrically far from the Ribbon, and apply a high-intensity percentile cut to suppress residual bright structures (70th percentile). The resulting sample is, therefore, a conservative off-Ribbon reference region used only to define the normalization baseline. $M_j = 1$ means that a pixel is included in the reference sample (is off-Ribbon) and $M_j=0$ denotes the excluded pixels.
We also define a set of good pixels
\[
\mathcal{I} = \{\, j \mid F_j \in \mathbb{R},~E_j > 0 \,\}
\]
and then select only those that are off–ribbon:
\[
\mathcal{O} = \{\, j \in \mathcal{I} \mid M_j = 1 \,\},
\]
as well as their intensities and weights (the exposure times):
\[
F_\mathcal{O} = \{F_j\}_{j\in\mathcal{O}}, \qquad 
W_\mathcal{O} = \{E_j\}_{j\in\mathcal{O}}.
\]
The baseline is defined as the exposure-weighted median of the off-ribbon intensity distribution.  
Because the off-ribbon intensity distribution is typically asymmetric and can have long tails, the median provides a more robust measure of the characteristic reference level than the mean.  

Sorting the pairs $(F_j,E_j)$ by increasing $F_j$ and computing the cumulative exposure
\[
C_k = \sum_{j=1}^{k} E_{(j)}, \qquad 
C_\mathrm{tot} = \sum_{j=1}^{n} E_{(j)},
\]
the baseline intensity (exposure-weighted median baseline) is the first value whose cumulative exposure exceeds half the total exposure:
\begin{equation}
s_\mathrm{ESA} =
\begin{cases}
F_{(m)} \;\text{such that}\;
\min \left\{ k :
\frac{C_k}{C_\mathrm{tot}} \ge \frac{1}{2}
\right\},
& \text{if valid exposures exist;}\\[10pt]
\mathrm{median}(F_\mathcal{O}),
& \text{otherwise.}
\end{cases}
\end{equation}

All pixels are then rescaled by this baseline:
\begin{equation}
F_j^{(\mathrm{GDF})} = \frac{F_j}{s_\mathrm{ESA}},
\end{equation}
producing a dimensionless map. By construction, the exposure-weighted median of the normalized off-ribbon reference sample is unity:
\begin{equation}
\operatorname{wmed}_{E}
\left(F^{(\mathrm{GDF})}_{\mathcal{O}}\right)=1.
\end{equation}

For maps kept in the native HEALPix discretization, no additional solid-angle weighting is applied in the baseline calculation because HEALPix pixels are equal-area. For maps transformed to the rectangular comparison grid, the same procedure is applied within the gridded representation: the physical intensity and exposure maps are first transformed to the common grid, the off-ribbon reference mask is constructed on that same grid, and the exposure-weighted median baseline is computed from the same gridded pixels that are normalized. This avoids mixing baselines between discretizations.

\underline{Diagnostics}

The primary diagnostic for the GDF normalization is the exposure-weighted median of the normalized off-ribbon reference sample,
\[
\operatorname{wmed}_{E}
\left(F^{(\mathrm{GDF})}_{\mathcal{O}}\right),
\]
which should equal unity by construction. We also report the unweighted mean,
\[
\frac{1}{|\mathcal{O}|}
\sum_{j\in\mathcal{O}} F_j^{(\mathrm{GDF})},
\]
and the exposure-weighted mean,
\[
\frac{\sum_{j\in\mathcal{O}} E_j F_j^{(\mathrm{GDF})}}
{\sum_{j\in\mathcal{O}} E_j},
\]
as descriptive statistics. These means are not expected to be equal to 1 because the GDF normalization is based on the exposure-weighted median rather than on an arithmetic or exposure-weighted mean. Significant deviations of the exposure-weighted median from unity would indicate a baseline mismatch, mask inconsistency, or rescaling error.

\subsection{Grid transformation}
\label{subsec:gridtransformation}

Because we project HEALPix (equal-area) data onto a grid composed of unequal-area cells, each cell must represent the correct solid-angle contribution. The transformation uses area-weighted averages to ensure that the resulting intensity density or rate retains its proper surface-brightness meaning and that the full-sky normalization is conserved.
In this paper we focus on rectangular grids, though the procedure is general and can be extended to arbitrary cell geometries (e.g., triangular, hexagonal). The definition of the cell shape enters through the classification function that assigns coordinates to regions of the sphere. For this reason, the pipeline includes a modular functionality allowing transformations from HEALPix maps to arbitrary grid geometries, enabling visualization with standard Python packages such as \textit{matplotlib.pyplot}.

For the rate and intensity maps, we use area-weighted averages when transforming already-computed maps. For products derived from counts and exposures, we instead transform the extensive quantities first and recompute the rate or intensity afterward.

\subsubsection{Rectangular grid geometry}

For a rectangular uniform grid, the all-sky matrix has dimensions $(m \times n)$ defined by:
\begin{equation}
\left(m,n\right) = \left(\frac{180}{\mathrm{resolution}_a},\,\frac{360}{\mathrm{resolution}_b}\right),
\end{equation}
where $\mathrm{resolution}_{a,b}$ are the latitudinal and longitudinal angular resolutions of a single grid cell. The total number of cells is thus
\begin{equation}
N_\mathrm{cell} = m \times n.
\end{equation}

Since the cell areas vary with latitude but remain constant within a given latitude strip, we calculate the area for each strip prior to computing the intensity densities. The number of latitudinal strips is $m$. Each strip is composed of cells with solid angles $d\omega$, which vary as a function of latitude $\theta$ according to:
\begin{equation} \label{weights}
\forall l \in \{1,2,...,N_\mathrm{cell}\}, \qquad
d\omega_l = \sin{\left(\theta_l\right)}\, d\theta\, d\phi,
\end{equation}
where $\theta_l$ is the colatitude of the $l$-th cell center, $\theta$ is the colatitude and $(d\theta,d\phi)$ are defined by $(\mathrm{resolution}_a,\mathrm{resolution}_b)$ respectively.  

\begin{figure}[htb!]
  \centering
  \includegraphics[width=0.7\textwidth]{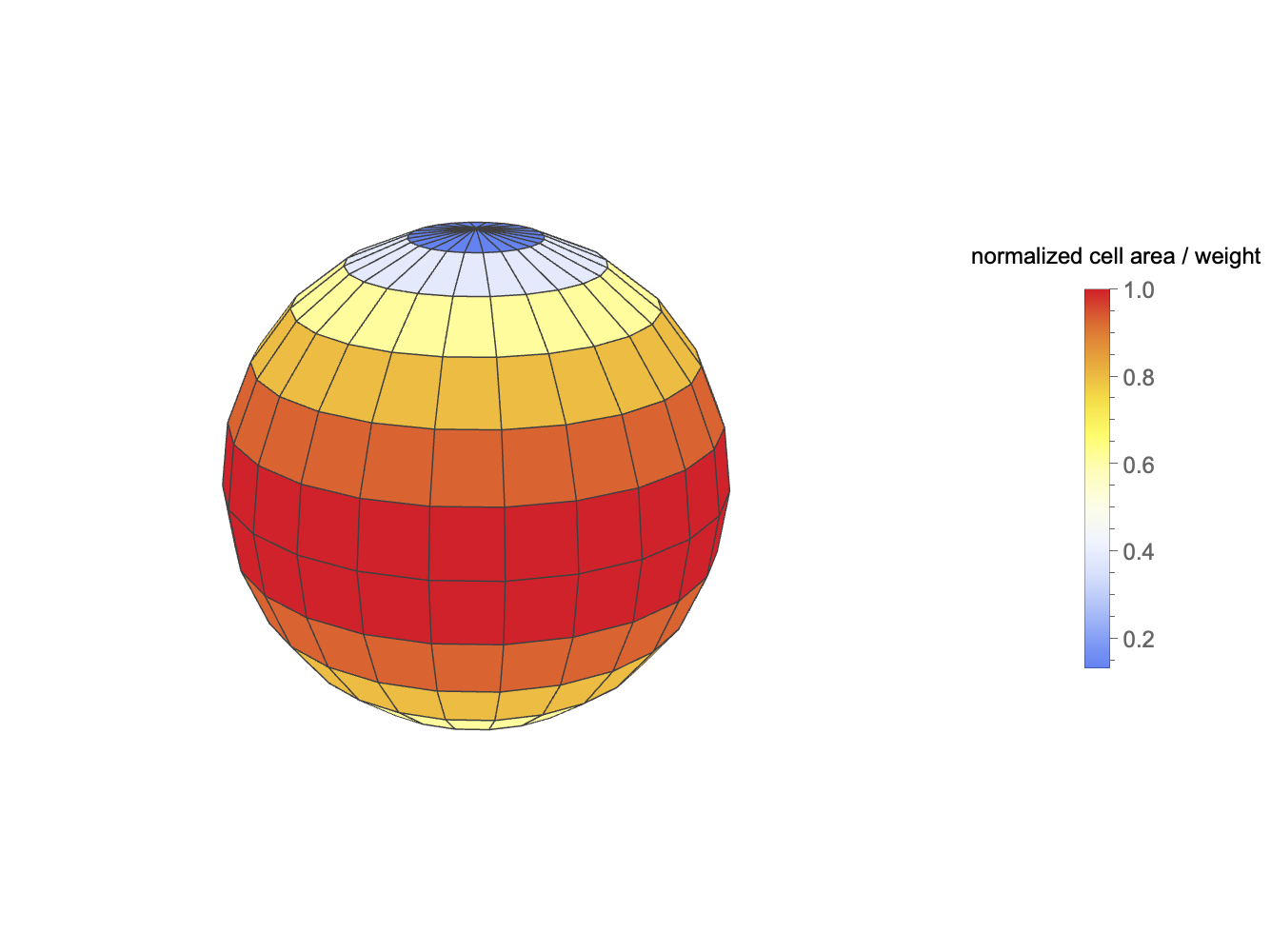}
  \caption{The cells are colored by their normalized solid-angle weights (cell areas divided by the maximum cell area). This illustrates that although the grid spacing is uniform in angular coordinates, the cell area is latitude-dependent: it is largest near the equator and decreases toward the poles.}
  \label{fig:spher}
\end{figure}

Attaching the $d\omega_l$ values horizontally creates an $(m\times n)$ matrix of the solid-angle weights of the cells, visualized schematically in Figure~\ref{fig:spher}.

\subsubsection{Aggregation of intensive and extensive quantities}

HEALPix pixels are assigned to these rectangular cells using a classification function that maps each pixel’s coordinates $(\theta_j,\phi_j)$ to its corresponding cell indices $(r,c)$ on the grid:
\begin{equation}
r = 1 + \left\lfloor\frac{\theta_j}{\Delta\theta}\right\rfloor, \qquad
c = 1 + \left\lfloor\frac{\phi_j}{\Delta\phi}\right\rfloor,
\end{equation}
with $(\Delta\theta,\Delta\phi)$ being the latitudinal and longitudinal grid spacings.  
This mapping partitions the HEALPix pixel set into disjoint subsets $\mathcal{I}_{r,c}$ (representing cells), each containing all pixels whose centers fall within the cell $(r,c)$.

For equal-area HEALPix input, each pixel has the same solid angle
\begin{equation}
\omega_\mathrm{pix} = \frac{\pi}{3 N_{\mathrm{side}}^2},
\end{equation}
hence for the intensity or rate maps, the rectangular-cell value reduces to the simple arithmetic mean of all included pixels:
\begin{equation}
\forall (r,c), \quad
M_{r,c} = \frac{1}{N_{r,c}} \sum_{j \in \mathcal{I}_{r,c}} f_j,
\end{equation}
where $f_j$ are the pixel values (intensity or rate), $N_{r,c}=|\mathcal{I}_{r,c}|$, and it is already an area-consistent average, i.e., it already gives an area-consistent estimate of the surface brightness.
For extensive quantities such as exposure, counts, or signal, the cell value is instead given by
\[
M_{r,c}^{(\mathrm{sum})}
=
\sum_{j\in\mathcal{I}_{r,c}} f_j.
\]

For binned IBEX or \textsc{Theseus} data, the input bins may represent finite sky areas that vary with latitude. In that case, the intensity or rate values are aggregated using the source-bin solid angles \(\omega_j\):
\[
M_{r,c} =
\frac{\sum_{j\in\mathcal{I}_{r,c}} f_j\,\omega_j}
     {\sum_{j\in\mathcal{I}_{r,c}} \omega_j}.
\]
The exposure is summed as follows:
\[
E_{r,c} =
\sum_{j\in\mathcal{I}_{r,c}} E_j.
\]
We define a \textit{target grid/cell} as the rectangular grid/grid cell, and an \textit{input grid/cell} as an HP/IBEX/\textsc{THESEUS} gird/ grid cell.
The target-grid solid angles \(\Delta\Omega_{r,c}\) are then used for global diagnostics and for relative normalization of the completed rectangular map, not as additional weights in the target-cell aggregation. 

\underline{Area-weighted mean and normalization} \label{subsec:areaweight}

Because cells closer to the poles represent smaller surface areas, we compute their solid-angle weights using:
\begin{equation}
\Delta\Omega_{r,c} \simeq \sin{\theta_r}\,\Delta\theta\,\Delta\phi,
\label{eq:solidangle}
\end{equation}
where $\theta_r$ is the central latitude of the $r$-th latitudinal row.  
The area-weighted global mean of a map $M_{r,c}$ is then defined as:
\begin{equation}
\overline{M}_\Omega = 
\frac{\sum_{r,c} M_{r,c}\,\Delta\Omega_{r,c}}
     {\sum_{r,c} \Delta\Omega_{r,c}}.
\end{equation}
If a relative map is desired (with the mean equal to unity), we normalize as:
\begin{equation}
M^{(\mathrm{rel})}_{r,c} = 
\frac{M_{r,c}}{\overline{M}_\Omega}.
\end{equation} \label{eq:rel_norm}
In the pipeline, grid transformation and normalization are performed as separate operations. The gridding functions preserve the physical scale of the input map, while the relative-normalization routine applies Eq.~(\ref{eq:rel_norm}) afterward using the rectangular-grid solid-angle weights.

\subsubsection{Extension to \textsc{Theseus} and IBEX binned data}

The same idea applies to the \textsc{Theseus} or IBEX binned observations, where the input data are the binned points $(\theta_j,\phi_j,v_j)$. Note that the \textsc{Theseus} pixels contain the average intensity integrated over a pixel, rather than the centered intensity. The algorithm fits a continuous 2D surface function to the binned data and averages it over the pixel's solid angle. The classification and aggregation steps are identical, but the aggregation operation differs depending on the physical quantity:
\begin{equation}
A_\mathrm{op}(\{v_j\}_{j\in\mathcal{I}_{r,c}}) =
\begin{cases}
\dfrac{\sum\limits_{j\in\mathcal{I}_{r,c}} v_j\,\omega_j}
      {\sum\limits_{j\in\mathcal{I}_{r,c}} \omega_j},
& \text{for the intensity/rate}, \\[1.0em]
\sum\limits_{j\in\mathcal{I}_{r,c}} v_j,
& \text{for the exposure, counts, or signal},
\end{cases}
\end{equation}
where $\omega_j$ is the source-bin solid angle. For equal-area source pixels, such as native HEALPix pixels, the first expression reduces to the arithmetic mean.

The target-grid solid angles $\Delta\Omega_{r,c}$ are used for global diagnostics and for relative normalization of the final rectangular map. They are not used as weights inside the target-cell aggregation. When multiple finite input bins contribute to a target cell, their source-bin solid angles $\omega_j$ are used to compute the source-area-weighted mean intensity.

\subsubsection{Diagnostic checks}

The gridding routines report diagnostic quantities including the total target solid angle, the fractional coverage relative to $4\pi$, the painted (observed) target area, source-conserved means, painted-target means, and cell-coverage statistics, where appropriate.
For all datasets, the integrated solid angle equals $4\pi$ within $0.05\%$, and the normalization check yields $1.000\pm0.001$. These diagnostics confirm that the rectangular-grid geometry is represented consistently. Conservation of the intensity scale is checked separately by comparing native, source-conserved, and painted-target means for the intensity maps, and by comparing exposure-weighted global fluxes before and after gridding.
The global means reported by the gridding diagnostics are consistency checks, not normalization targets. They should remain stable between equivalent transformations, but they are not expected to equal unity unless an explicit relative normalization has been applied. Exposure maps are physical observing-time maps and, therefore, are checked for geometric consistency and conservation under gridding, not for unit normalization.

\section{Results} \label{subsec:results}

In this Section, we use the postprocessing definitions from Section~\ref{sec:postprocessing} to compare the maps produced by \textsc{Capra} with the baseline IBEX maps and with the \textsc{THESEUS} reconstruction. 
We show three types of the intensity maps. The un-postprocessed maps retain the physical intensity scale and are used to check the direct output of the reconstruction. 
The relative maps, defined in Section~\ref{subsec:relnorm}, divide each map by its solid-angle-weighted mean and are therefore useful for comparing the morphologies independent of absolute normalization. The GDF-normalized maps, defined in Section~5.1.2, divide each map by its own off-Ribbon reference level and are used to compare the Ribbon-to-background contrasts.

\textsc{Capra} maps are shown either on the native HEALPix grid or after transformation to a common $6^\circ \times 6^\circ$ rectangular grid. 
The native HEALPix representation is used when showing the direct output of the pipeline, while the rectangular grid is used when we do direct pixel-wise comparison with IBEX and \textsc{THESEUS} maps. All \textsc{Capra} maps shown in this section were generated using the Gaussian smoothing kernel described in Section~\ref{sec:smoothing}. The code has been made open-source \citep{Bukowiecka_Capra}.

\subsection{\textsc{Capra} maps across energy channels}

\begin{figure*}[ht!]
  \centering
  \setlength{\tabcolsep}{2pt}

  \begin{subfigure}[b]{0.28\textwidth}
    \includegraphics[width=\textwidth]{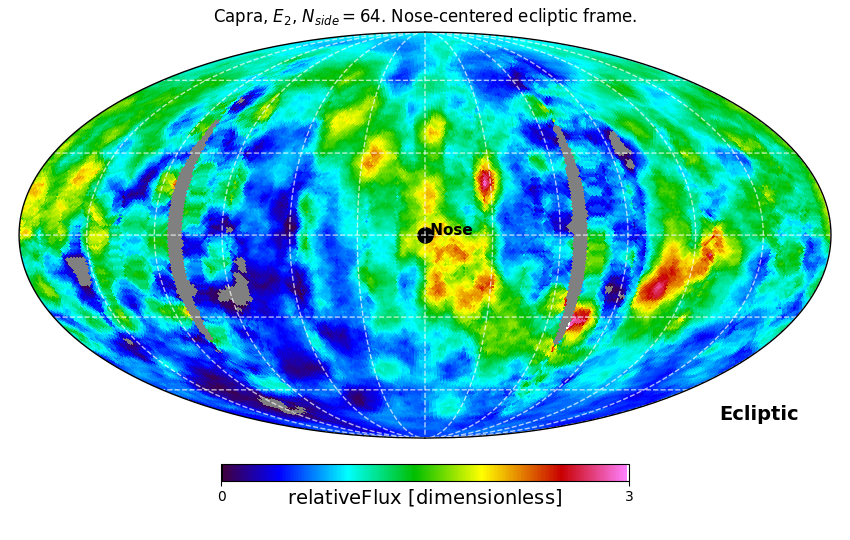}
    \caption{\textsc{Capra}, ESA 2}
  \end{subfigure}\hfill
  \begin{subfigure}[b]{0.28\textwidth}
    \includegraphics[width=\textwidth]{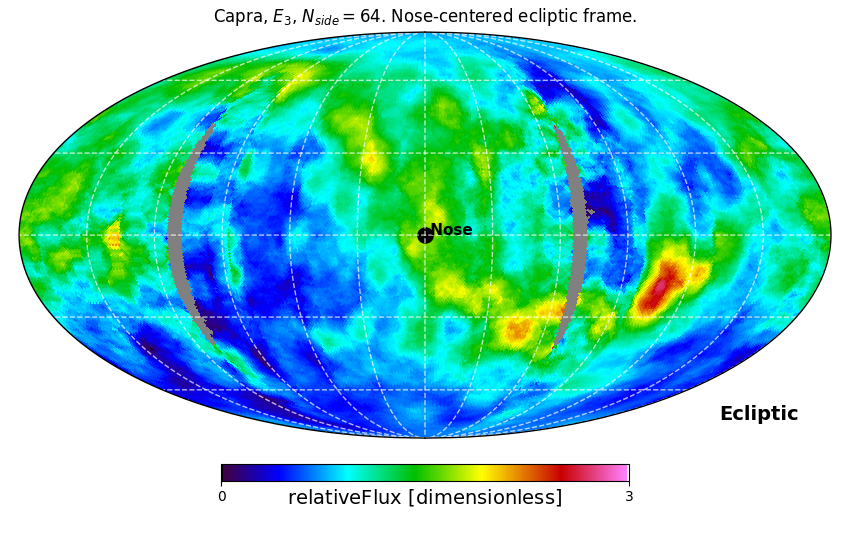}
    \caption{\textsc{Capra}, ESA 3}
  \end{subfigure}\hfill
  \begin{subfigure}[b]{0.28\textwidth}
    \includegraphics[width=\textwidth]{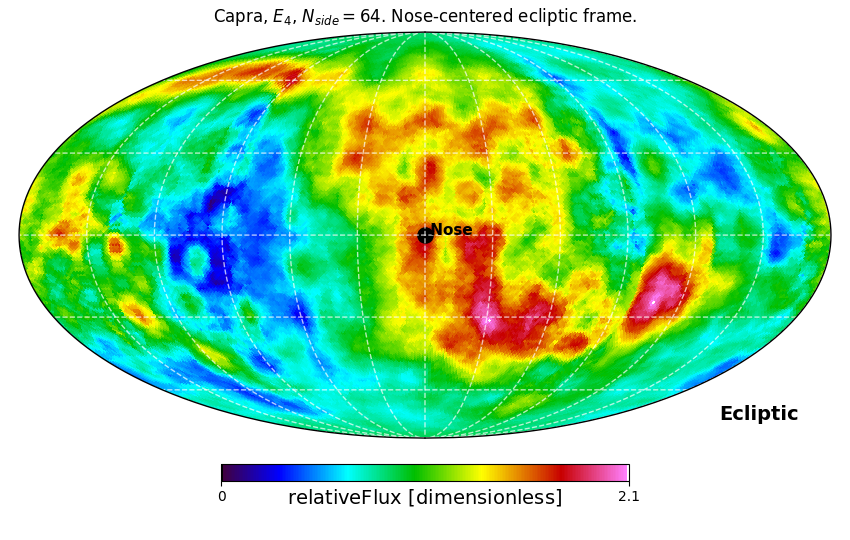}
    \caption{\textsc{Capra}, ESA 4}
  \end{subfigure}

  \vspace{-0.3em}

  \makebox[\textwidth]{%
    \begin{subfigure}[b]{0.28\textwidth}
      \includegraphics[width=\textwidth]{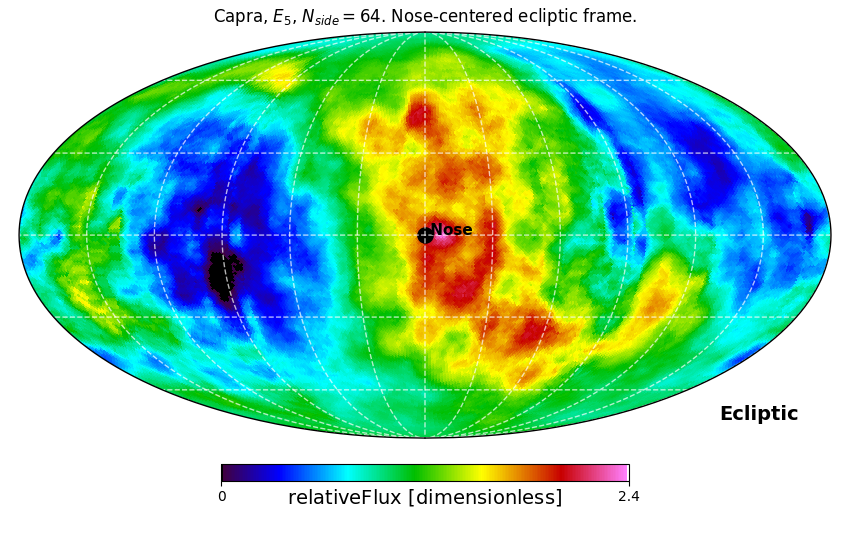}
      \caption{\textsc{Capra}, ESA 5}
    \end{subfigure}
    \hspace{2em}%
    \begin{subfigure}[b]{0.28\textwidth}
      \includegraphics[width=\textwidth]{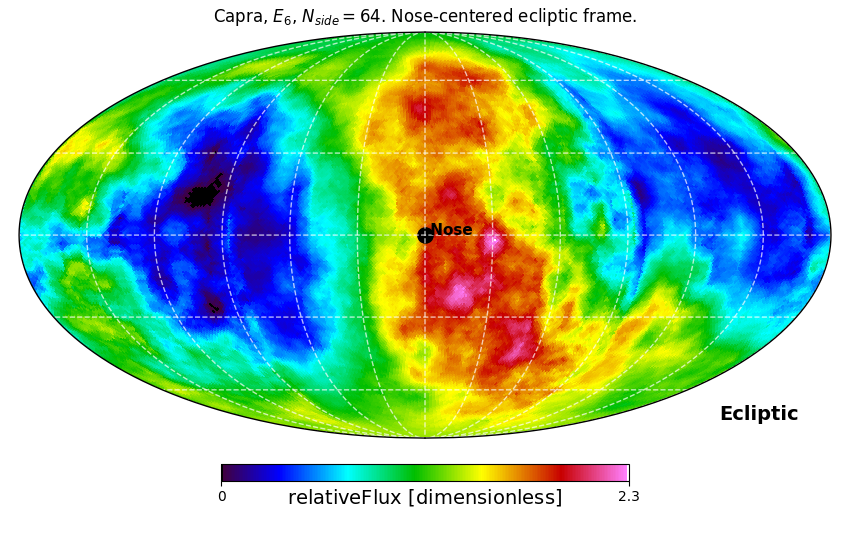}
      \caption{\textsc{Capra}, ESA 6}
    \end{subfigure}
  }

  \caption{
    All-sky \textsc{HP} \textit{relative normalized} intensity maps for ESA steps 2–6.
    Each row corresponds to a distinct IBEX-Hi energy channel
    (0.71–4.29 keV). All panels are in ecliptic nose-centered coordinates
    (Mollweide projection); masked (grey) regions indicate unobserved pixels.
  }
  \label{fig:hp_allesa}
\end{figure*}

Figure~\ref{fig:hp_allesa} shows the \textsc{Capra} relative intensity maps for ESA steps 2--6. 
We use relative normalization here because the goal is to compare the large-scale morphology across energy channels, rather than the absolute intensity scale of each channel. 
This normalization keeps the spatial structure and contrast within each map while removing the channel-dependent mean level.

The reconstructed maps recover the main large-scale ENA structures, including the Ribbon and the broad nose-centered enhancement. The Ribbon morphology remains visible across the energy range, while its contrast and detailed appearance vary with ESA step, as expected for IBEX-Hi ENA maps. The masked regions correspond to sky regions with invalid or unobserved pixels.

\subsection{Comparison on a common rectangular grid}

To compare \textsc{Capra} directly with the baseline IBEX maps and the \textsc{THESEUS} products, all three datasets are transformed to the same $6^\circ \times 6^\circ$ rectangular grid. 
This removes differences due only to the plotting grid and allows the remaining differences to be interpreted as differences between the input products or reconstruction methods.

\subsubsection{Un-postprocessed intensity}

\begin{figure*}[p]
\centering

\small
\textbf{\textsc{Capra} HP} \hfill \textsc{THESEUS} \hfill \textsc{IBEX}

\vspace{2pt}

\begin{subfigure}[t]{0.315\textwidth}
\centering
\mapimg{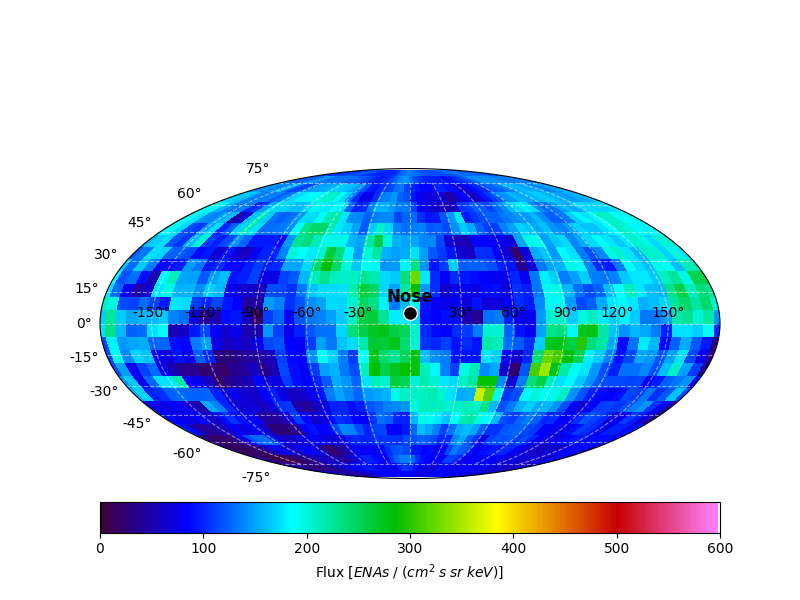}
\end{subfigure}\hfill
\begin{subfigure}[t]{0.315\textwidth}
\centering
\mapimg{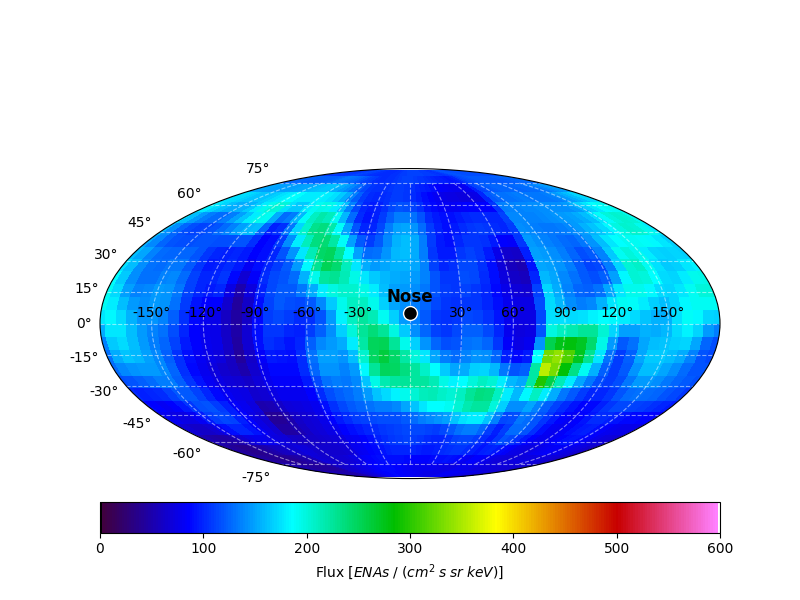}
\end{subfigure}\hfill
\begin{subfigure}[t]{0.315\textwidth}
\centering
\mapimg{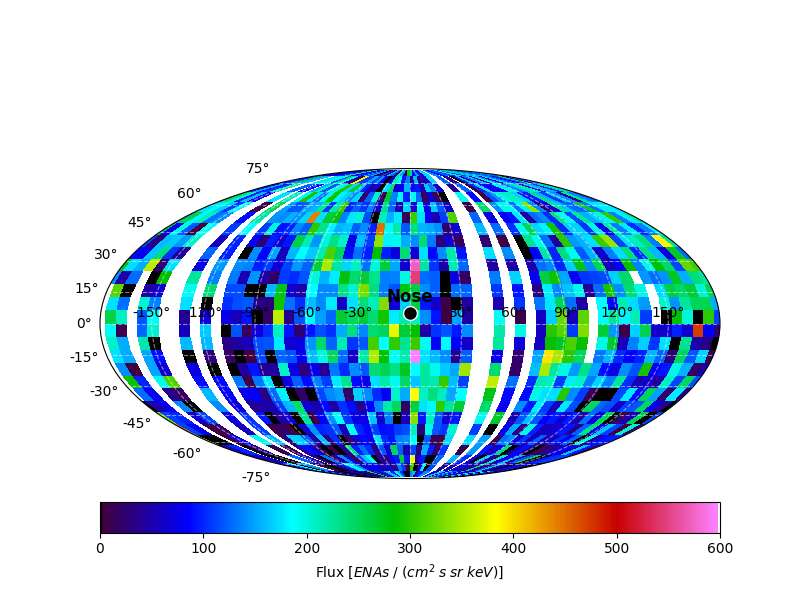}
\end{subfigure}

\vspace{1pt}

\begin{subfigure}[t]{0.315\textwidth}
\centering
\mapimg{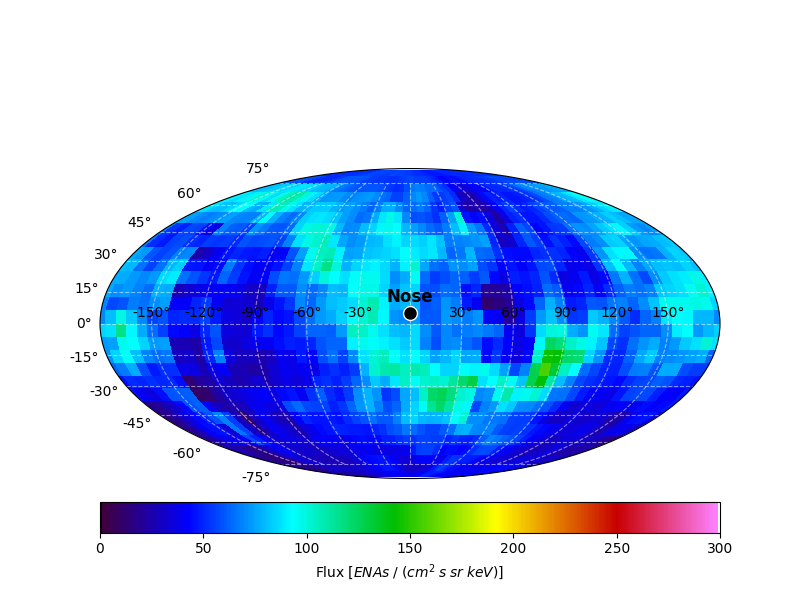}
\end{subfigure}\hfill
\begin{subfigure}[t]{0.315\textwidth}
\centering
\mapimg{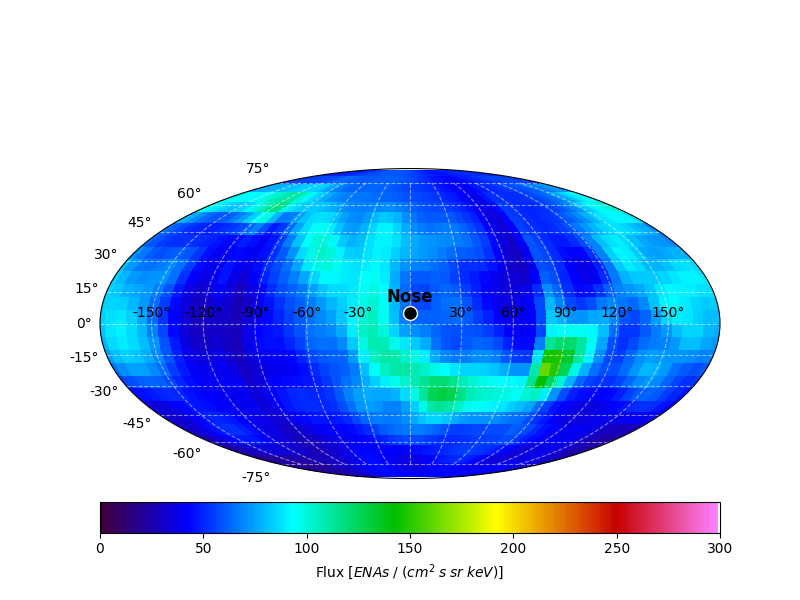}
\end{subfigure}\hfill
\begin{subfigure}[t]{0.315\textwidth}
\centering
\mapimg{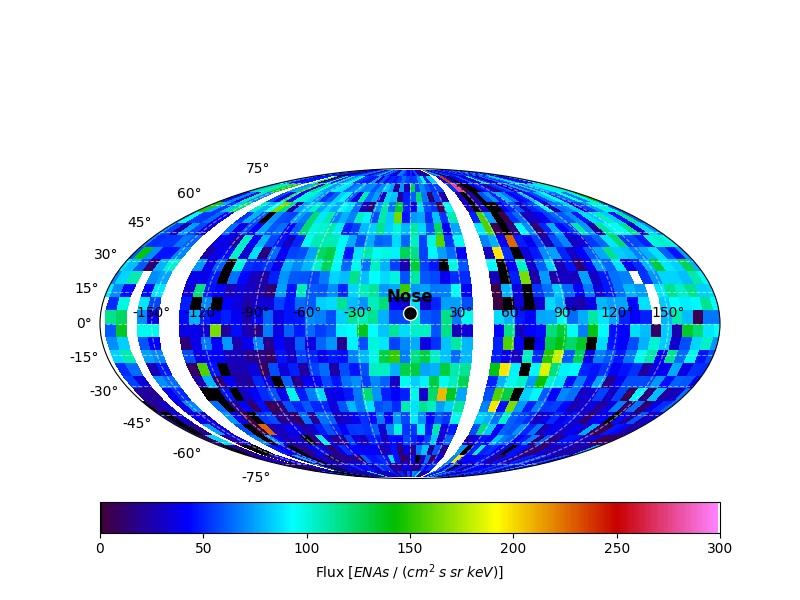}
\end{subfigure}

\vspace{1pt}

\begin{subfigure}[t]{0.315\textwidth}
\centering
\mapimg{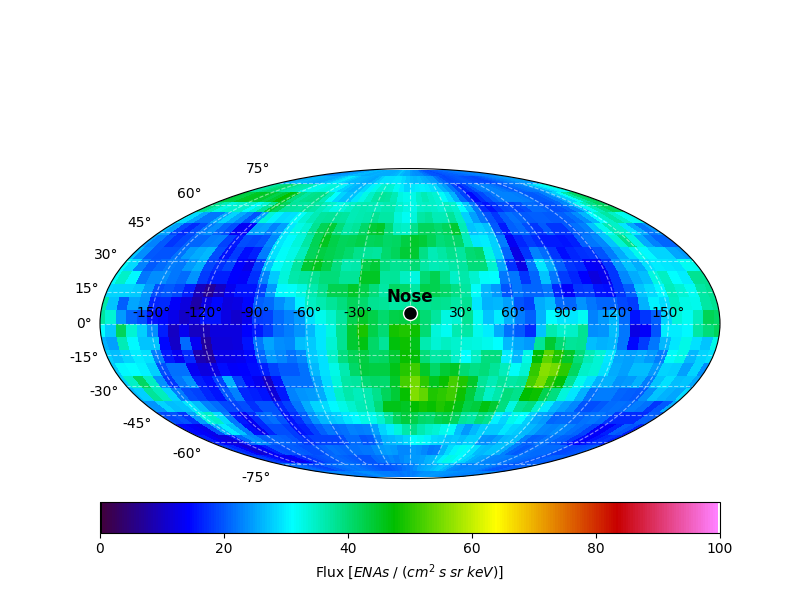}
\end{subfigure}\hfill
\begin{subfigure}[t]{0.315\textwidth}
\centering
\mapimg{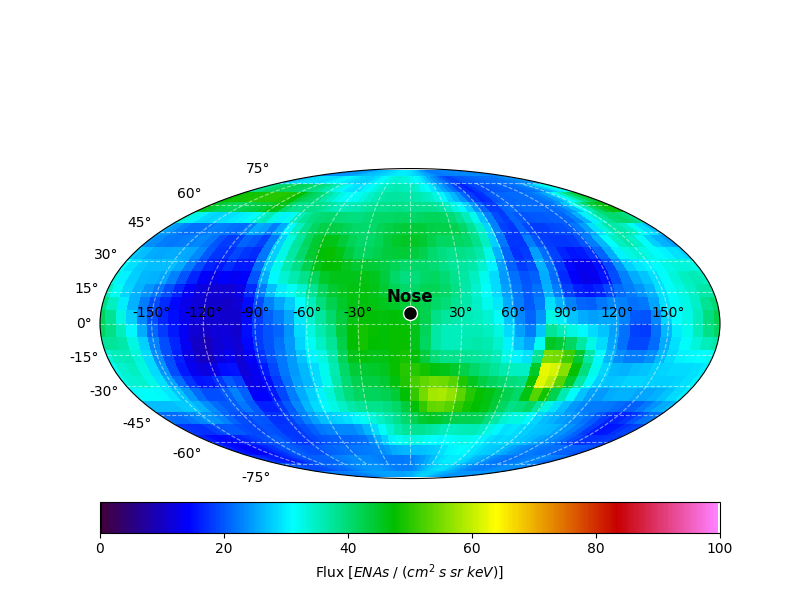}
\end{subfigure}\hfill
\begin{subfigure}[t]{0.315\textwidth}
\centering
\mapimg{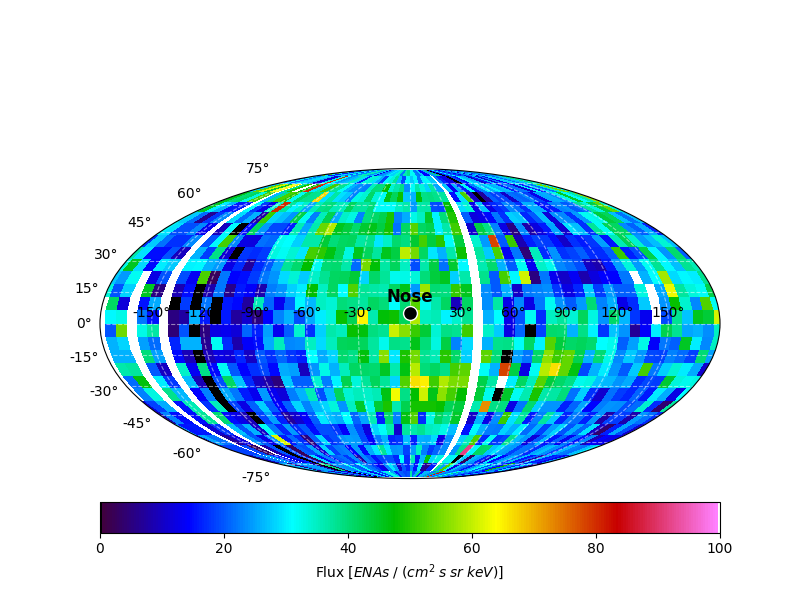}
\end{subfigure}

\vspace{1pt}

\begin{subfigure}[t]{0.315\textwidth}
\centering
\mapimg{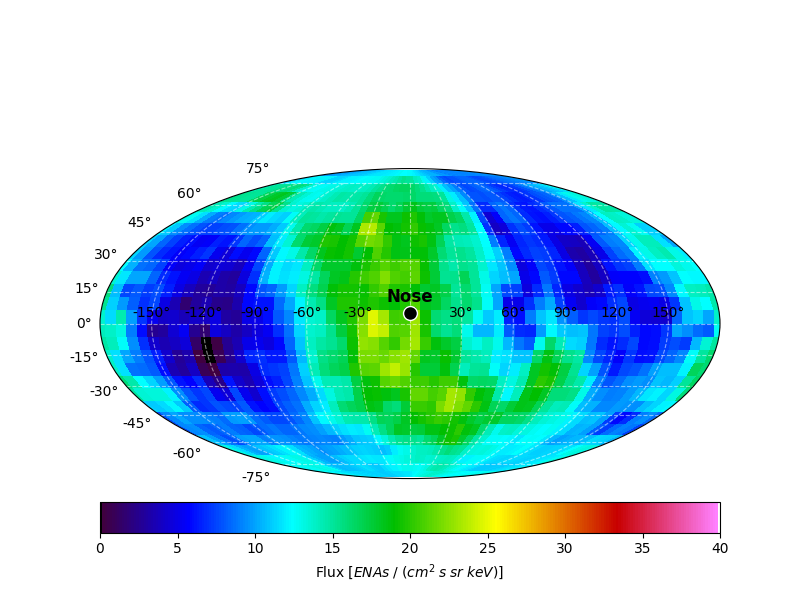}
\end{subfigure}\hfill
\begin{subfigure}[t]{0.315\textwidth}
\centering
\mapimg{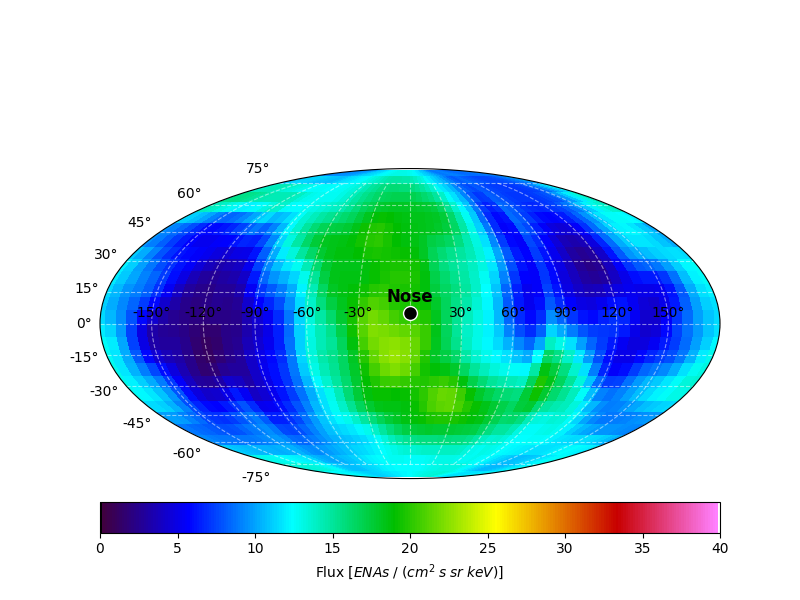}
\end{subfigure}\hfill
\begin{subfigure}[t]{0.315\textwidth}
\centering
\mapimg{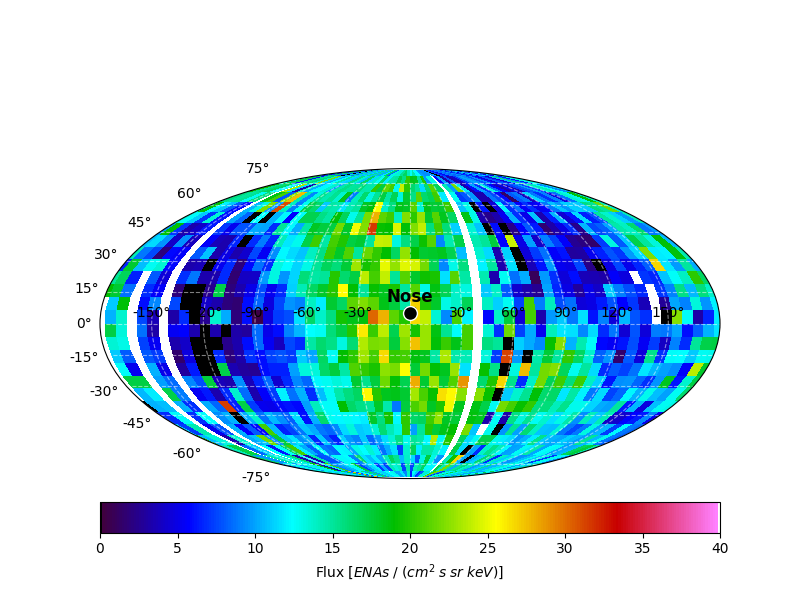}
\end{subfigure}

\vspace{1pt}

\begin{subfigure}[t]{0.315\textwidth}
\centering
\mapimg{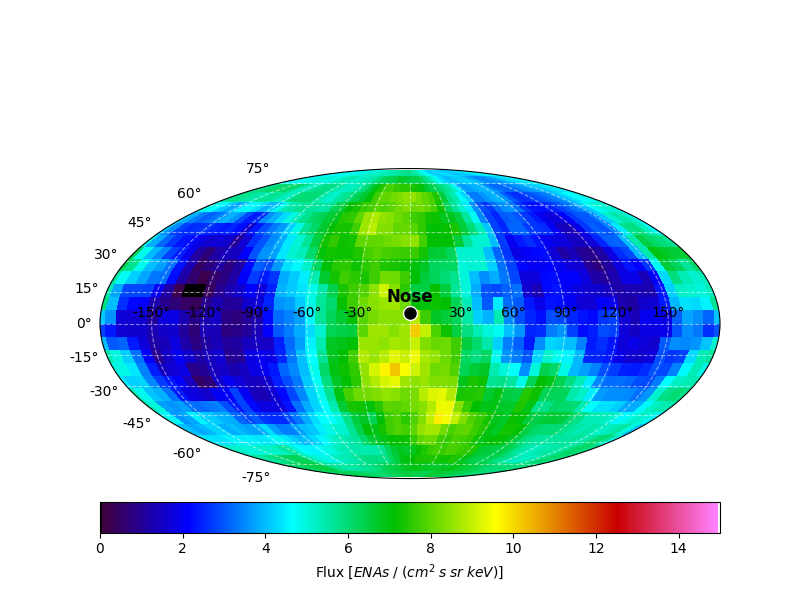}
\end{subfigure}\hfill
\begin{subfigure}[t]{0.315\textwidth}
\centering
\mapimg{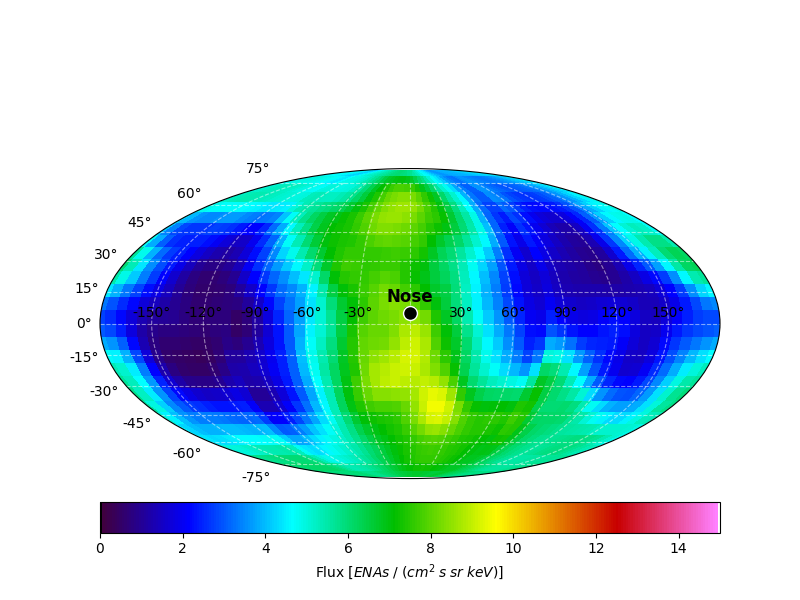}
\end{subfigure}\hfill
\begin{subfigure}[t]{0.315\textwidth}
\centering
\mapimg{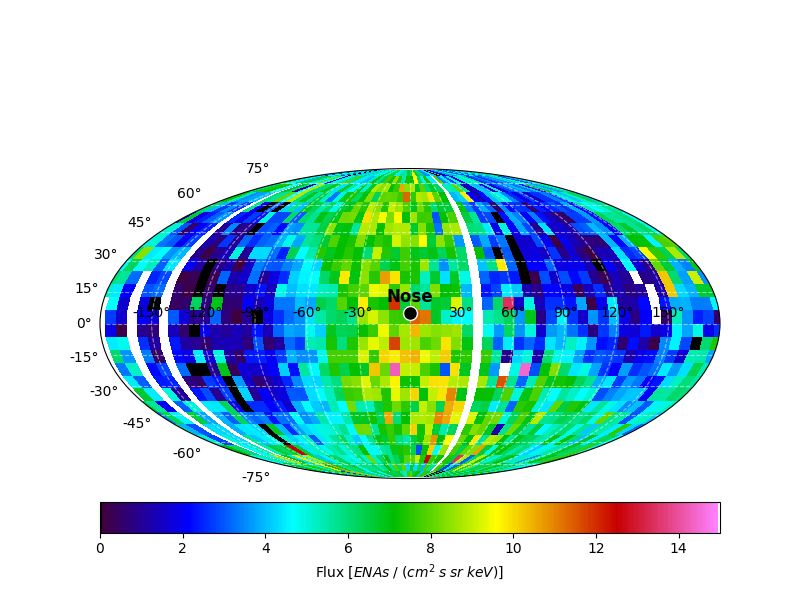}
\end{subfigure}

\vspace{-2pt}

\caption{
Comparison of intensity maps for ESA steps 2--6. Columns show, from left to right, the \textbf{\textsc{Capra} HP} reconstructions, \textsc{THESEUS} products, and baseline \textsc{IBEX} maps. Rows correspond to ESA steps 2--6 from top to bottom. All panels use the same Mollweide projection, nose-centered ecliptic frame, and colormap. Masked white regions indicate unobserved pixels.
}
\label{fig:part0}
\end{figure*}

Figure~\ref{fig:part0} compares the un-postprocessed intensity maps for IBEX, \textsc{Capra}, and \textsc{THESEUS}. This comparison is useful because it shows the maps before any normalization has changed their intensity scale. It checks whether \textsc{Capra} preserves the global structure present in the underlying IBEX data.

The main sky features appear in the same broad regions in all three products. 
\textsc{Capra} follows the large-scale structure of the baseline IBEX maps, while reducing some of the small-scale patchiness associated with sparse exposure. We retain negative pixels in the baseline IBEX maps where they occur, because clipping them would bias later conservation and global-intensity checks.

\subsubsection{Relative intensity}

\begin{figure*}[p]
\centering

\small
\textbf{\textsc{Capra} HP} \hfill \textsc{THESEUS} \hfill \textsc{IBEX}

\vspace{2pt}

\begin{subfigure}[t]{0.315\textwidth}
\centering
\mapimg{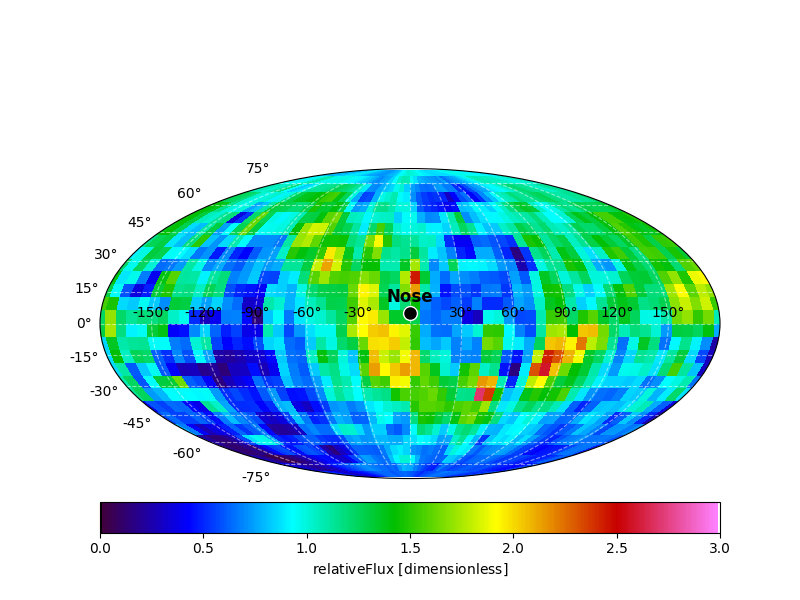}
\end{subfigure}\hfill
\begin{subfigure}[t]{0.315\textwidth}
\centering
\mapimg{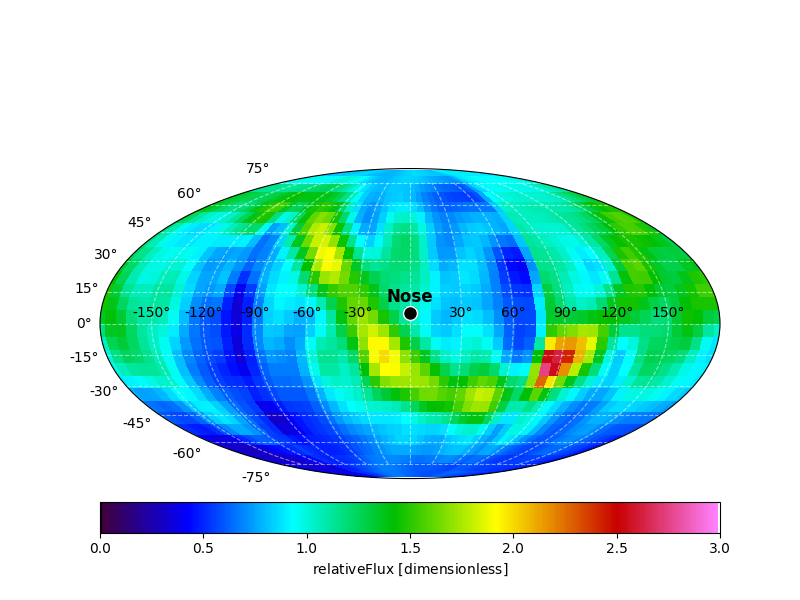}
\end{subfigure}\hfill
\begin{subfigure}[t]{0.315\textwidth}
\centering
\mapimg{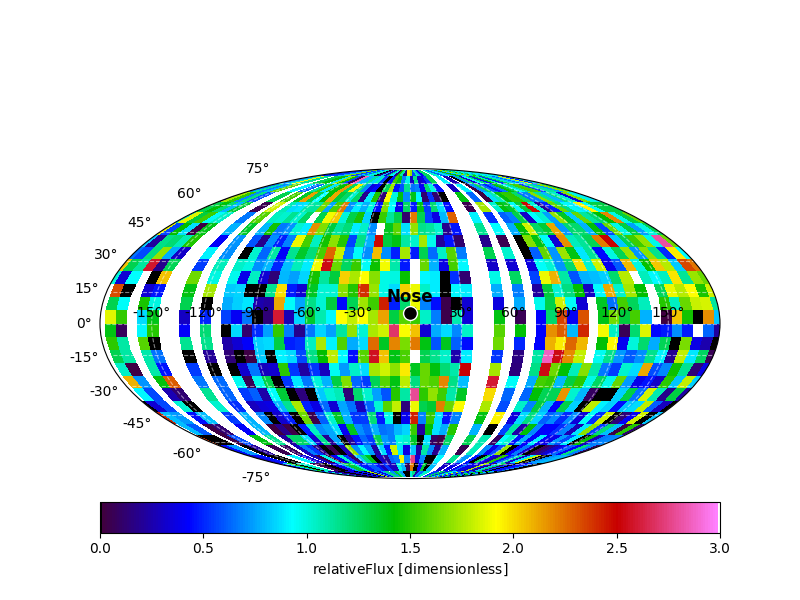}
\end{subfigure}

\vspace{1pt}

\begin{subfigure}[t]{0.315\textwidth}
\centering
\mapimg{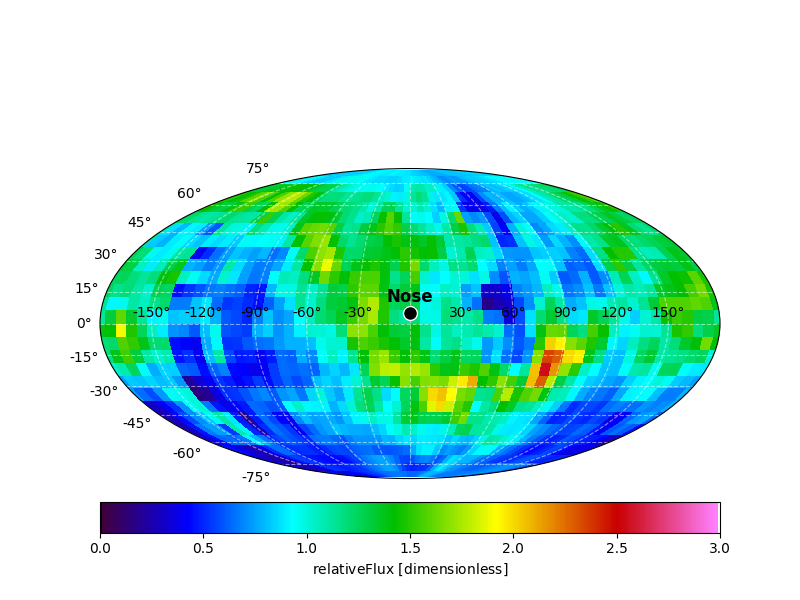}
\end{subfigure}\hfill
\begin{subfigure}[t]{0.315\textwidth}
\centering
\mapimg{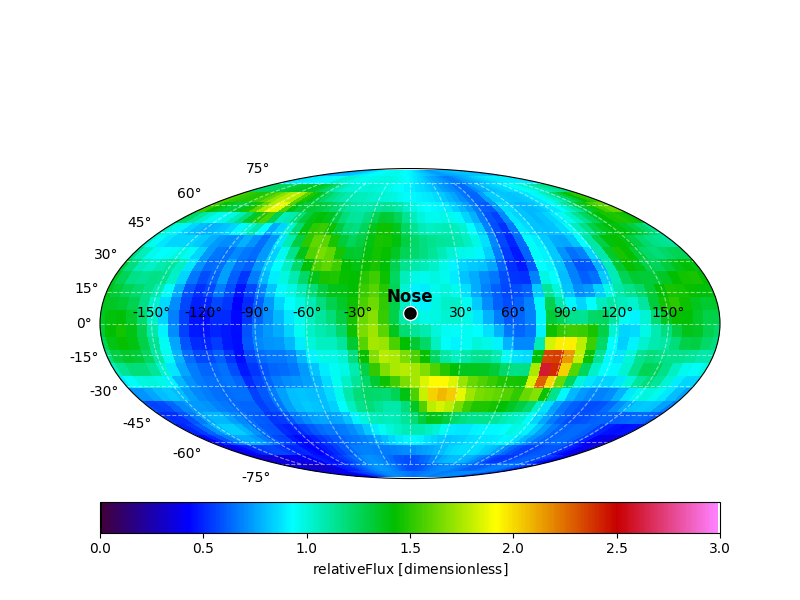}
\end{subfigure}\hfill
\begin{subfigure}[t]{0.315\textwidth}
\centering
\mapimg{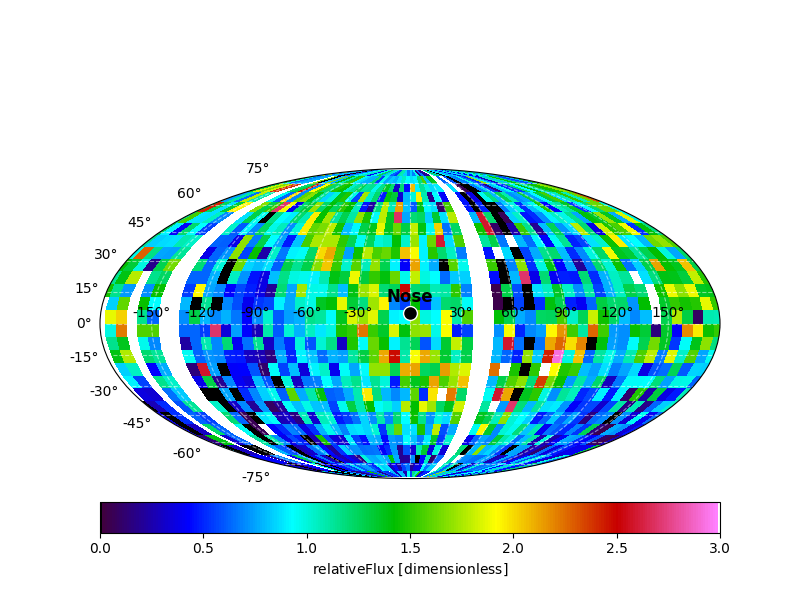}
\end{subfigure}

\vspace{1pt}

\begin{subfigure}[t]{0.315\textwidth}
\centering
\mapimg{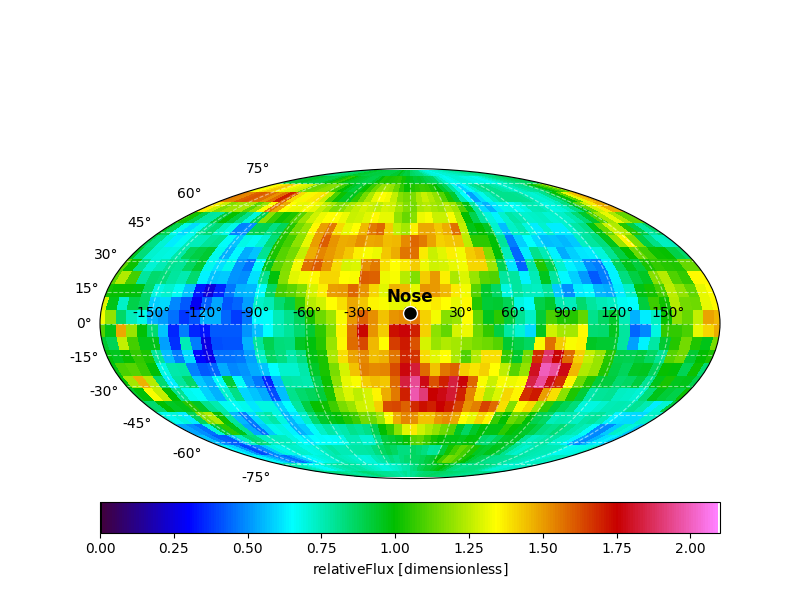}
\end{subfigure}\hfill
\begin{subfigure}[t]{0.315\textwidth}
\centering
\mapimg{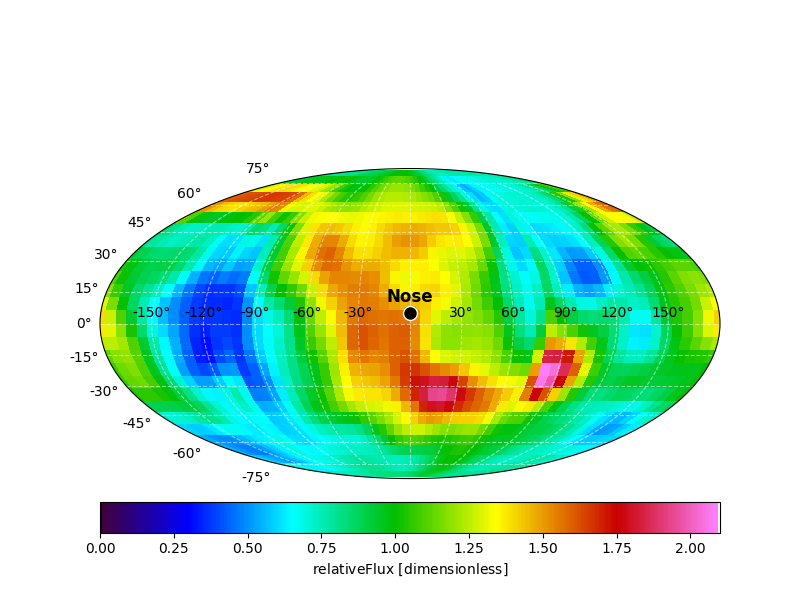}
\end{subfigure}\hfill
\begin{subfigure}[t]{0.315\textwidth}
\centering
\mapimg{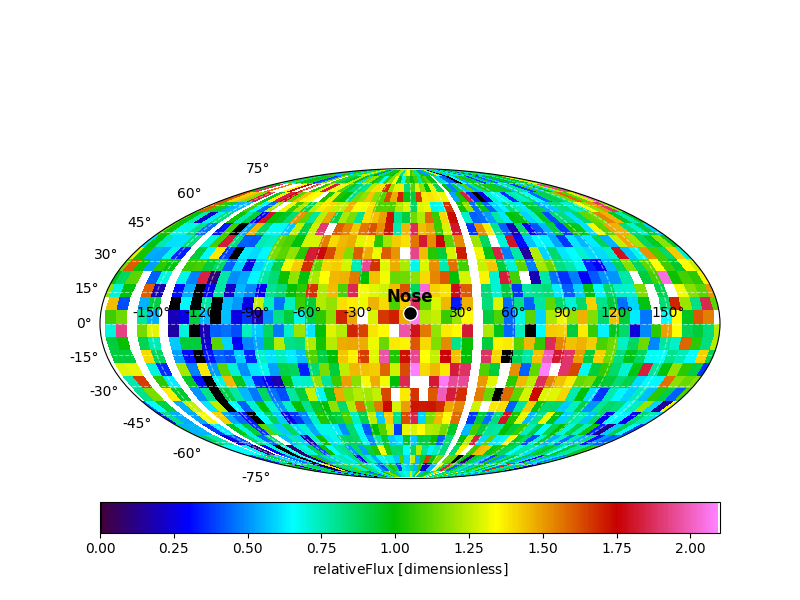}
\end{subfigure}

\vspace{1pt}

\begin{subfigure}[t]{0.315\textwidth}
\centering
\mapimg{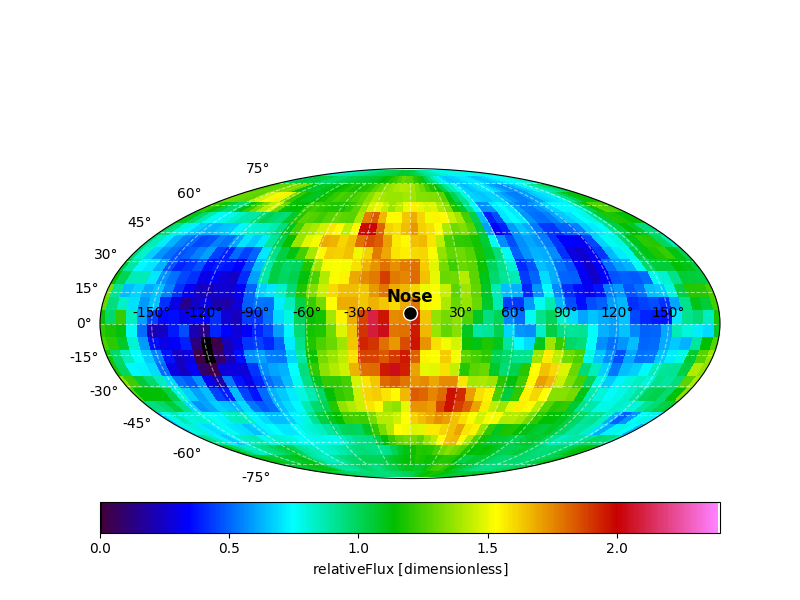}
\end{subfigure}\hfill
\begin{subfigure}[t]{0.315\textwidth}
\centering
\mapimg{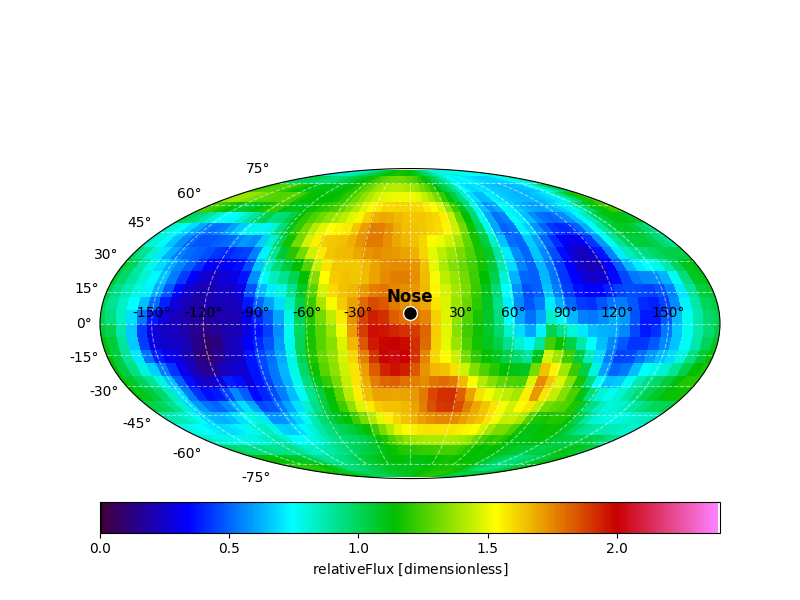}
\end{subfigure}\hfill
\begin{subfigure}[t]{0.315\textwidth}
\centering
\mapimg{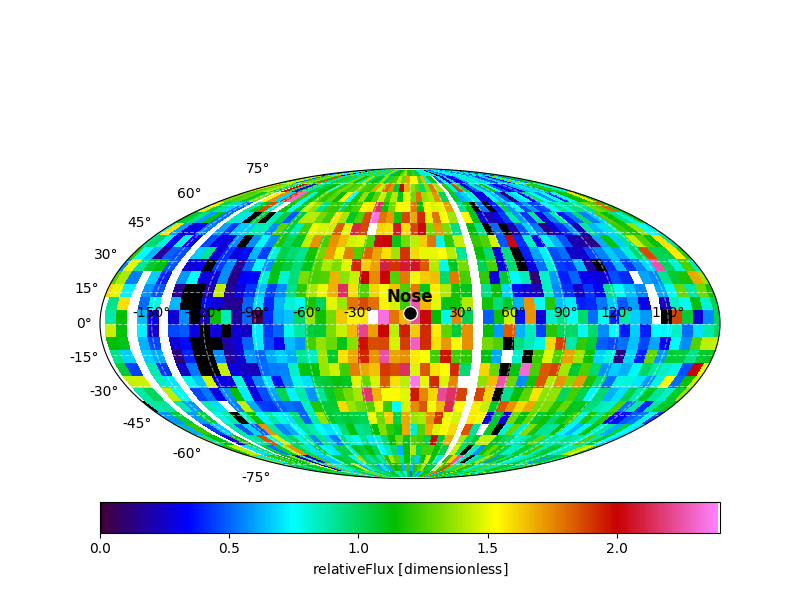}
\end{subfigure}

\vspace{1pt}

\begin{subfigure}[t]{0.315\textwidth}
\centering
\mapimg{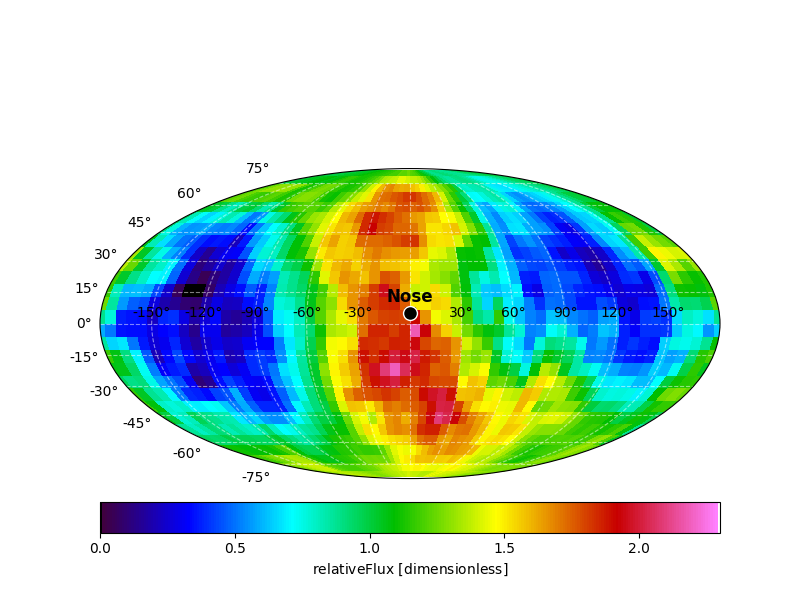}
\end{subfigure}\hfill
\begin{subfigure}[t]{0.315\textwidth}
\centering
\mapimg{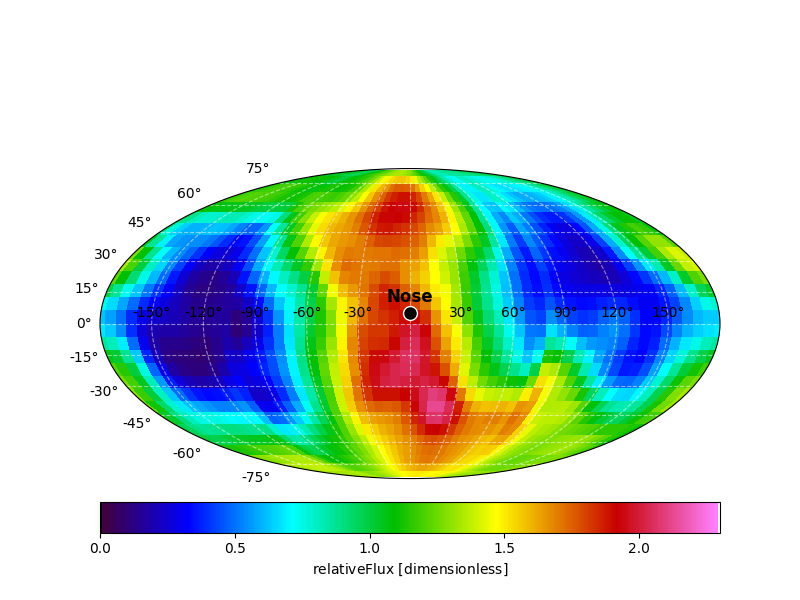}
\end{subfigure}\hfill
\begin{subfigure}[t]{0.315\textwidth}
\centering
\mapimg{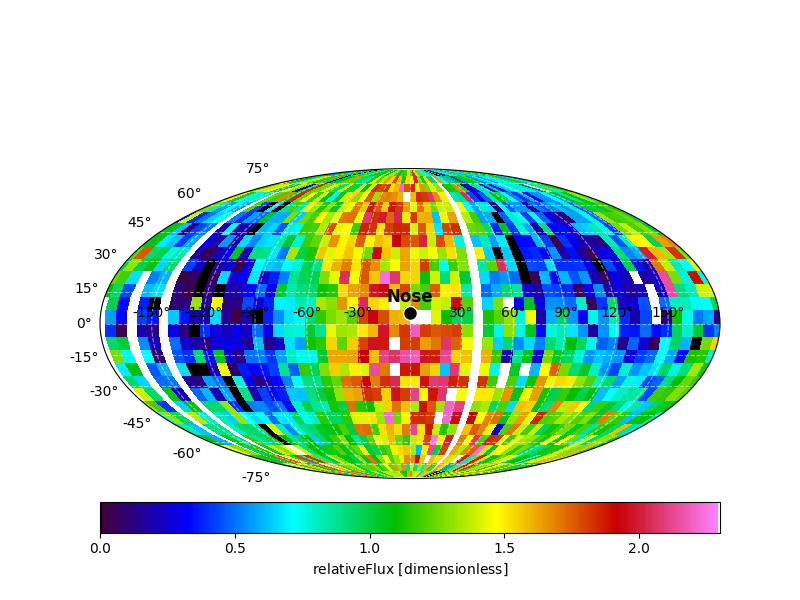}
\end{subfigure}

\vspace{-2pt}

\caption{
Comparison of relative normalized intensity maps for ESA steps 2--6. Columns show, from left to right, the \textbf{\textsc{Capra} HP} reconstructions, \textsc{THESEUS} products, and baseline \textsc{IBEX} maps. Rows correspond to ESA steps 2--6 from top to bottom. All panels use the same Mollweide projection, nose-centered ecliptic frame, and colormap. Masked white regions indicate unobserved pixels.
}
\label{fig:part1}
\end{figure*}

Figure~\ref{fig:part1} compares the same three datasets after relative normalization. 
We do this to verify that the methods recover the same morphology, independent of their absolute calibration or mean intensity level.

After this normalization, the Ribbon and large-scale gradients are broadly consistent between the three maps. The agreement is strongest at the level of global morphology: the main enhanced and depleted regions appear in similar locations. Differences are mainly seen in small-scale texture and contrast, reflecting the different treatments of sparse sampling, smoothing, and reconstruction.

\subsubsection{GDF-normalized intensity}

\begin{figure*}[p]
\centering
\small

\textbf{\textsc{Capra} HP}

\vspace{2pt}

\makebox[0.31\textwidth]{ESA 2}\hfill
\makebox[0.31\textwidth]{ESA 3}\hfill
\makebox[0.31\textwidth]{ESA 4}

\vspace{1pt}

\begin{subfigure}[t]{0.31\textwidth}
\centering
\mapimg{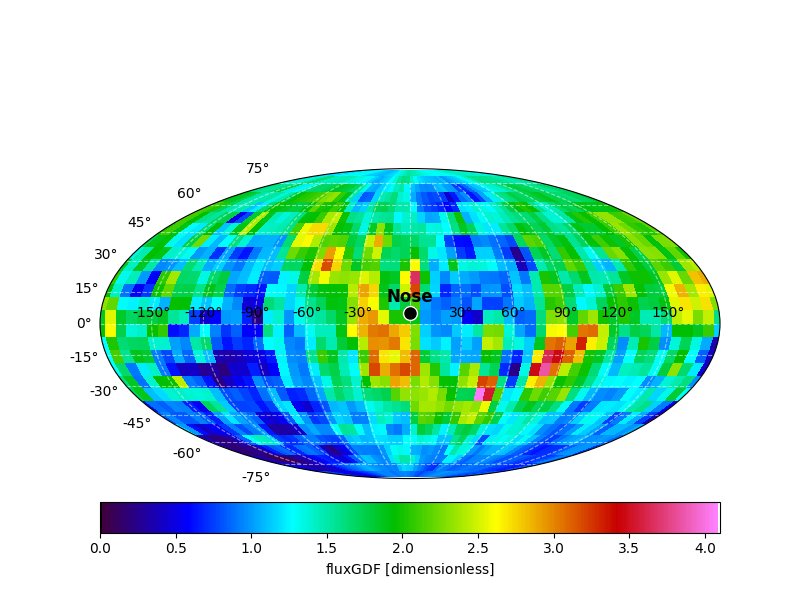}
\end{subfigure}\hfill
\begin{subfigure}[t]{0.31\textwidth}
\centering
\mapimg{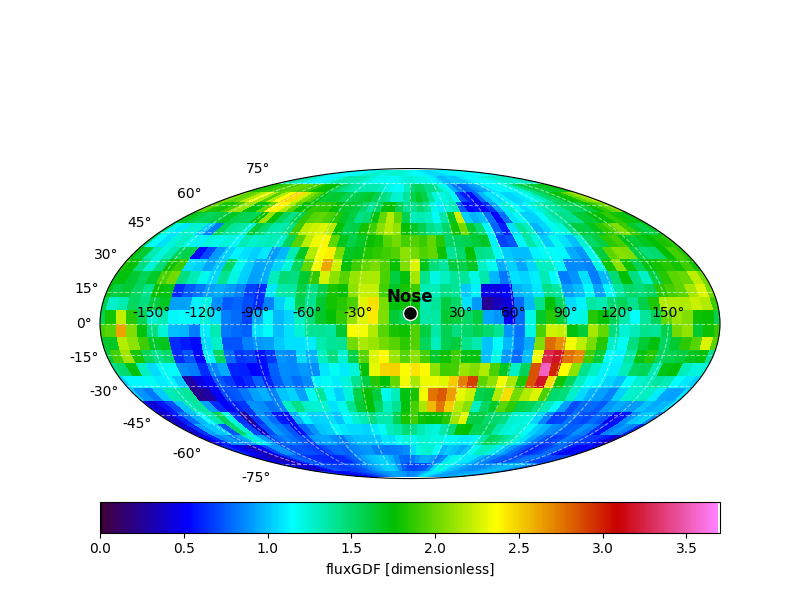}
\end{subfigure}\hfill
\begin{subfigure}[t]{0.31\textwidth}
\centering
\mapimg{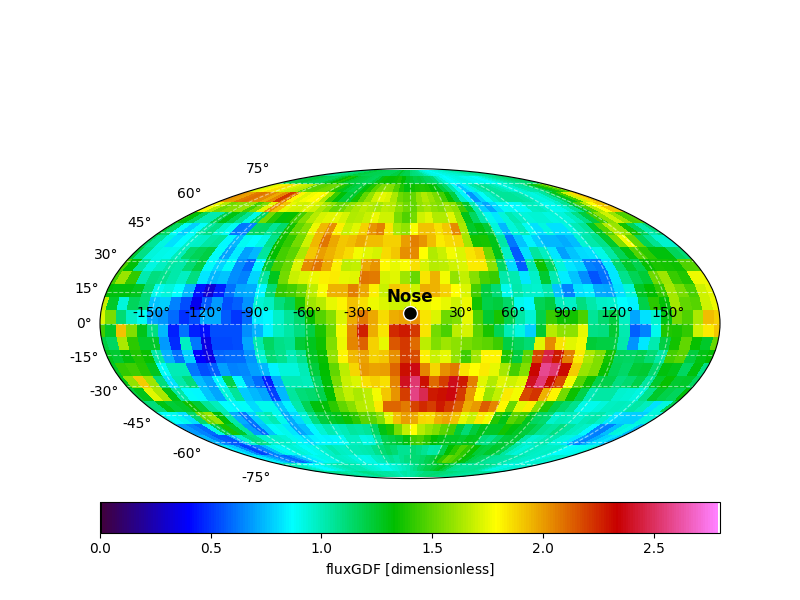}
\end{subfigure}

\vspace{1pt}

\makebox[0.31\textwidth]{ESA 5}\hfill
\makebox[0.31\textwidth]{ESA 6}\hfill
\makebox[0.31\textwidth]{}

\vspace{1pt}

\begin{subfigure}[t]{0.31\textwidth}
\centering
\mapimg{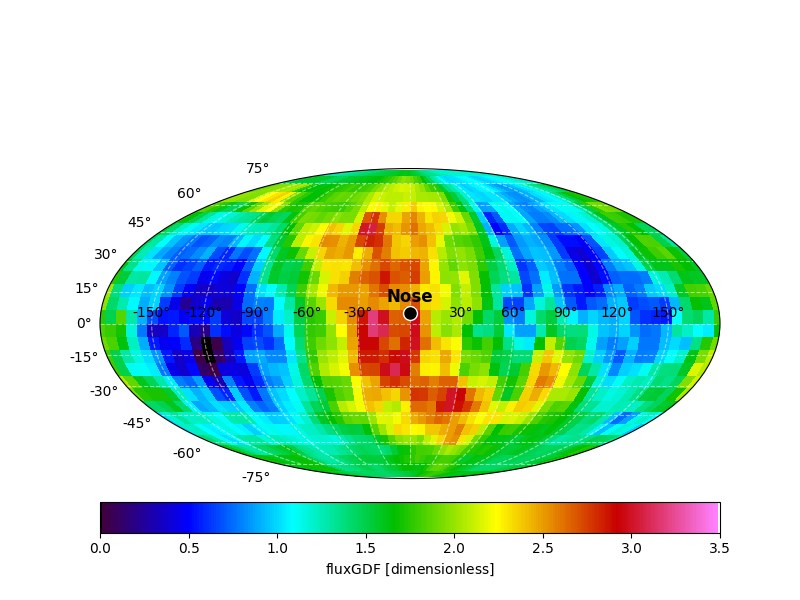}
\end{subfigure}\hfill
\begin{subfigure}[t]{0.31\textwidth}
\centering
\mapimg{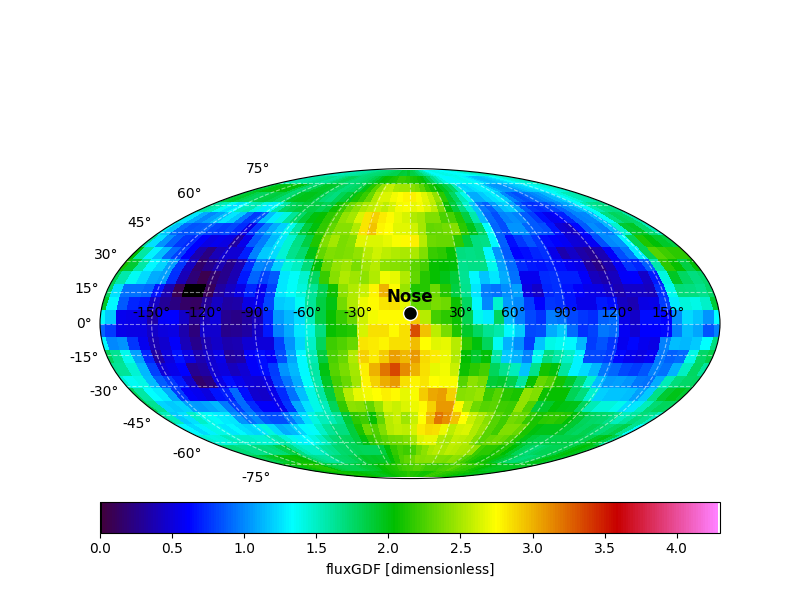}
\end{subfigure}\hfill
\begin{subfigure}[t]{0.31\textwidth}
\centering
\mbox{}
\end{subfigure}

\vspace{5pt}

\textsc{IBEX}

\vspace{2pt}

\makebox[0.31\textwidth]{ESA 2}\hfill
\makebox[0.31\textwidth]{ESA 3}\hfill
\makebox[0.31\textwidth]{ESA 4}

\vspace{1pt}

\begin{subfigure}[t]{0.31\textwidth}
\centering
\mapimg{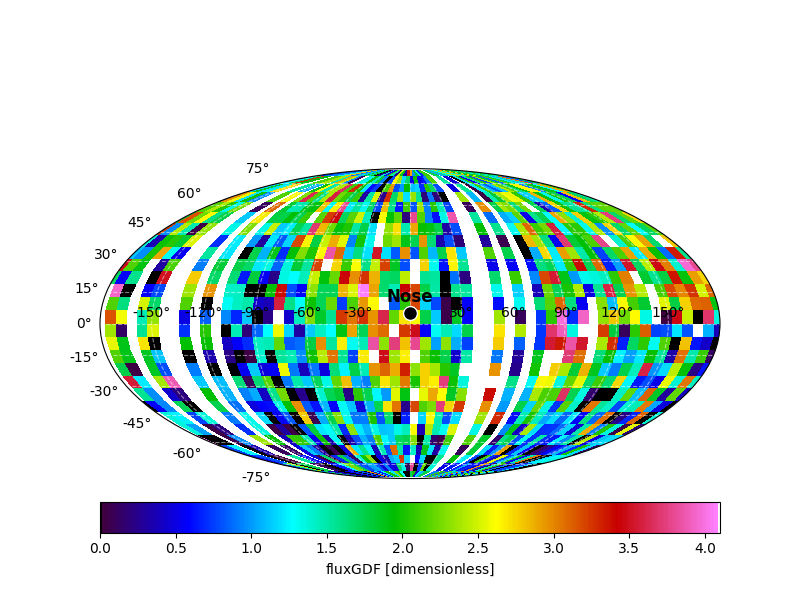}
\end{subfigure}\hfill
\begin{subfigure}[t]{0.31\textwidth}
\centering
\mapimg{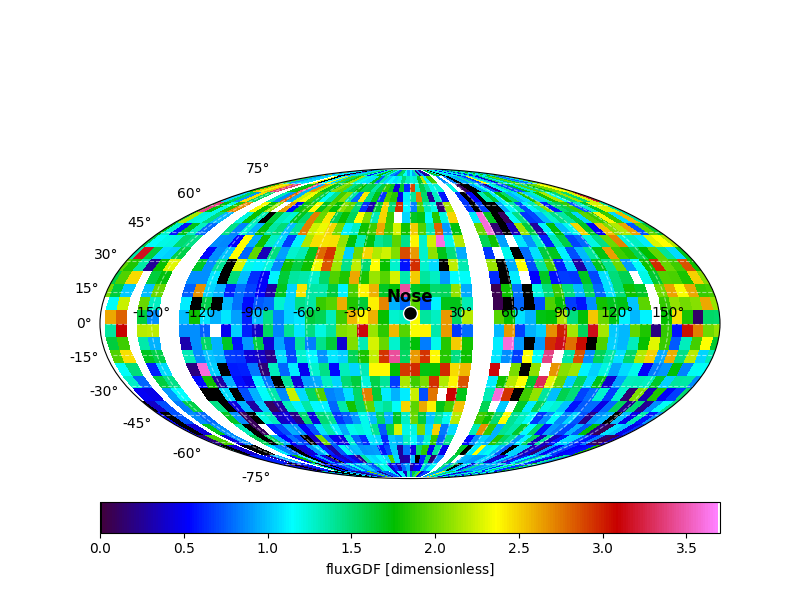}
\end{subfigure}\hfill
\begin{subfigure}[t]{0.31\textwidth}
\centering
\mapimg{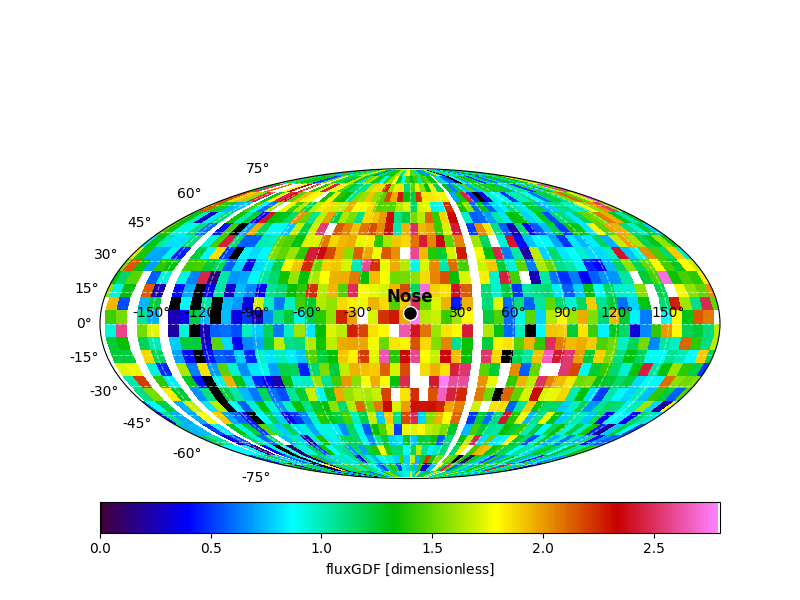}
\end{subfigure}

\vspace{1pt}

\makebox[0.31\textwidth]{ESA 5}\hfill
\makebox[0.31\textwidth]{ESA 6}\hfill
\makebox[0.31\textwidth]{}

\vspace{1pt}

\begin{subfigure}[t]{0.31\textwidth}
\centering
\mapimg{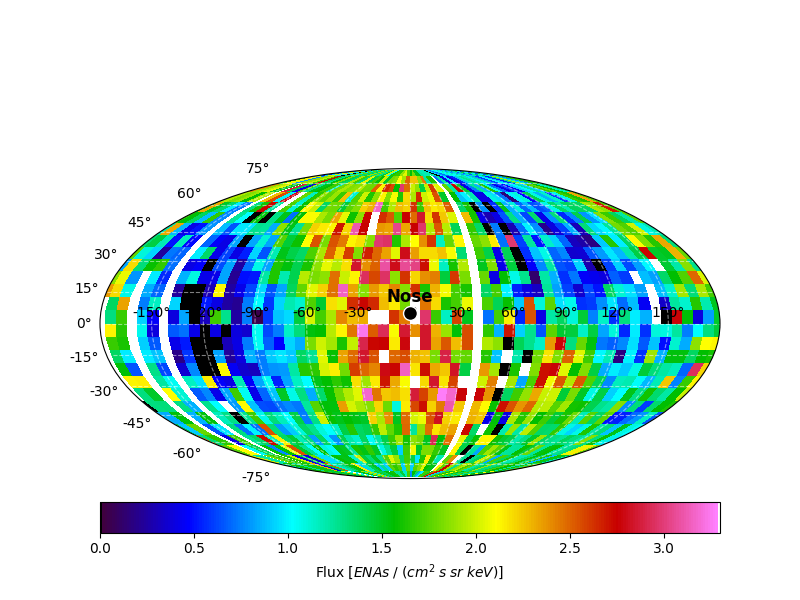}
\end{subfigure}\hfill
\begin{subfigure}[t]{0.31\textwidth}
\centering
\mapimg{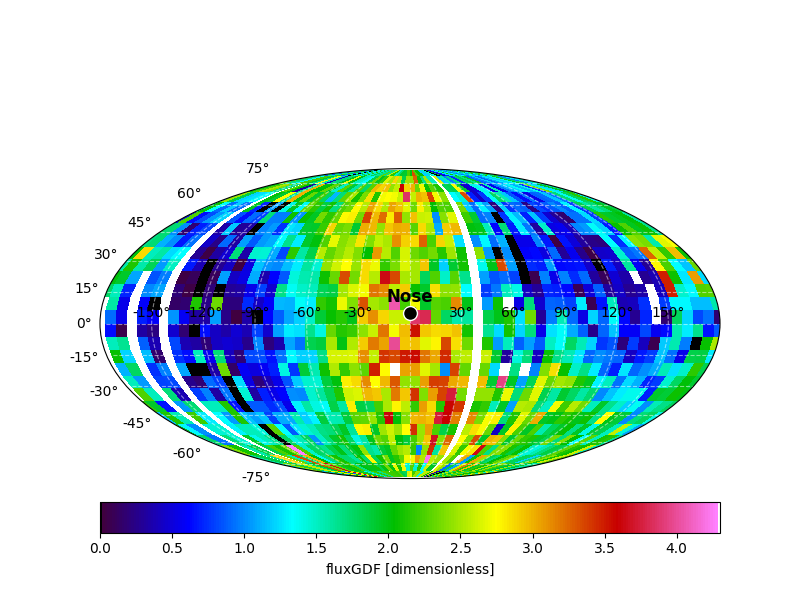}
\end{subfigure}\hfill
\begin{subfigure}[t]{0.31\textwidth}
\centering
\mbox{}
\end{subfigure}

\vspace{-2pt}

\caption{
Comparison of GDF-normalized intensity maps for ESA steps 2--6. The upper block shows the \textbf{\textsc{Capra} HP} reconstructions, and the lower block shows the baseline \textsc{IBEX} maps. Within each block, ESA steps 2--4 are shown in the first row and ESA steps 5--6 in the second row. All panels use the same Mollweide projection, nose-centered ecliptic frame, and colormap. Masked white regions indicate unobserved pixels.
}
\label{fig:part2}
\end{figure*}

Figure~\ref{fig:part2} compares the GDF-normalized baseline IBEX
and \textsc{Capra} maps. This normalization measures each map relative to its own off-Ribbon reference level and therefore emphasizes the Ribbon/background contrast rather than the absolute intensity scale.

The GDF-normalized maps show that \textsc{Capra} preserves the main Ribbon enhancement and the broad nose/anti-nose structure present in the baseline IBEX maps. Compared with the direct IBEX gridding, Capra reduces some of the sampling-driven mottling, especially in low-exposure regions, without introducing new large-scale structures.
Together with the preservation tests in Appendix~\ref{sec:conservation}, it shows that \textsc{Capra} provides a stable reconstruction of the large-scale IBEX intensity distribution while reducing bin-level sampling artifacts.

\subsection{Comparison with \textsc{THESEUS}}

\begin{figure*}[p]
\centering

\small
\textbf{\textsc{Capra} HP64} \hfill \textbf{\textsc{Capra} HP16} \hfill \textsc{THESEUS}$\rightarrow$HP16

\vspace{2pt}

\begin{subfigure}[t]{0.315\textwidth}
\centering
\mapimg{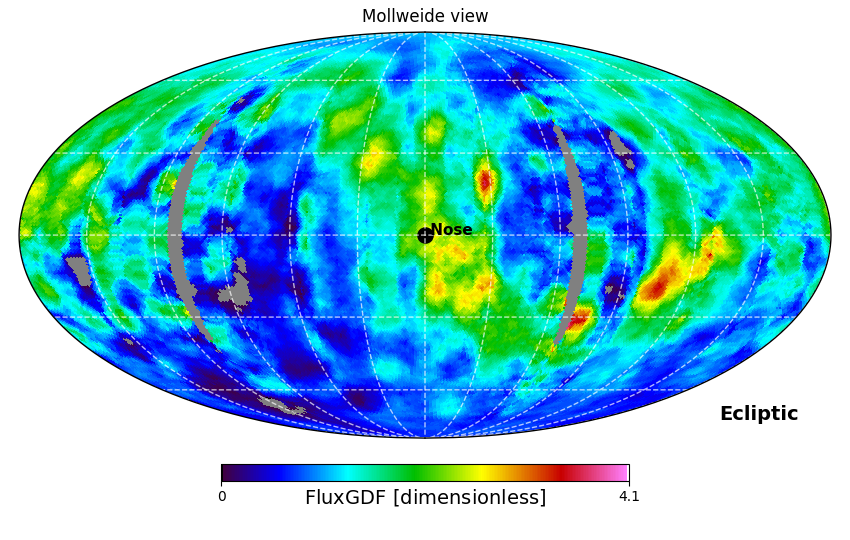}
\end{subfigure}\hfill
\begin{subfigure}[t]{0.315\textwidth}
\centering
\mapimg{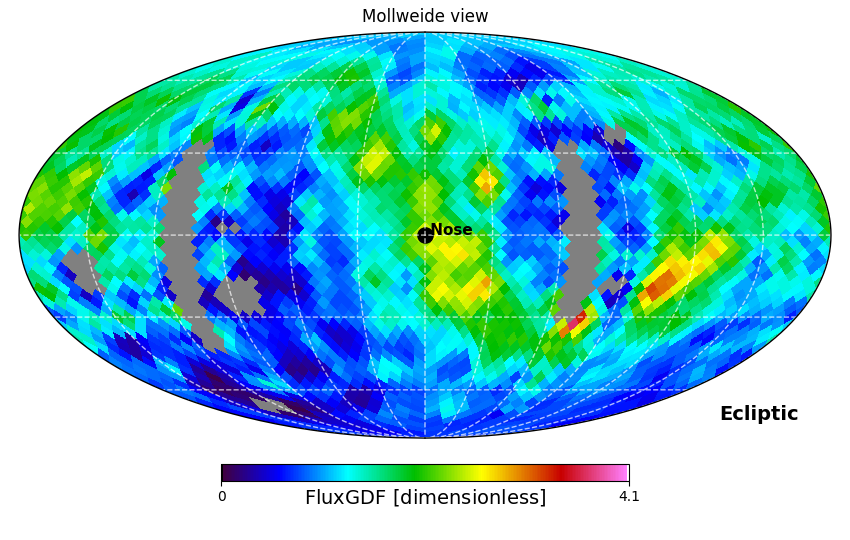}
\end{subfigure}\hfill
\begin{subfigure}[t]{0.315\textwidth}
\centering
\mapimg{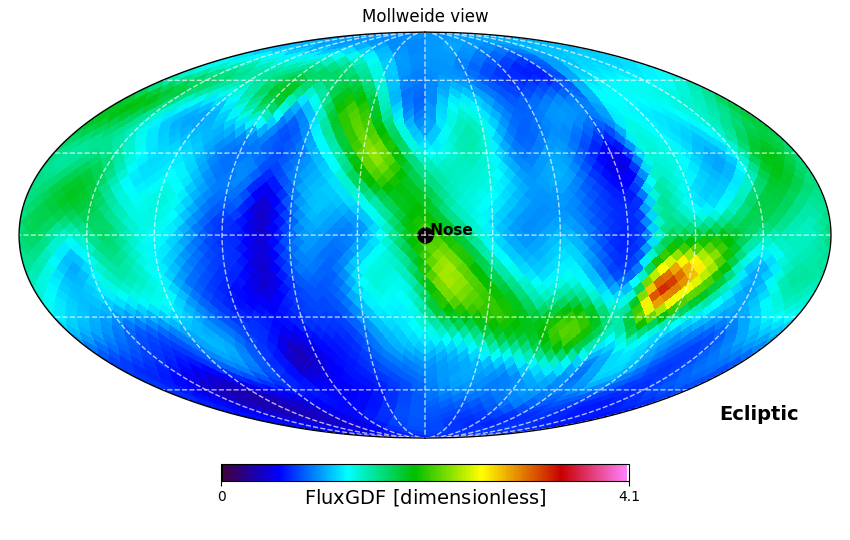}
\end{subfigure}

\vspace{1pt}

\begin{subfigure}[t]{0.315\textwidth}
\centering
\mapimg{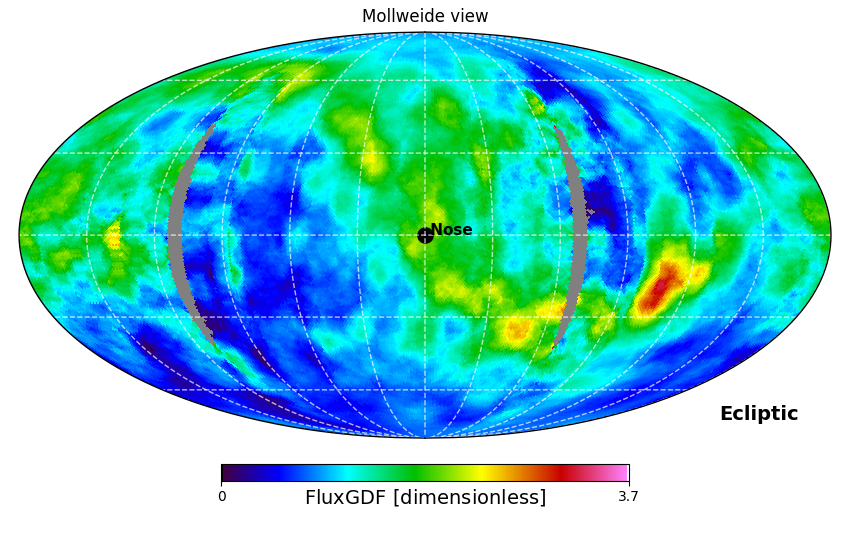}
\end{subfigure}\hfill
\begin{subfigure}[t]{0.315\textwidth}
\centering
\mapimg{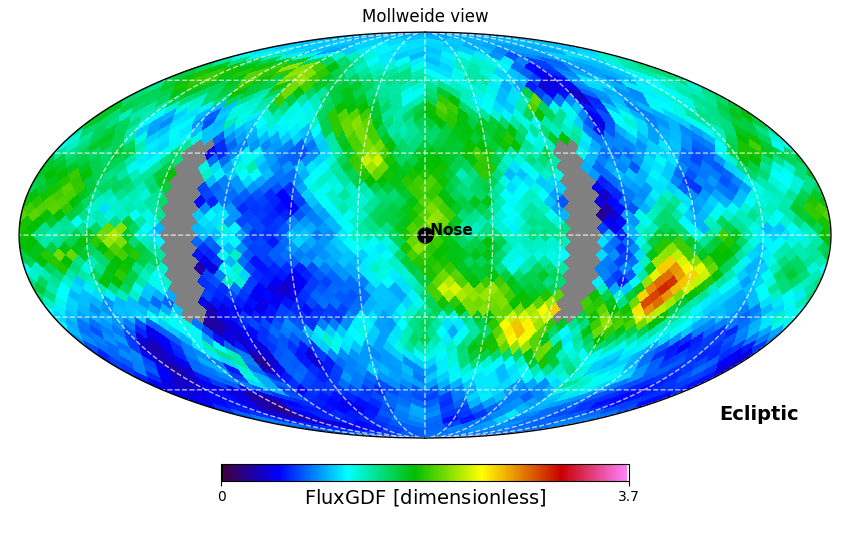}
\end{subfigure}\hfill
\begin{subfigure}[t]{0.315\textwidth}
\centering
\mapimg{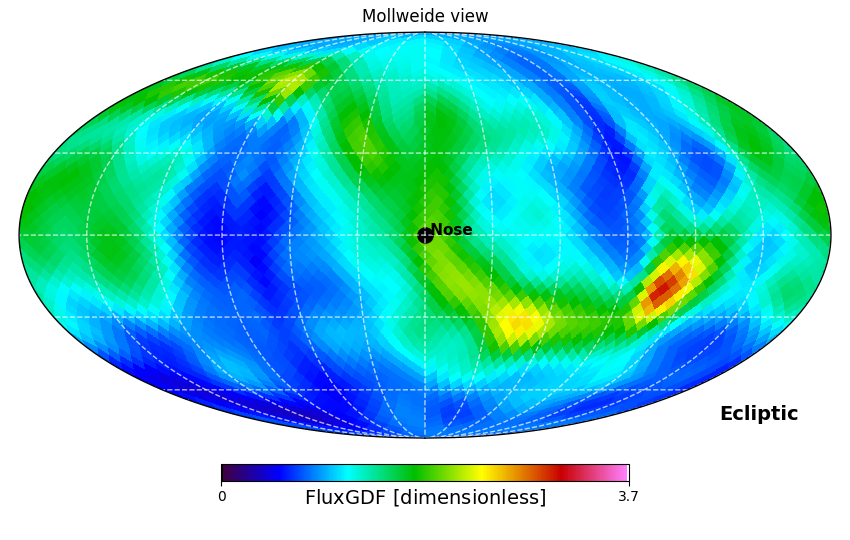}
\end{subfigure}

\vspace{1pt}

\begin{subfigure}[t]{0.315\textwidth}
\centering
\mapimg{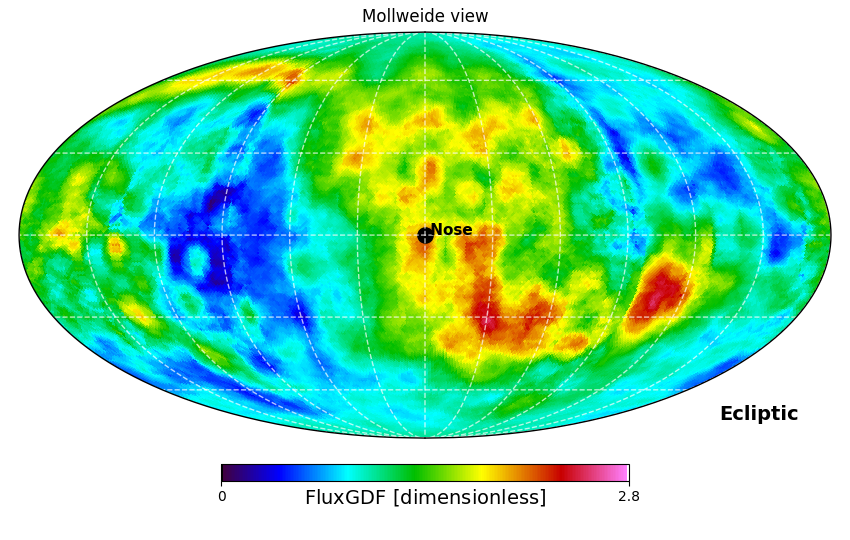}
\end{subfigure}\hfill
\begin{subfigure}[t]{0.315\textwidth}
\centering
\mapimg{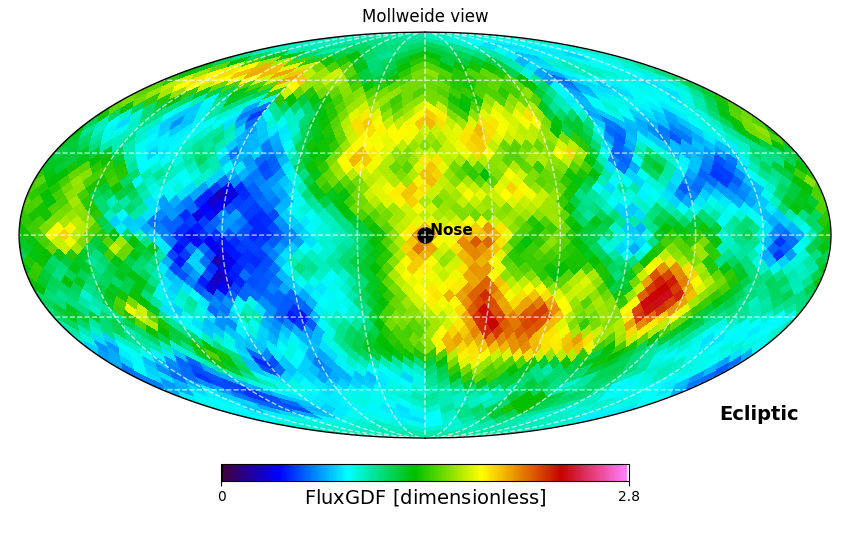}
\end{subfigure}\hfill
\begin{subfigure}[t]{0.315\textwidth}
\centering
\mapimg{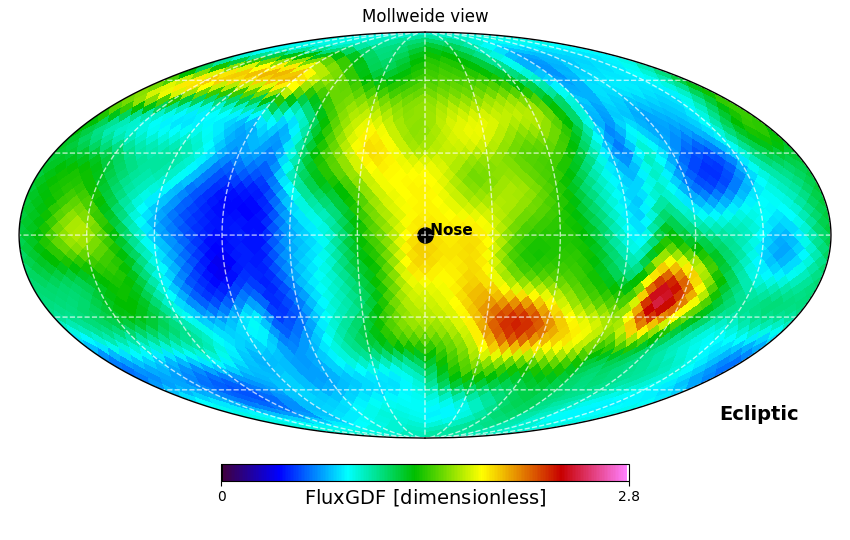}
\end{subfigure}

\vspace{1pt}

\begin{subfigure}[t]{0.315\textwidth}
\centering
\mapimg{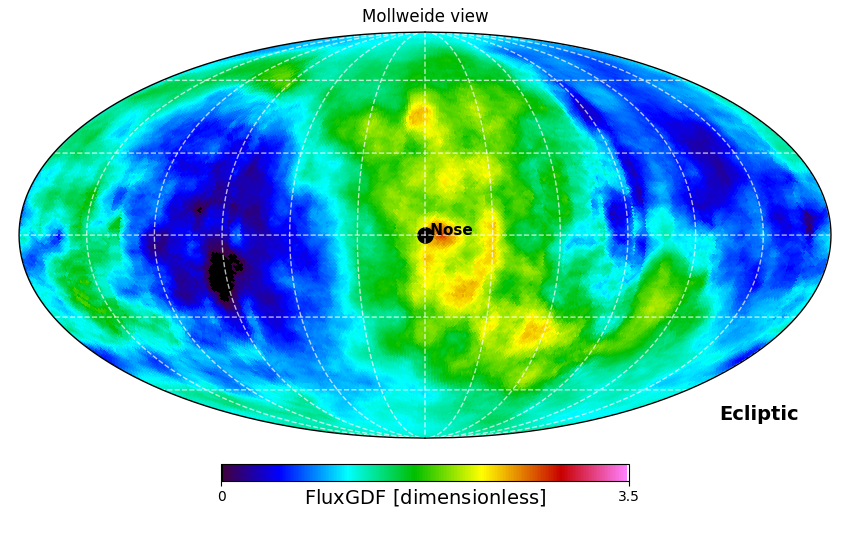}
\end{subfigure}\hfill
\begin{subfigure}[t]{0.315\textwidth}
\centering
\mapimg{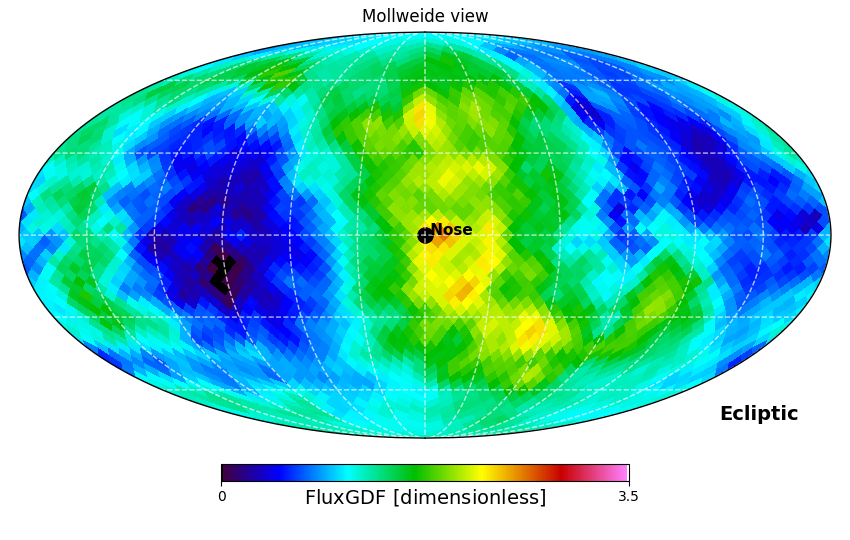}
\end{subfigure}\hfill
\begin{subfigure}[t]{0.315\textwidth}
\centering
\mapimg{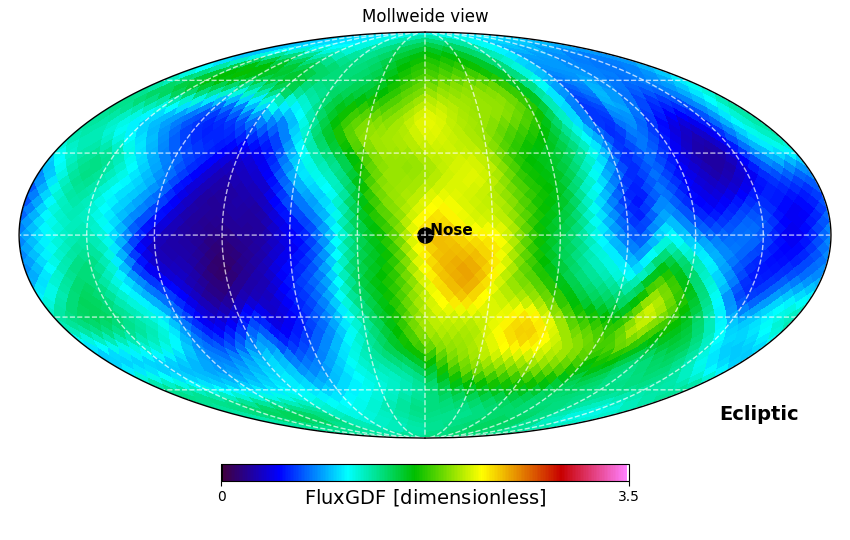}
\end{subfigure}

\vspace{1pt}

\begin{subfigure}[t]{0.315\textwidth}
\centering
\mapimg{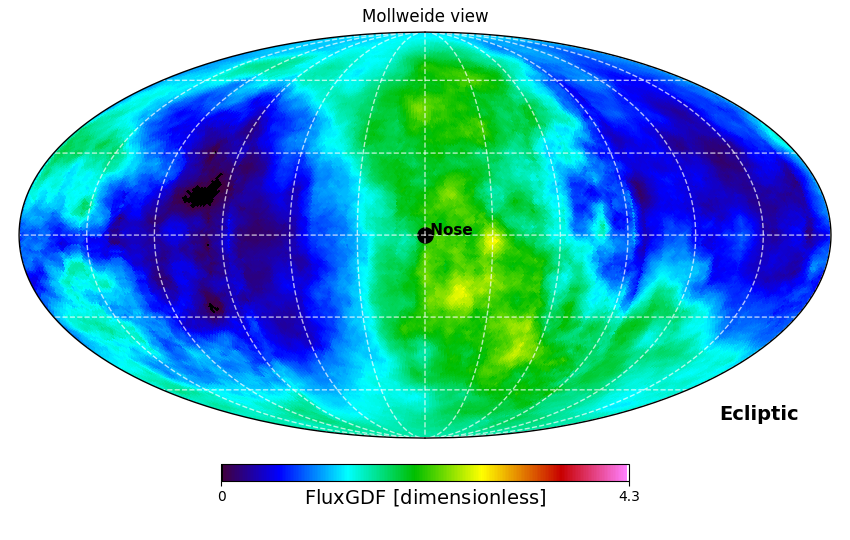}
\end{subfigure}\hfill
\begin{subfigure}[t]{0.315\textwidth}
\centering
\mapimg{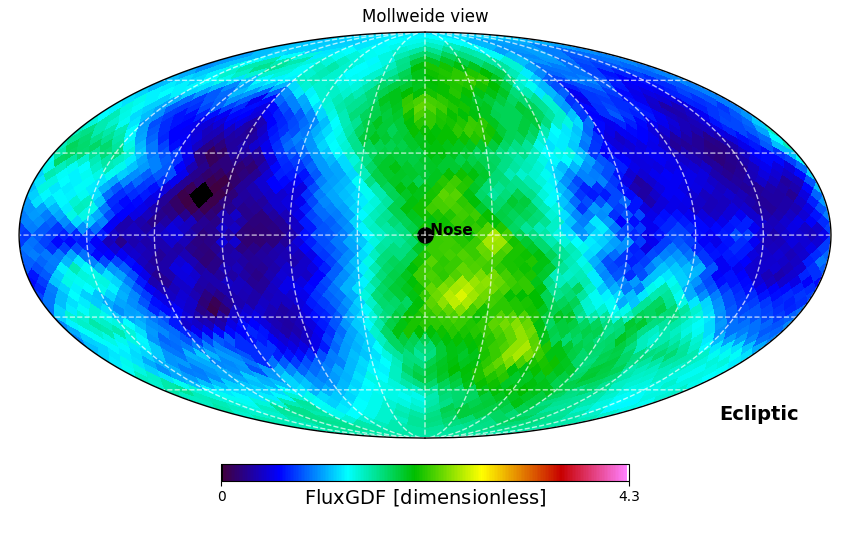}
\end{subfigure}\hfill
\begin{subfigure}[t]{0.315\textwidth}
\centering
\mapimg{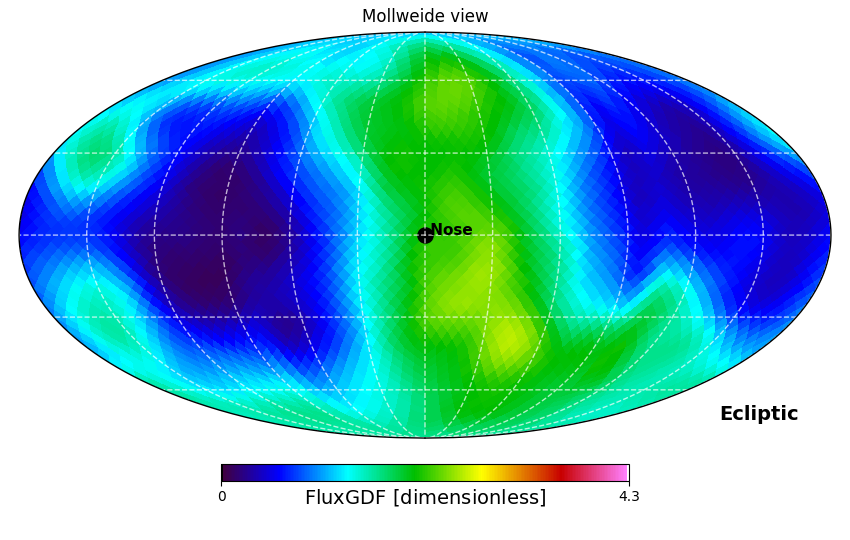}
\end{subfigure}

\vspace{-2pt}

\caption{
Comparison of GDF-normalized intensity maps for ESA steps 2--6. Columns show, from left to right, the \textbf{\textsc{Capra} HP64} maps, the \textbf{\textsc{Capra} HP16} maps, and the \textsc{THESEUS}$\rightarrow$HP16 maps. Rows correspond to ESA steps 2--6 from top to bottom. All maps are shown in nose-centered ecliptic coordinates using a shared color scale, allowing direct visual comparison of Ribbon morphology and the large-scale background structure.
}
\label{fig:thes_hp_gdf_comparison}
\end{figure*}

Figure~\ref{fig:thes_hp_gdf_comparison} compares \textsc{Capra} and \textsc{THESEUS} more directly using GDF-normalized intensity maps. We use this normalization because \textsc{THESEUS} and \textsc{Capra} are different reconstruction methods, and the most useful physical comparison is therefore not the absolute intensity scale but the contrast of the Ribbon relative to each method's own off-Ribbon baseline.

The large-scale morphology is broadly consistent between \textsc{Capra} and \textsc{THESEUS}. 
Both methods recover the Ribbon arc and the low-intensity anti-nose region in similar locations. The main differences are methodological rather than positional. \textsc{Capra} retains more small-scale texture associated with the direct reconstruction of the IBEX sampling pattern, even after the Gaussian smoothing step. THESEUS appears visually smoother because its statistical reconstruction denoises the map where the data allow it, but its PSF deconvolution can recover intrinsically narrower large-scale structures, such as the Ribbon. Therefore, visual smoothness should not be interpreted as lower angular resolving power.

The comparison between \textsc{Capra} at $N_{\rm side}=16$ and $N_{\rm side}=64$ resolutions shows that increasing the \textsc{Capra} resolution does not create new large-scale features. Instead, the higher-resolution maps refine the same global morphology already present at lower resolution. The higher-resolution \textsc{Capra} maps preserve more localized texture, while the lower-resolution maps provide a closer visual match to the coarse \textsc{THESEUS} representation. This behavior is expected: the tessellation controls the angular sampling of the reconstructed field, whereas the remaining
differences with \textsc{THESEUS} reflect the different reconstruction assumptions, denoising, and PSF treatment.

Because these distinctions are difficult to see in the compact multi-channel
layout, we examine one representative channel in more detail below.

\subsubsection{Enlarged ESA 4 comparison}

\begin{figure*}
    \centering
    \includegraphics[width=0.48\textwidth]{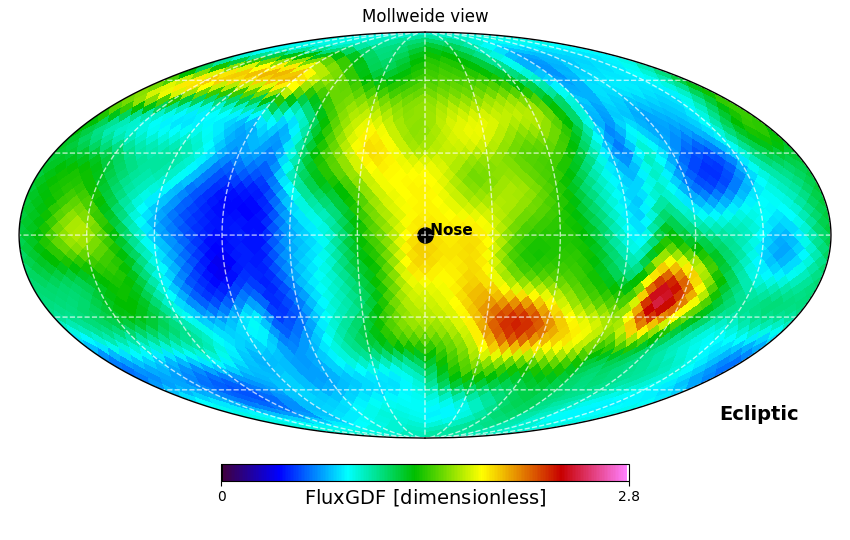}
    \hfill
    \includegraphics[width=0.48\textwidth]{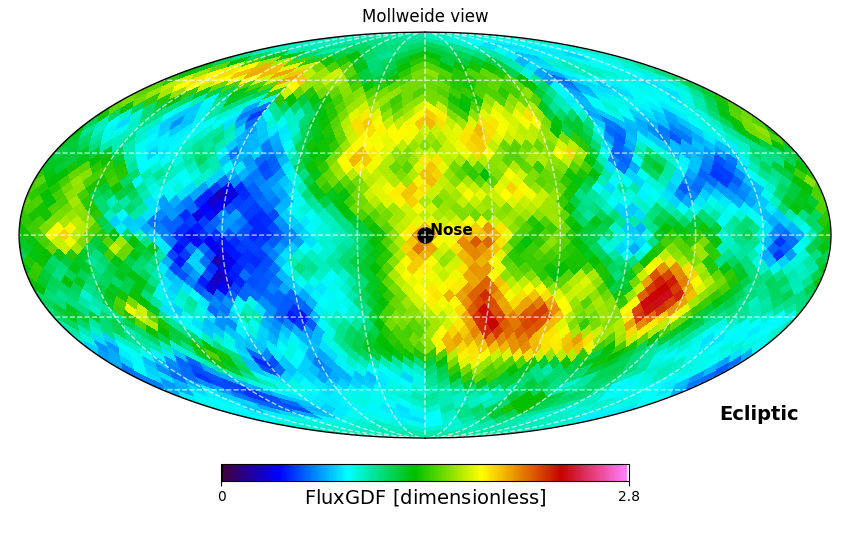}

    \caption{Enlarged comparison of the \textit{GDF-normalized} ESA 4 intensity maps from \textsc{THESEUS} (left) and \textsc{Capra} (right), both shown at
    $N_{\rm side}=16$ in nose-centered ecliptic coordinates and with the same color scale. }
    \label{fig:esa4_enlarged}
\end{figure*}

The multi-channel comparison in Figure~\ref{fig:thes_hp_gdf_comparison} shows that \textsc{Capra} and \textsc{THESEUS} recover the same dominant large-scale structures, but the small panels make it difficult to distinguish two separate effects: the local map texture and the sharpness of the reconstructed physical features. To clarify this distinction, Figure~\ref{fig:esa4_enlarged} shows an enlarged side-by-side comparison for ESA 4, which is representative of the channels where the Ribbon and the broad nose/anti-nose contrast are both clearly visible.

The enlarged maps show that the main morphology is consistent between the two
reconstructions. Both maps recover the low-intensity anti-nose region, the broad
enhancement near the nose, and the brighter Ribbon-related structures at
similar sky locations. The most visible difference is in the texture of the maps: the \textsc{Capra} reconstruction retains more pixel-scale variability, associated with the direct reconstruction of the sampled IBEX data, whereas
\textsc{THESEUS} appears smoother because its statistical reconstruction denoises the map where the data allow it. As discussed before, this visual smoothness should not be interpreted as a loss of angular resolving power. Because \textsc{THESEUS} includes PSF deconvolution, it can still recover intrinsically narrower reconstructed structures, even when the full-sky map appears smoother by eye.

In summary, the \textsc{Capra} and \textsc{THESEUS} maps agree well in the location of the dominant large-scale structures, while differing mainly in texture, smoothing, and the sharpness of reconstructed Ribbon contrast.

\section{\textbf{HEALPix-native map transformations}} \label{sec:dat_manip}

The HEALPix representation makes it possible to manipulate the reconstructed maps after the main reconstruction step. This is useful because high-resolution maps are more expensive to generate than to transform. Once a map has been constructed at sufficiently high resolution, it can be rotated, rebinned, or projected onto a different grid without returning to the original orbit-arc data.

In this section, we illustrate three such operations: reference-frame transformations, changes in the HEALPix tessellation, and projection onto rectangular grids with a user-defined angular resolution. 
These operations are not separate reconstruction methods; they are postprocessing tools that preserve the physical meaning of the reconstructed intensity field while adapting the map to a given visualization or comparison task.

\subsection{Reference-frame transformations}

Changing the reference frame does not alter the reconstructed intensity values. We demonstrate the preservation of intensities under coordinate and tessellation transformations in Appendix~ \ref{sec:conservation}.  Switching to a different reference frame changes only the coordinates in which the same sky distribution is represented. This is useful when a physical structure is more naturally described in a frame other than ecliptic coordinates. For example, a Ribbon-centered frame  \citep{ribbon, Dayeh_2023} emphasizes the geometry of the IBEX Ribbon by placing the fitted Ribbon center at the pole, while the Galactic frame shows how the same structure is oriented with respect to the Milky Way.

\underline{Ribbon-centered frame}\label{appendix:ribbon}

Let the Ribbon-center direction in ecliptic coordinates be
$(\lambda_R,\beta_R)=(221^\circ,39^\circ)$, where $\lambda_R$ is longitude and
$\beta_R$ is latitude. Let the upwind direction be
$(\lambda_{\rm up},\beta_{\rm up})=(255^\circ,5.14^\circ)$. In the implementation
we first rotate by $R_z(\lambda_R)$ and then by $R_y(90^\circ-\beta_R)$, so that
the Ribbon-center direction is mapped to the pole of the new coordinate system.
The transformed upwind vector is then used to define the remaining rotation about
the new $z$ axis, placing the upwind direction on the zero meridian.

To transform the Sun-centered ecliptic coordinates to the Ribbon-centered ones, we need to define an appropriate rotation matrix. 
Let the Ribbon-center direction in ecliptic coordinates be
$(\lambda_R,\beta_R)=(221^\circ,39^\circ)$, where $\lambda_R$ is longitude and
$\beta_R$ is latitude. Let the upwind direction be $(\lambda_{\rm up},\beta_{\rm up})=(255^\circ,5.14^\circ)$. In the implementation we first rotate by $R_z(\lambda_R)$ and then by $R_y(90^\circ-\beta_R)$, so that the Ribbon-center direction is mapped to the pole of the new coordinate system. 
\begin{align}
    rotation_1 &= 
\begin{bmatrix}
\cos{\theta_r}\cos{l_r} & \cos{\theta_r}\sin{l_r} & -\sin{\theta_r}\\
-\sin{l_r} & \cos{l_r} & 0\\
\cos{l_r}\sin{\theta_r} & \sin{\theta_r}\sin{l_r} & \cos{\theta_r}
\end{bmatrix}.
\end{align}
The transformed upwind vector $(\theta_{up}, l_{up})$ is then used to define the remaining rotation about the new $z$ axis, placing the upwind direction on the zero meridian.

\begin{align}
    rotation_2 &= 
\begin{bmatrix}
\cos{l_t} & \sin{l_t} & 0\\
-\sin{l_t} & \cos{l_t} & 0\\
0 & 0 & 1
\end{bmatrix}
\end{align}
We apply this transformation to the HEALPix coordinates in the ecliptic reference frame to transform them into the Ribbon centered one.
We can then get the new HEALPix coordinates from the transformed vectors and use the HEALPix mapping function to obtain pixel indices, which will allow us to sort the values and visualize them. This could be done in \emph{Mathematica} and visualized in \emph{Python}, or just performed directly in a \emph{Python} script. We provide both alternatives, but in this paper present results obtained from the \emph{Python} script.

\begin{figure}[hbt!]
  \centering
  \begin{subfigure}[b]{0.45\textwidth}
    \centering
    \includegraphics[height=5cm]{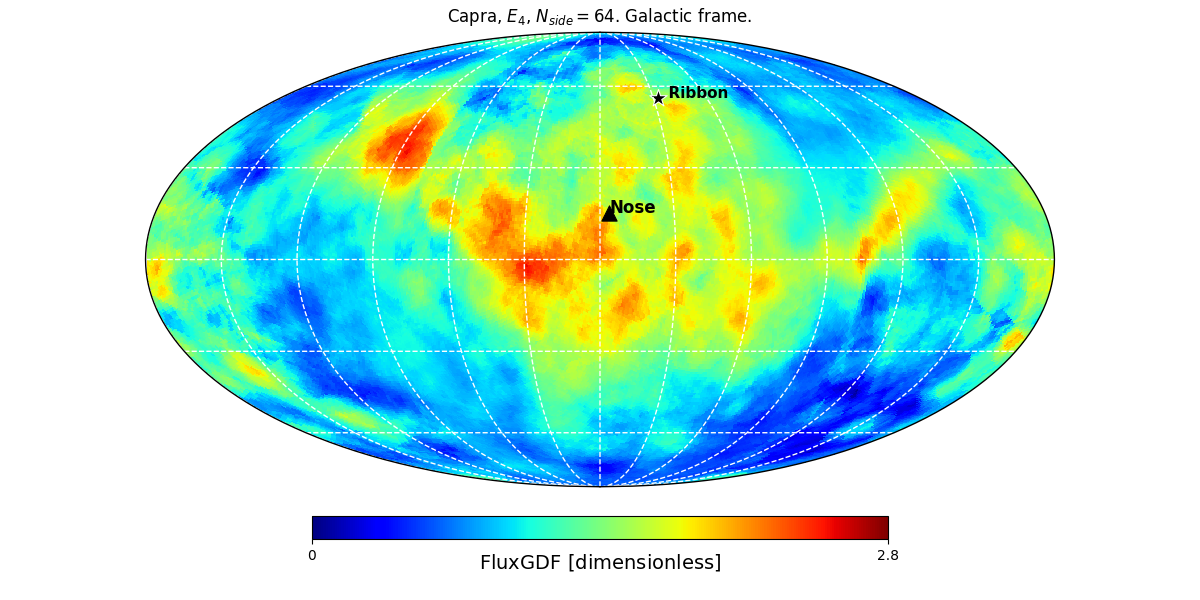}
    \caption{$\mathrm{Galactic \, reference\, frame}$}
  \end{subfigure}
  \hfill
  \begin{subfigure}[b]{0.45\textwidth}
    \centering
    \includegraphics[height=5cm]{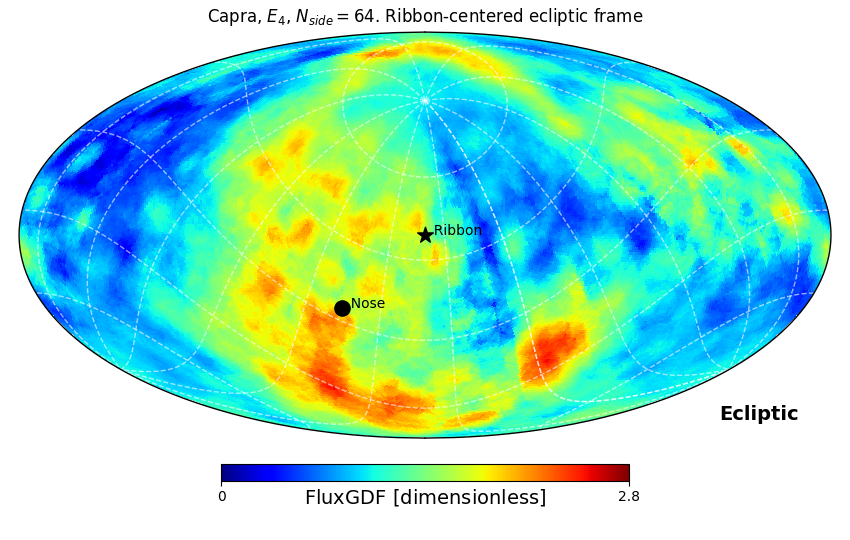}
    \caption{$\mathrm{Ribbon\text{-}centered \, ecliptic \, reference\, frame}$}
  \end{subfigure}

  \caption{Distribution of the GDF-normalized intensity density in the Galactic (left) and Ribbon–centered ecliptic (right) reference frames. Both maps are coordinate transformations of Figure~\ref{fig:thes_hp_gdf_comparison} (m).}
  \label{fig:persp}
\end{figure}

As shown in Figure~\ref{fig:persp}, in the Galactic frame (left), the Ribbon appears as a broad, nearly circular band projected across a wide portion of the sky. Although the Ribbon center lies at high Galactic latitude, the Ribbon itself is located at a large angular distance from that center, with an angular radius of order $(70^\circ)$. Consequently, part of the projected band passes close to the Galactic equator. This visual proximity to the Galactic plane is therefore a consequence of the chosen coordinate projection and of the Ribbon's large angular radius; it does not indicate that the Ribbon is physically associated with the structure of the Milky Way.

In the Ribbon-centered frame (right), the Ribbon center is placed at the pole of the coordinate system. In this representation, an ideal circular Ribbon would appear as a band of approximately constant angular distance from the pole, rather than as a great circle or a meridian. This projection therefore highlights the approximate circularity and coherence of the Ribbon morphology and clarifies its geometry relative to the heliospheric nose and tail, independent of either the Galactic or ecliptic coordinate systems.

\subsection{Tessellation change}

These transformations motivate constructing the native \textsc{Capra} products on a sufficiently fine base HEALPix tessellation: the map can always be downsampled afterward, either over the full sky or only in selected regions, whereas information lost by starting from an unnecessarily coarse tessellation cannot be recovered. The fine base tessellation should still be chosen consistently with the instrument sampling, counting statistics, and computational cost; it should not be interpreted as increasing the physical angular resolution of the data. Since HEALPix pixels are equi-areal, coarser tessellations can be obtained by averaging the finer resolution pixels that fall within the larger one. In the context of this section, we only deal with the intensities, so we perform the simple arithmetic mean; if we were dealing with the counts, signal or rates, we would perform exposure-weighted mean. This preserves the global normalization and geometric consistency of the intensity distribution (see Section \ref{subsec:relnorm}). 

We let a scalar field $F_j$ be defined over a HEALPix grid with a tessellation $N_\mathrm{side}$.
The goal is to transform it to a new grid of tessellation $N'_\mathrm{side}$, either \emph{uniformly} across the entire sky or \emph{partially} over a selected region. Each original pixel $j$ stores the intensity calculated in the pipeline as $F_j$.

\subsubsection{Uniform tessellation change}

For each pixel $j'$ in the coarser grid, we identify the set of original pixels $\mathcal{S}(j')$ that geometrically fall inside it. For intensity maps, which are density-like quantities, the coarse-pixel value is the area-weighted mean of the contributing fine pixels. Because all HEALPix pixels at a fixed $N_{\rm side}$ have equal areas, this reduces to the arithmetic mean, i.e., 
\begin{equation}
F'_{j'} =
\frac{1}{|S(j')|}
\sum_{j \in S(j')} F_j ,
\end{equation} \label{eq:uniform}
where $S(j')$ is the set of fine pixels contained in the coarse pixel $j'$.
This preserves the integrated intensity over the rebinned region,
\begin{equation}
\sum_{j'} F'_{j'} \Omega'_{j'} =
\sum_j F_j \Omega_j ,
\end{equation}
up to the treatment of masked pixels. If a coarse pixel contains both observed and unobserved fine pixels, the average is computed only from the observed contributors. The resulting coarse map can therefore reduce small masked gaps, but such pixels should still be interpreted as averages over partial coverage rather than as newly observed sky.

\subsubsection{Partial tessellation change}

In some cases, we will only need to reduce resolution over a selected region of the sky, for example to coarsen sparsely observed areas, while keeping the interesting regions highly resolved.
We select a rectangular region (in this example one, but can be easily generalized to more)  $\mathcal{R}_m$ centered at $(\theta_m,\phi_m)$ with latitude and longitude margins $(\Delta\theta_m,\Delta\phi_m)$,
\begin{equation}
\mathcal{R}_m =
\{ (\theta,\phi)\,|\, |\theta-\theta_m|\le \Delta\theta_m,\;
|\phi-\phi_m|\le \Delta\phi_m \},
\end{equation}
the tessellation change is applied only within $\mathcal{R}_1\cup\mathcal{R}_2$.  
The pixels outside these regions retain their original values.

For all pixels in the selected region,
\begin{equation}
F'_{j'} =
  \frac{1}{|\mathcal{S}(j')|}
  \sum_{j \in \mathcal{S}(j') \cap \mathcal{R}} F_j,
\label{eq:partial}
\end{equation}
and for $j'\notin\mathcal{R}$,
\[
F'_{j'} = F_{j'}.
\]
This allows for local coarsening of selected regions without altering regions of interest. Note, that the user may define a rectangular region, but the actual coarsened region is snapped to a set of complete target HEALPix pixels.

\begin{figure*}[ht!]
  \centering
  \begin{subfigure}[b]{0.48\textwidth}
    \includegraphics[width=\textwidth]{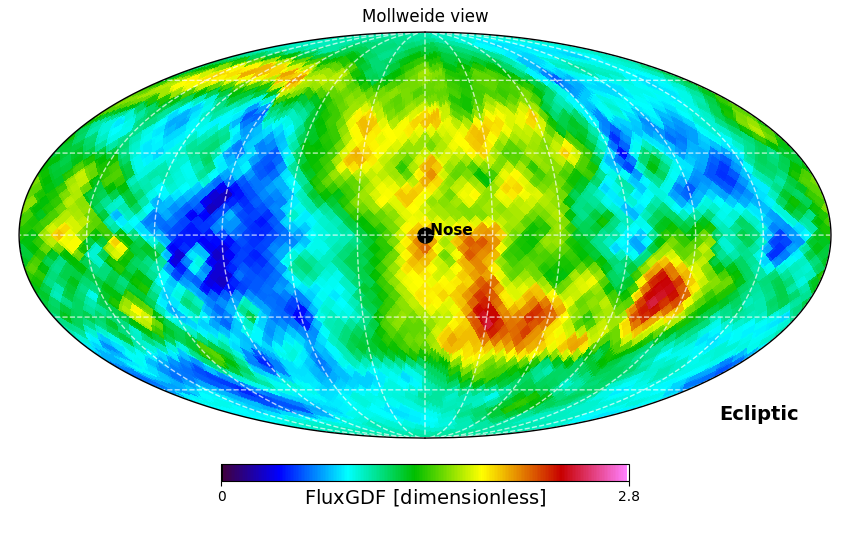}
    \caption{Uniform tessellation change, GDF-normalized map.  The entire map is rebinned to a lower $N_\mathrm{side}$ using Eq.~(\ref{eq:uniform}).}
  \end{subfigure}
  \hfill
  \begin{subfigure}[b]{0.48\textwidth}
    \includegraphics[width=\textwidth]{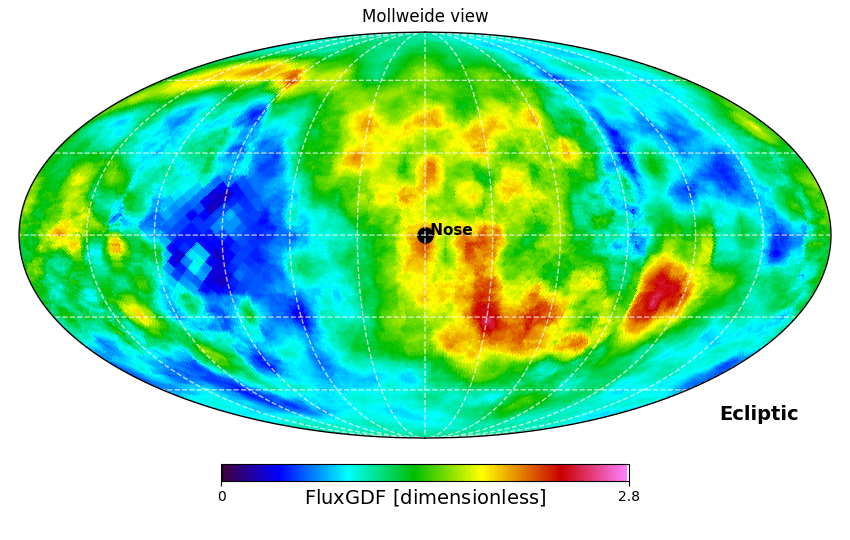}
    \caption{Partial tessellation change, GDF-normalized map.  Only the selected regions are rebinned according to Eq.~(\ref{eq:partial}).}
  \end{subfigure}
  \caption{Examples of uniform and partial tessellation transformations applied to GDF-normalized intensity maps from \textsc{Capra}.  
  In the partial case, one selected patch is coarsened while the rest of the map retains full resolution. The patch is defined in colatitude-longitude coordinates and is centered at $((\theta,\phi)=(90^\circ,0^\circ))$, with half-widths $(\Delta\theta=20^\circ)$ and $(\Delta\phi=40^\circ)$, corresponding to $(70^\circ \leq \theta \leq 110^\circ)$ and $(-40^\circ \leq \phi \leq 40^\circ)$. Unobserved (masked) pixels become integrated into the averaged fields, leading to almost continuous sky coverage (almost fully covered map).}
  \label{fig:tess-change}
\end{figure*}

Both transformations conserve the total intensity, as later shown as part of the diagnostics in Appendix \ref{sec:conservation}, e.g., Figure \ref{fig:diagnostics_all}. In both cases, the arithmetic mean ensures that pixel values remain density–like (intensity per steradian) rather than cumulative quantities. The corresponding implementations, \textsc{uniformTessellationChangeHP} (for the uniform tessellation change) and \textsc{partialTessellationChangeHP} (for the partial tessellation change), are provided within the \textsc{Capra} postprocessing package.

As mentioned in the beginning of this section, tessellation transformations are valuable because they allow us to adjust the map resolution to the needs of a given analysis without recomputing the full sky maps from the baseline spacecraft data. Uniform coarsening provides a fast and consistent way to generate
lower-resolution products for visualization or statistical comparisons.
Partial coarsening is particularly useful when only selected regions of
the sky require smoothing, for example, sparsely observed domains or
areas dominated by statistical noise, while scientifically important
regions, such as the Ribbon or the nose, retain their full
native angular detail.

These resolution controls are also essential when comparing maps produced
by different instruments or mapping algorithms, which often operate at
incompatible native tessellations.  
By uniformly or selectively downsampling the HEALPix products, we can
match the angular resolution of external datasets (e.g., \textsc{Theseus})
and ensure that statistics and diagnostics are performed
on a common geometry.

\subsection{Projection onto user-defined rectangular grids} \label{subsec:diff-res}

\begin{figure}[htb!]
  \centering
  \includegraphics[width=0.7\textwidth]{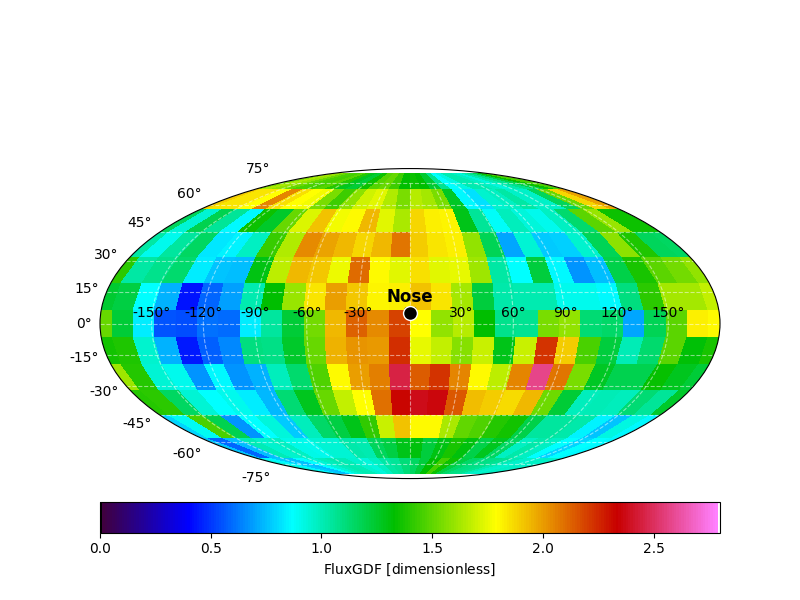}
  \caption{Example of a HEALPix to grid transformation using an alternative
rectangular layout with $12 \arcdeg \times 12 \arcdeg$ cells, GDF-normalized map.  
This shows that the framework is not limited to the
$6\arcdeg \times6 \arcdeg$ resolution adopted elsewhere in the paper, but can
produce maps at arbitrary user-defined angular scales.}
  \label{fig:diff-res}
\end{figure}

The same grid-transformation procedure can be applied at angular resolutions other than the $6^\circ \times 6^\circ$ grid used for the main comparisons. Figure~\ref{fig:diff-res} shows an example using $12^\circ \times 12^\circ$ cells. This coarser grid is useful for quick-look visualization or analyses where suppressing small-scale noise is more important than retaining angular detail. The choice of rectangular-grid resolution is a postprocessing choice, not a limitation of the reconstruction.

\section{\textbf{Conclusions}} \label{sec:conclusions}

In this work we have presented \textsc{Capra}, a HEALPix-based pipeline for reconstructing all-sky ENA intensity maps from binned event-counting spacecraft surveys. The method is designed to keep the reconstruction geometrically transparent: the measurements are projected directly on the sphere, the exposures are propagated together with the signal, and the final products can be rotated, rebinned, or projected onto comparison grids without rerunning the full reconstruction.

\textsc{Capra} provides two map products. The baseline product assigns each scan-angle bin to the corresponding HEALPix direction and therefore follows the sampling of the data as directly as possible. An optional smoothing step redistributes the signal over neighboring pixels using a user-defined kernel. In this work we used a Gaussian kernel to reduce the viewing strip boundaries and small-scale mottling in the relatively sparse IBEX-Hi test dataset. This smoothing is a reconstruction choice and not a part of the instrument response, and it trades angular resolution for improved visual continuity and lower sampling-driven noise.

Using IBEX-Hi data, we showed that \textsc{Capra} recovers the main large-scale ENA morphology, including the Ribbon and broad nose/anti-nose structure. Comparisons with baseline IBEX maps and \textsc{THESEUS} products show strong agreement in the location of the dominant sky features. The remaining differences are mainly methodological: \textsc{Capra} more directly preserves small-scale texture from the sampled IBEX data, whereas THESEUS uses statistical denoising and PSF deconvolution, producing maps that can appear smoother while also recovering narrower reconstructed Ribbon structure.

We also introduced a common postprocessing framework for relative and GDF-normalized intensity maps. The relative normalization is useful for comparing morphology independent of the absolute intensity scale, while the GDF normalization compares each map relative to its own off-Ribbon reference level. Together, these two normalizations separate algorithmic morphology comparisons from comparisons of Ribbon/background contrast.

The numerical tests we performed show that the reconstruction preserves the relevant quantities across coordinate transformations and changes in tessellation. Counts, background-subtracted signal, rates, and intensities remain consistent under the tested transformations, and the expected pixel-level relations between signal, exposure, rate, and intensity are recovered. 

The convergence tests indicate that, for the IBEX-Hi dataset used here, $N_{\rm side}=64$ provides a practical balance between angular resolution and computational cost.
We observed a near-linear speedup out to 8 cores, along with high efficiencies ($\approx 70\%$) out to 16 cores, which confirms that the algorithm parallelizes naturally due to the independence of pixel-level computations. We verified $O(N_{\mathrm{pix}})$ scaling and stable memory usage, which enables the efficient creation of high-resolution reconstructions ($N_{\mathrm{side}}\geq 64$) using standard HPC resources. The observed near-linear scaling at low core counts, together with bounded memory usage, makes \textsc{Capra} suitable for high-resolution all-sky reconstructions on standard HPC systems.

These results support \textsc{Capra} as a transparent and scalable reconstruction framework for current IBEX analyses, future IMAP ENA mapping and other event-counting astrophysical applications. Its HEALPix-native structure makes it well suited for workflows that require consistent treatment of spherical geometry, flexible resolution changes, reference-frame transformations, and direct comparison between reconstruction products.

\begin{acknowledgments}
The authors thank Jonathan Gorard for constructive discussions and careful proofreading that improved both the technical accuracy and readability of this work. The authors also thank Iwona Nowierska for constructing one of the figures included in this manuscript, which significantly improved the clarity of the presentation.   
N.B. acknowledges the hospitality of the Center for Computational Astrophysics-Flatiron Institute. N.B. also acknowledges support from the URI Institute for AI $\&$ Computational Research. The computations were performed on the UMass-URI UNITY high-performance computing (HPC) cluster hosted at the Massachusetts Green HPC Center. 
M.B. was supported by Polish National Science Centre grant 2023/51/B/ST9/01921.
\end{acknowledgments}

\begin{contribution}

N.B. led the analysis, software development and writing of the paper. D.R. acquired the observations and contributed relevant scientific expertise to the project as well as contributed to the preparation of this manuscript. M.B. suggested this study and oversaw the project progress. 


\end{contribution}

%



\appendix


\section{\textbf{HEALPix discretization scheme}} \label{sec:hp}
\noindent
To reconstruct sky maps, the system needs events count in the time domain, connected with the position of the instrument viewing axis in the sky, or histograms of the registered events, binned over the spin angle for all orbit arcs selected for the map reconstruction. 
The proposed system should have equal-area pixels with easily scalable dimensions, and regular shapes. For that purpose we have chosen the HEALPix (Hierarchical Equal Area iso-Latitude Pixelization) scheme \citep{gorski_etal:05a}. 

\begin{figure} [bht!]
  \centering
  \includegraphics[width=.4\textwidth]{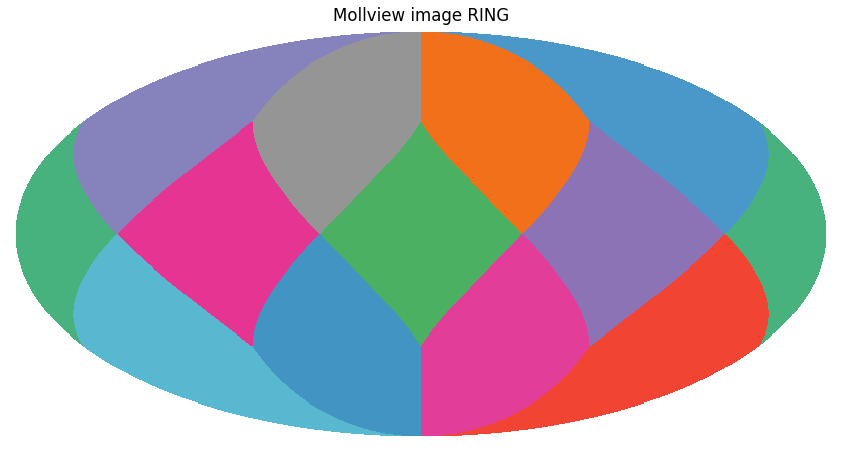}\includegraphics[width=.4\textwidth]{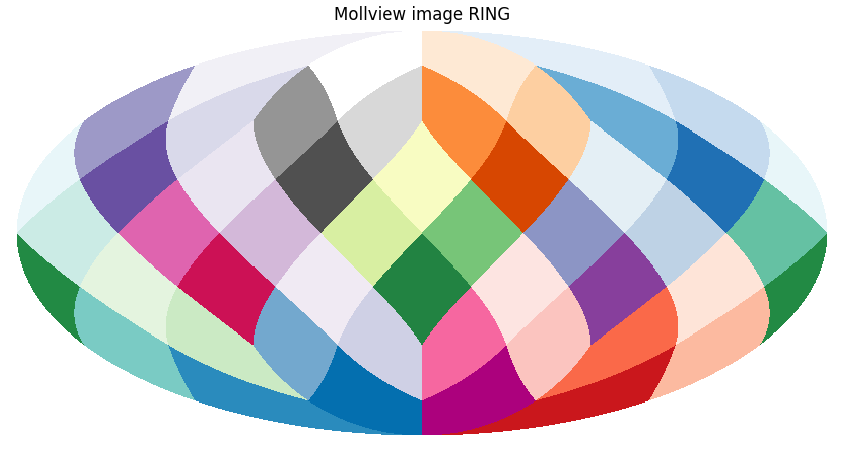}
  \includegraphics[width=.4\textwidth]{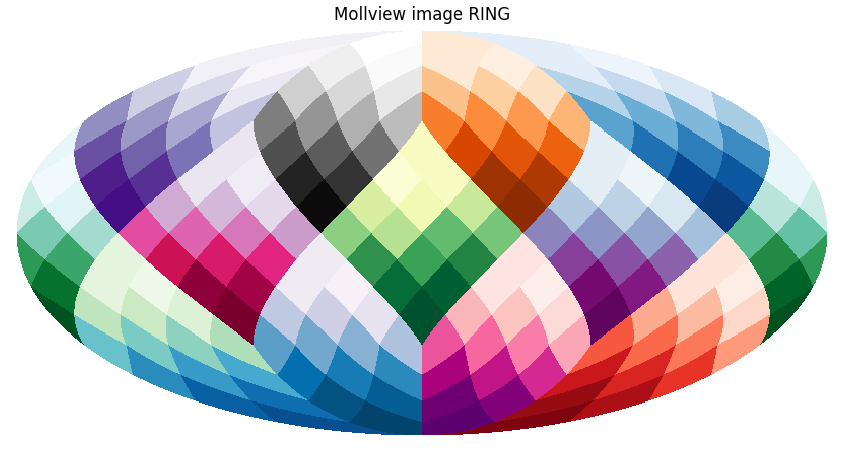}\includegraphics[width=.4\textwidth]{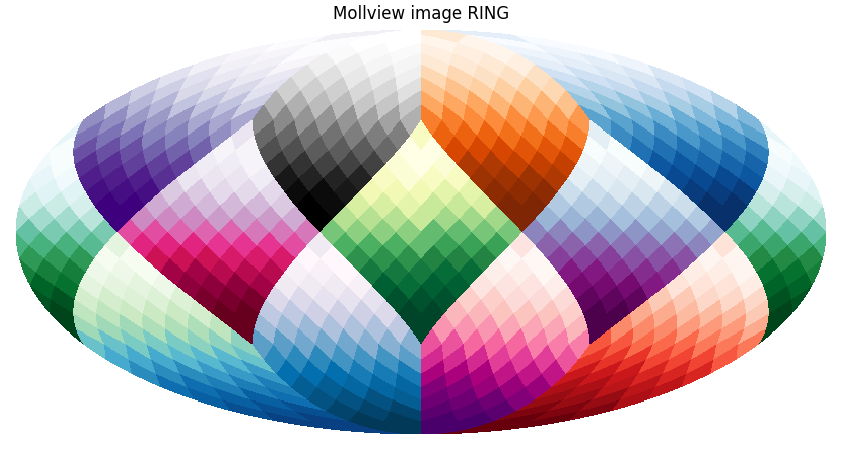}
  \caption{View of example HEALPix partitions of the sphere. Individual colors mark twelve identical (area wise) base-resolution pixels (polar and equatorial). Inside every base-resolution pixel different gradients mark smaller identical (area wise) sub-pixels, which depend on the tessellation. Ordering of the spheres (from top left clockwise) corresponds to the tessellations $N_{\mathrm{side}} \in \{ 1, 2, 4, 8\}$. }
  \label{fig:healpix-partition}
\end{figure}

In this system, the sky is divided into polar caps and and equatorial belt (Figure \ref{fig:healpix-partition}, upper left panel), each of them divided into equal-area fields in the azimuthal direction. This is characterized by the two parameters:

- $N_\Theta$ - a number of the base-resolution pixel layers between the north and south poles,

- $N_\Phi$ - a multiplicity of the meridional cuts, or the number of equatorial, or circum-polar base-resolutions pixels.

For the purpose of this paper, $N_\Theta = 3$ and $N_\Phi = 4$ were adopted, hence our sky is divided into two polar caps and a polar belt ($N_\Theta$ = 3), and each of them is divided into $N_\Phi$ = 4 fields. 
Each of the baseline lobes can be subdivided into 4 equal-area pixels, and each of these into 4 smaller areas. This results in an increasingly fine tessellation of the sky. The level of subdivisions is parameterized with the parameter $N_{\mathrm{side}}$, called the tessellation. Figure \ref{fig:healpix-partition} presents Mollweide projections of the spheres partitioned with $N_{\mathrm{side}} = 1, 2, 4$ and $8$. The different hues mark the original lobes, and the shades within the lobes mark the pixel numbering. The subdivision process can go until a desired resolution is obtained \citep{gorski_etal:05a}. 

There are two alternative ways of numbering the pixels that HEALPix offers. 
The first is the ring scheme, which counts the pixels moving down from the north to the south pole along each iso-latitude rings. The second one is the nested scheme, which allows to replicate the tree structure of pixel numbering. 
For the purpose of this paper we have chosen the ring-scheme. HEALPix maps are organized as a 1D lists of pixels that are identified with a sorting number $j$, where $j \in \{1,2,...,N_{\mathrm{pix}} \}$ and $N_{\mathrm{pix}} = 12N_{\mathrm{side}}^2$. Throughout the algorithm, sorted Cartesian coordinates are stored and often appended to various quantites, which helps to maintain the correct order and scheme.

After choosing the HEALPix tessellation $N_{\mathrm{side}}$, one can calculate the pixel coordinates. The fact that HEALPix pixels are equal-area greatly facilitates numerical integration over partitions of the sky of functions projected on the HEALPix-tessellated sky.

\section{\textbf{Validation and testing}} \label{sec:conservation}

\subsection{Numerical verification of preservation}
To verify that the pipeline preserves physical quantities within different tessellations and energy channels, we implemented a suite of \textit{preservation verifications}. These diagnostics ensure that quantities like counts, signal rates, and intensities remain internally consistent when transformed between the detector space and HEALPix map representations. In other words, we check their invariance under coordinate transformations. We show the summary of the quantities, their interpretation and invariance in Table \ref{tab:conservation_summary}.

\subsubsection{Formulation}

For each pixel $j$, the count, exposure, signal, and rate are denoted as
\[
C_j,\quad E_j,\quad S_j,\quad R_j,
\]
with the solid angle per pixel
\[
\Omega = \frac{\pi}{3\,N_\mathrm{side}^2},
\]
where $N_\mathrm{side}$ is the HEALPix tessellation.  

\underline{Expected rates}

The \textit{expected} signal rate per pixel, i.e., the surface brightness obtained directly from the signal and exposure, is
\begin{equation}
R_j^{\mathrm{exp}} = 
  \frac{S_j}{E_j},
\end{equation}
and the local signal rate difference is
\begin{equation}
\Delta R_j = R_j - R_j^{\mathrm{exp}}.
\end{equation}
$\Delta R_j$ should be 0 across all $N_{\mathrm{side}}$ and all energy channels.

\underline{Global checks}

The total signal and the map-integrated signal are
\[
S_{\mathrm{tot}} = \sum_j S_j, 
\qquad
M_{\mathrm{tot}} = \sum_j R_j\,E_j\, ,
\]
and the corresponding signal rate normalization factor is
\begin{equation}
K_R = 
  \frac{M_{\mathrm{tot}}}{S_{\mathrm{tot}}}.
\end{equation}
$K_R$ should be 1 across all $N_{\mathrm{side}}$ and ESA (ideal conservation).
Global mean signal rates derived from the signal and from the reconstructed map are
\[
R_{\mathrm{sig}} = 
  \frac{S_{\mathrm{tot}}\Omega}{E_{\mathrm{tot}}\,\Omega},
\qquad
R_{\mathrm{map}} =
  \frac{M_{\mathrm{tot}}\Omega}{E_{\mathrm{tot}}\,\Omega},
\]
where $E_{\mathrm{tot}} = \sum_i E_i$ is the total exposure.
$R_{\mathrm{sig}}$ should be constant across $N_{\mathrm{side}}$, because counts and exposures scale with $N_{\mathrm{pix}}$, which makes this invariant with $N_{\mathrm{side}}$, but it changes with ESA, because physical count rates differ with energy channels. The same logic applies to the $R_{\mathrm{map}}$. Mathematically, these two are equivalent. However, if $R_j$ has been filtered/masked/interpolated differently than $S$ and $E$, then we will see the difference, since $R_{\mathrm{map}}$ reflects exposure weighted mean of the constructed rate map, while $R_{\mathrm{sig}}$ is the aggregate physical rate derived from the signal and exposure.

\underline{Signal reconstruction}

To check that signals correspond to rate–exposure products, we compute
\begin{equation}
\Delta S = 
  \sum_j (R_j\,E_j - S_j),
\end{equation}
which should vanish within numerical precision, across all $N_{sides}$ and all ESA.

\underline{Bright-region stability}

For the pixels above the 95th percentile of the nonzero valid rate map, the mean rate is
\begin{equation}
R_{\mathrm{ROI}} =
  \frac{\sum\limits_{j \in \mathrm{ROI}} R_j\, E_j}
       {\sum\limits_{j \in \mathrm{ROI}} E_j\,  }, 
\end{equation}
where \textit{ROI} refers to the ``region of interest'' and $R_{\mathrm{ROI}}$ provides a measure of stability in high-signal regions with different resolutions, because this mean rate should be stable across different $N_{\mathrm{side}}$, but change with ESA, because we encounter different intensity regimes. The ROI is defined as the set of nonzero valid pixels with \(R_j\) above the 95th percentile of the nonzero valid rate distribution.

\underline{Rate–intensity connection}

The conversion from rate to intensity follows directly from geometric and energetic scaling:
\begin{equation}
F_j = 
  \frac{R_j}{G\,C_e},
\end{equation}
where $G$ is the geometric factor and $C_e$ is the central energy of the ESA step (listed in Table \ref{tab:G_E_values}).
Substituting the expression for $R_j^{\mathrm{exp}}$ gives
\begin{equation}
F_j^{\mathrm{exp}} = 
  \frac{S_j}{E_j\,G\,C_e},
\end{equation}
with per-pixel difference
\begin{equation}
\Delta F_j = F_j - F_j^{\mathrm{exp}}.
\end{equation}
The global intensity normalization coefficient, analogous to $K_R$, is
\begin{equation}
K_F =
  \frac{\sum_j F_j E_j \, G \, C_e}
       {\sum_i S_j}.
\end{equation}
$K_F$ should be 1 across all $N_{\mathrm{side}}$ and all ESA.
Global mean intensities derived from the map and from the expected signal are
\[
F_{\mathrm{map}} =
  \frac{\sum_j F_j E_j \Omega}{E_{\mathrm{tot}}\Omega}, \qquad
F_{\mathrm{exp}} =
  \frac{\sum_j F_j^{\mathrm{exp}} E_j \Omega}{E_{\mathrm{tot}}\Omega}.
\]
They both should be stable across all $N_{\mathrm{side}}$, since it is the same sky-integrated intensity, but vary with ESA, because the sky spectrum varies, or in other words physical intensity scales with energy. 

\underline{Implementation notes}

In practice, null or masked pixels were replaced by zeros, ensuring that missing data did not propagate through the global sums. All computations were performed in double precision (\textsc{N@}) and truncated for rounding error using \textsc{Chop}.  
The pixel solid angle $\Omega = \pi/(3\,N_\mathrm{side}^2)$ corresponds to the exact area of an equal-area HEALPix cell in steradians.

\subsubsection{Results and interpretation}

\begin{figure*}[ht!]
  \centering

  \begin{subfigure}[b]{0.19\textwidth}
    \centering
    \includegraphics[width=\textwidth]{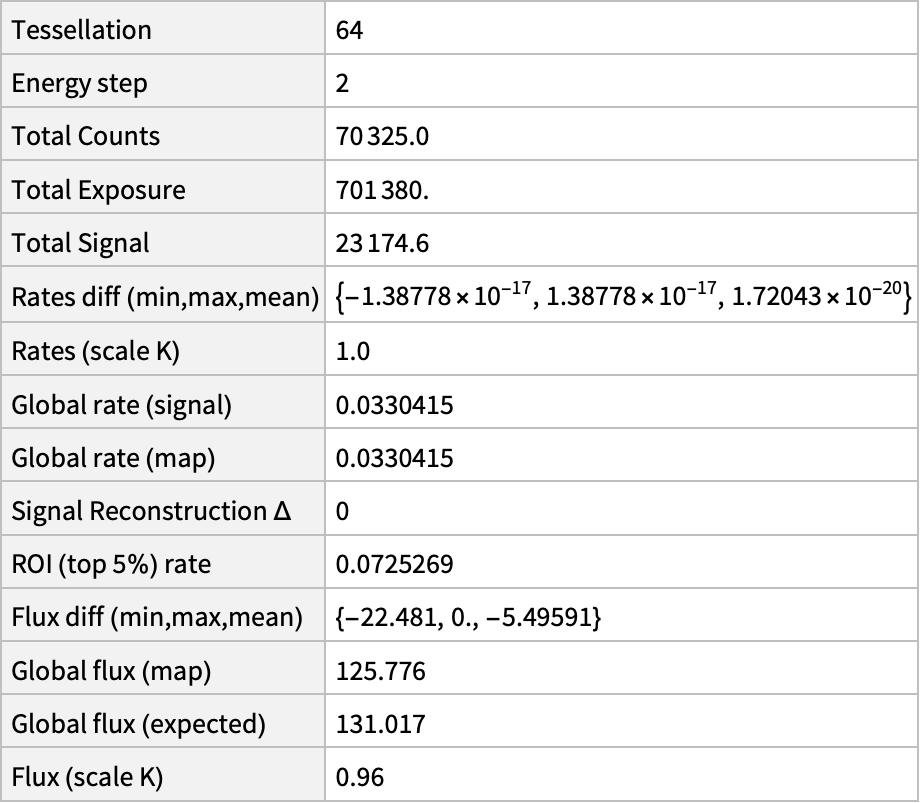}
    \caption{ESA 2, $N_\mathrm{side}=64$}
  \end{subfigure}\hfill
  \begin{subfigure}[b]{0.19\textwidth}
    \centering
    \includegraphics[width=\textwidth]{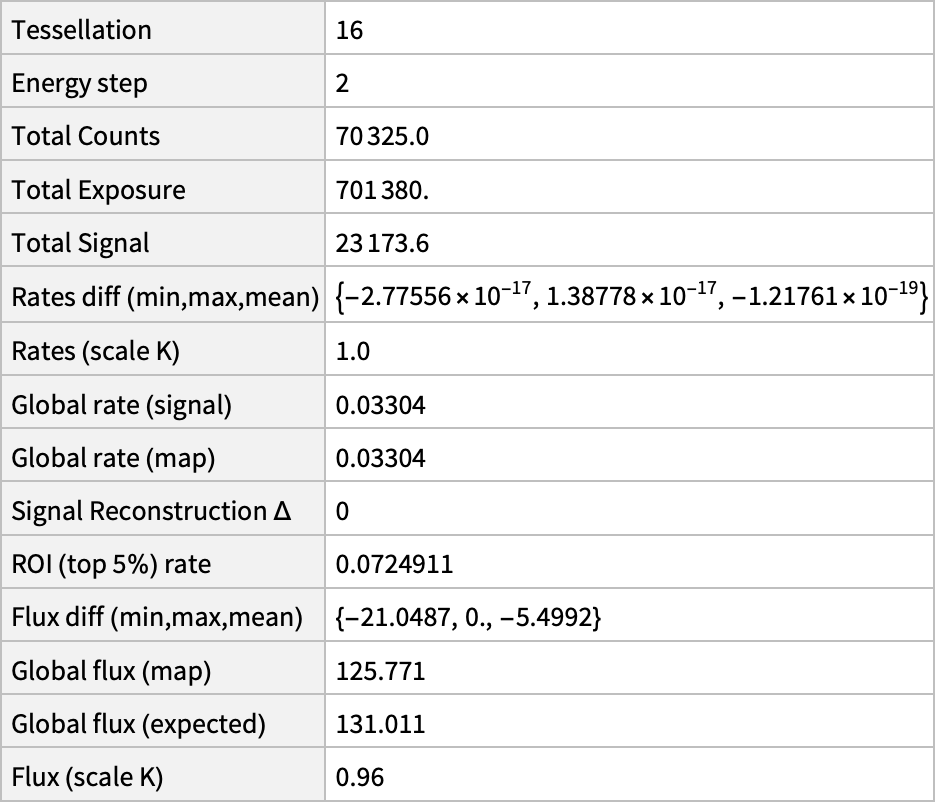}
    \caption{ESA 2, $N_\mathrm{side}=16$}
  \end{subfigure}\hfill
  \begin{subfigure}[b]{0.19\textwidth}
    \centering
    \includegraphics[width=\textwidth]{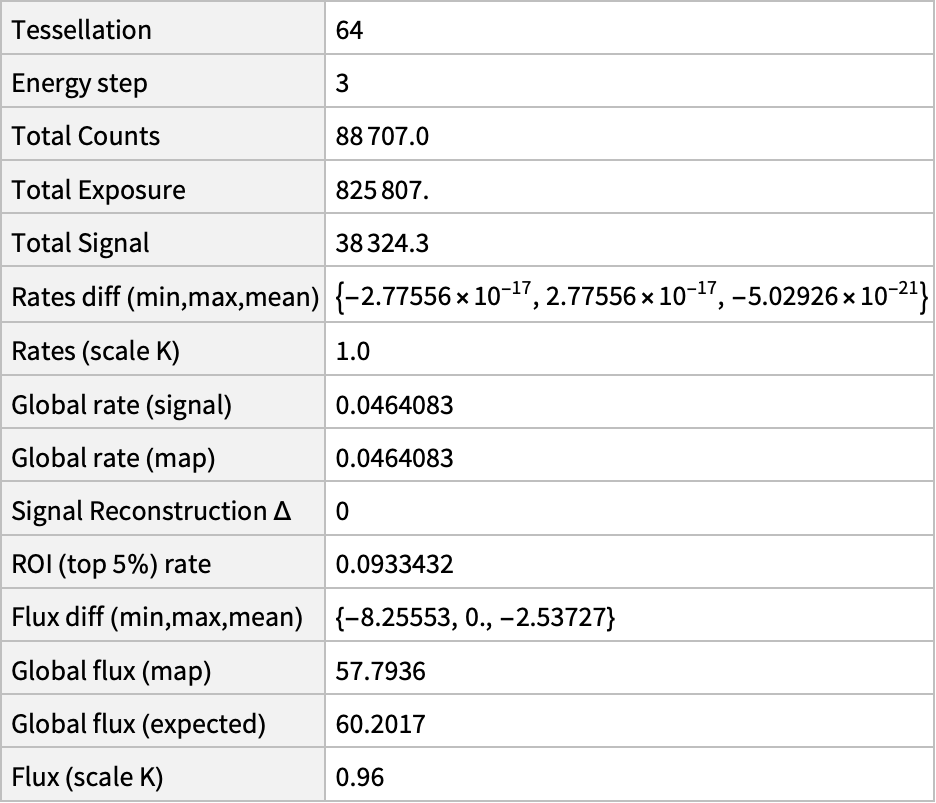}
    \caption{ESA 3, $N_\mathrm{side}=64$}
  \end{subfigure}\hfill
  \begin{subfigure}[b]{0.19\textwidth}
    \centering
    \includegraphics[width=\textwidth]{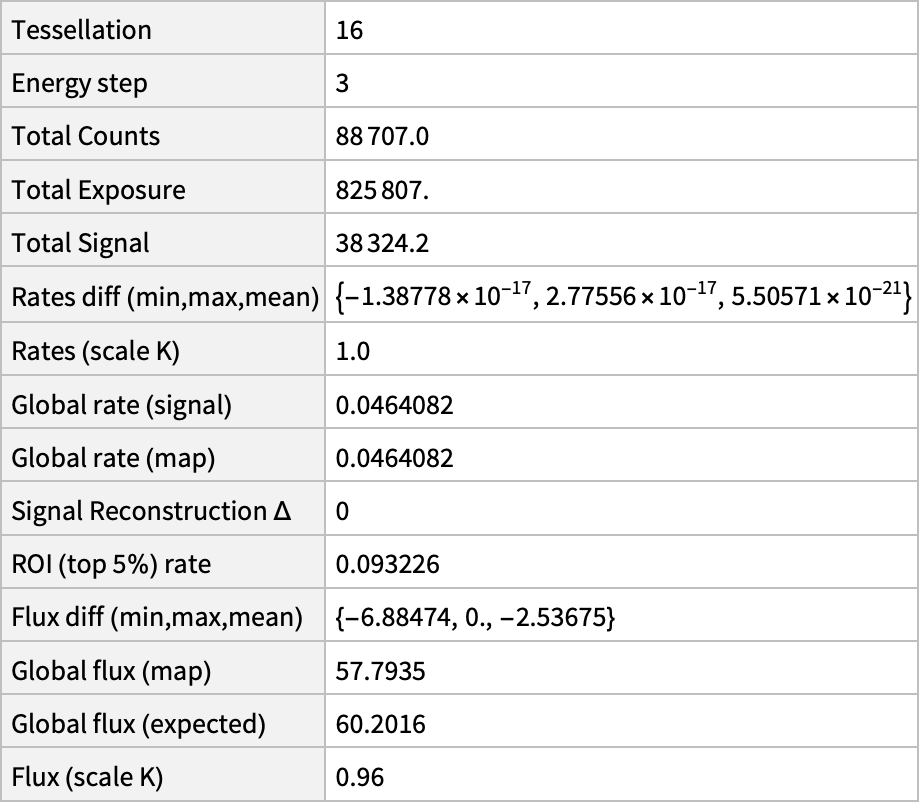}
    \caption{ESA 3, $N_\mathrm{side}=16$}
  \end{subfigure}\hfill
  \begin{subfigure}[b]{0.19\textwidth}
    \centering
    \includegraphics[width=\textwidth]{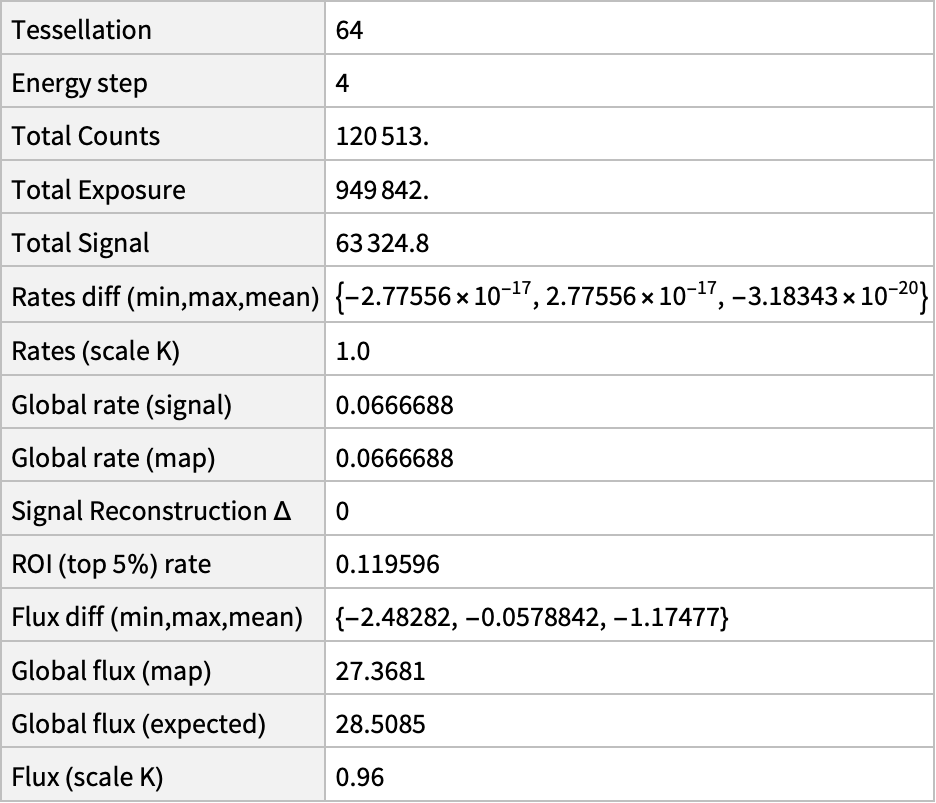}
    \caption{ESA 4, $N_\mathrm{side}=64$}
  \end{subfigure}

  \vspace{0.8em}

  \begin{subfigure}[b]{0.19\textwidth}
    \centering
    \includegraphics[width=\textwidth]{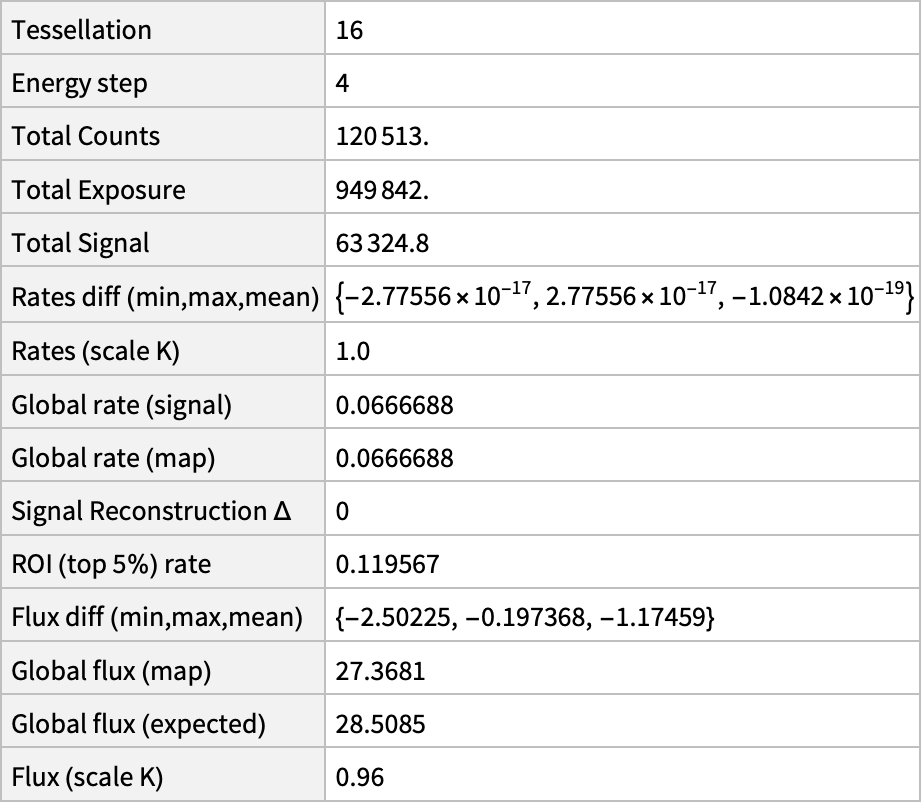}
    \caption{ESA 4, $N_\mathrm{side}=16$}
  \end{subfigure}\hfill
  \begin{subfigure}[b]{0.19\textwidth}
    \centering
    \includegraphics[width=\textwidth]{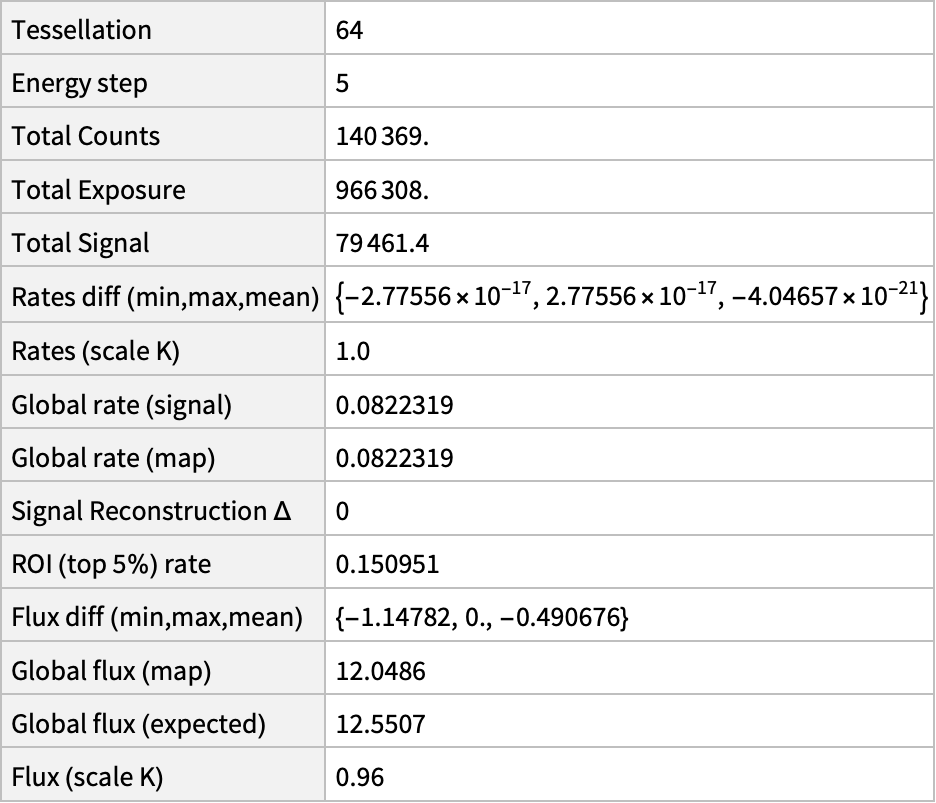}
    \caption{ESA 5, $N_\mathrm{side}=64$}
  \end{subfigure}\hfill
  \begin{subfigure}[b]{0.19\textwidth}
    \centering
    \includegraphics[width=\textwidth]{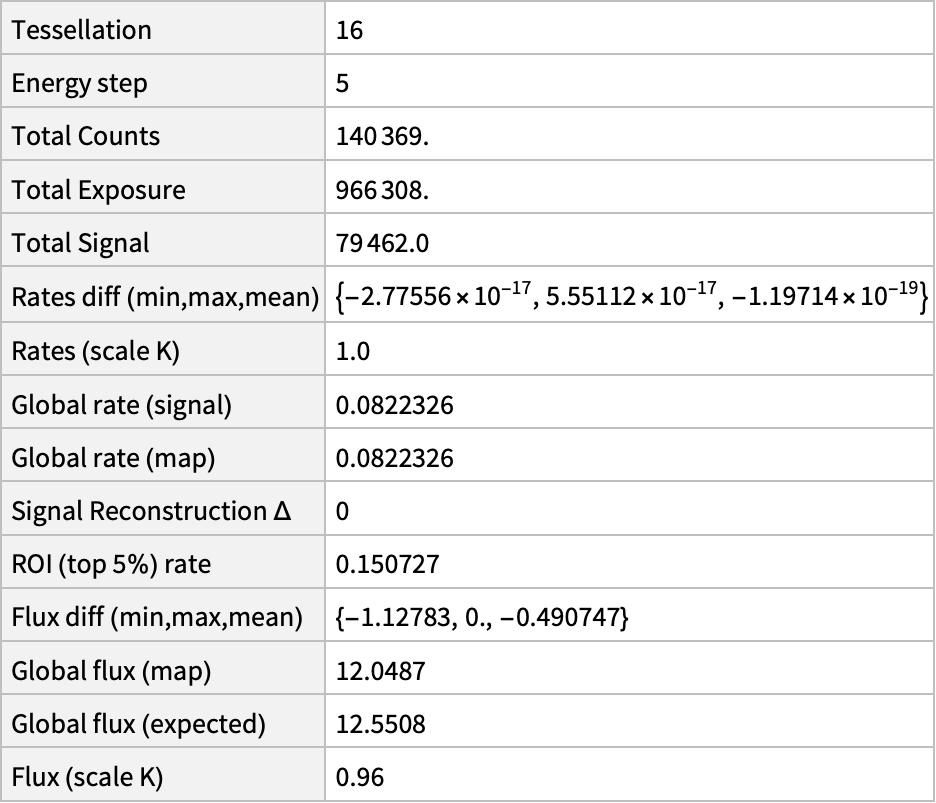}
    \caption{ESA 5, $N_\mathrm{side}=16$}
  \end{subfigure}\hfill
  \begin{subfigure}[b]{0.19\textwidth}
    \centering
    \includegraphics[width=\textwidth]{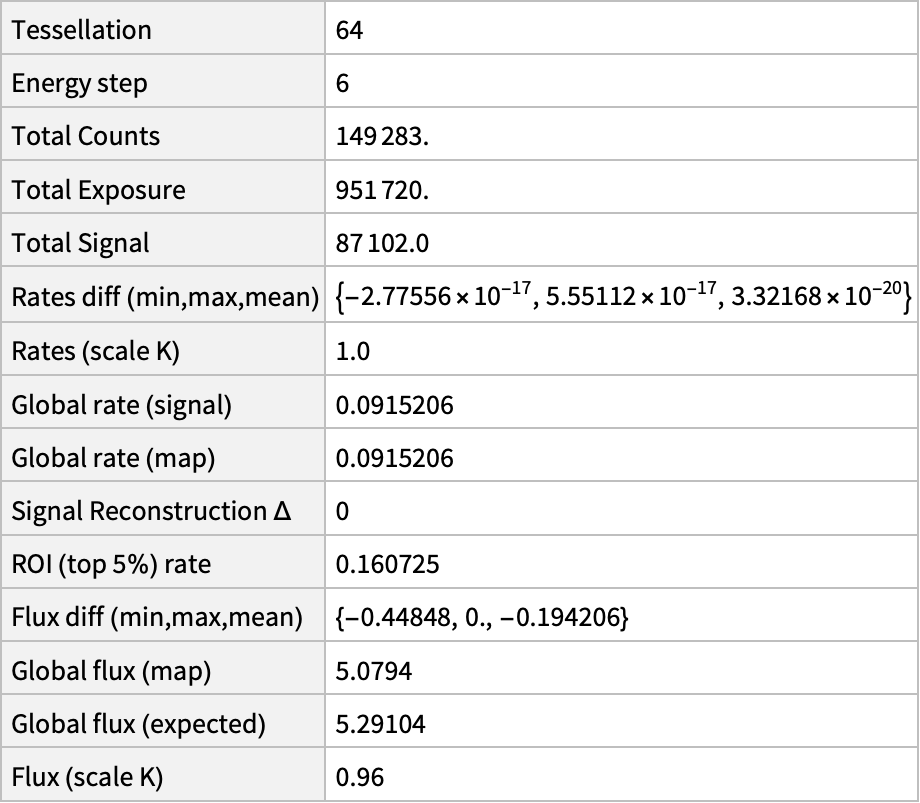}
    \caption{ESA 6, $N_\mathrm{side}=64$}
  \end{subfigure}\hfill
  \begin{subfigure}[b]{0.19\textwidth}
    \centering
    \includegraphics[width=\textwidth]{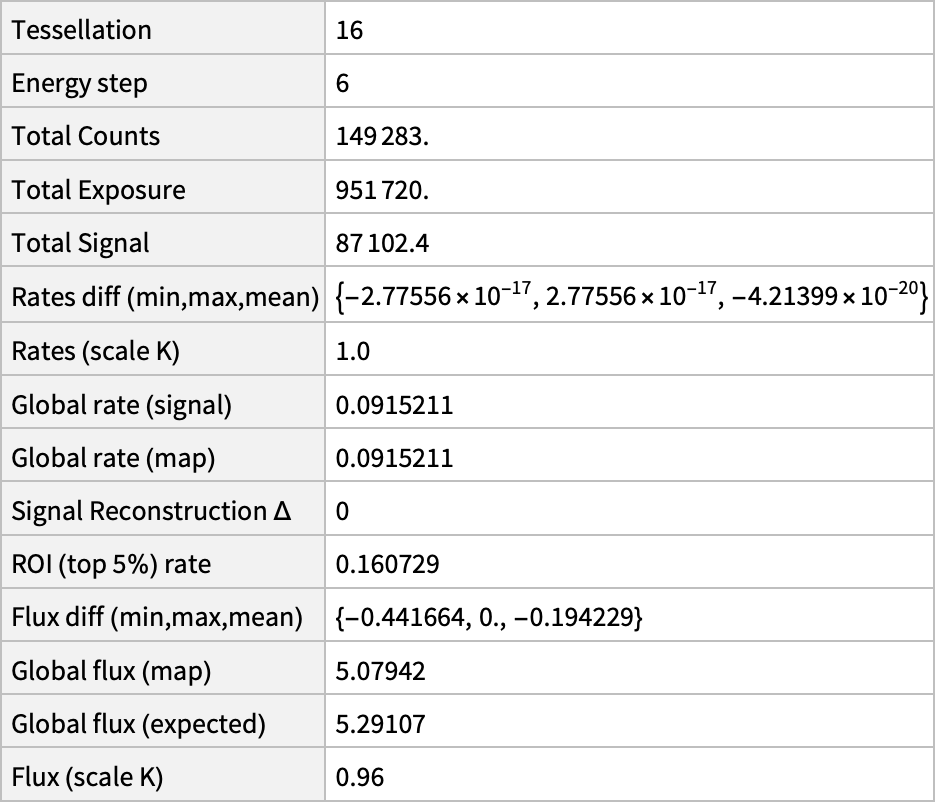}
    \caption{ESA 6, $N_\mathrm{side}=16$}
  \end{subfigure}

  \caption{
  Conservation diagnostics for $N_\mathrm{side}=64$ and $N_\mathrm{side}=16$ across ESA steps~2–6.  
  Each pair of panels shows global and local consistency metrics for the count, rate, and intensity pathways at two resolutions.  
  The near-unity normalization factors and vanishing residuals confirm conservation across all ESA channels and both resolutions.}
  \label{fig:diagnostics_all}
\end{figure*}

In Figure \ref{fig:diagnostics_all}, we present the preservation diagnostics for ESA steps~2-–6 at $N_\mathrm{side}=64$ and $N_\mathrm{side}=16$. Across all energy channels, the rate and intensity normalization coefficients ($K_R$, $K_F$) are $\approx$ 1, confirming that both the rate and intensity pathways conserve the total signal and exposure-weighted quantities.
Pixel-wise differences ($\Delta R_j$, $\Delta F_j$) are typically $ \lesssim 10^{-11}$ or smaller.

The global rates $R_{\mathrm{sig}}$ and $R_{\mathrm{map}}$ are identical (within rounding) and vary only with ESA step, which follows from the expected decline of particle intensity with the increase of energy.

Signal reconstruction differences $\Delta S$ are below $10^{-6}$ counts, which confirms that $S_j \approx R_j E_j$ holds.  Similarly, the fact that $F_{\mathrm{map}} = F_{\mathrm{exp}}$ shows that the geometric-factor and energy scaling are implemented correctly.

The top 5\% ROI rate decreases between ESA~2 and~6, tracing the energy-dependent brightness of the Ribbon region. Those variations are physical, not numerical and align with the expected ENA spectral behavior. Table \ref{tab:conservation_summary} summaries that and gives a brief explanation. 

\begin{table}[h]
\centering
\caption{Summary of preservation diagnostics and their constancy across tessellations and energy channels.}
\begin{tabular}{lccl}
\hline
\textbf{Quantity} & \textbf{Across Tessellations} & \textbf{Across Energy Channels} & \textbf{Interpretation} \\
\hline
Tessellation & varies & varies & Defines map resolution ($N_\mathrm{side}$). \\
Scale $K$ & constant ($\approx1$) & constant ($\approx1$) & Verifies global normalization $\sum M = \sum S$. \\
Pixel differences & $\to0$ & $\to0$ & Local rate/intensity identity; purely numerical check. \\
Global rate (signal) & constant & varies & Same data summed; energy dependence from spectrum. \\
Global rate (map) & constant & varies & Exposure-weighted mean; follows spectral shape. \\
Total counts & constant & varies & Count conservation; differs by ESA channel. \\
Total exposure & constant & nearly constant & Driven by pointing/time; weak energy dependence. \\
Total signal & constant & varies & Conserved under rebinning; reflects intensity spectrum. \\
Signal reconstruction $\Delta$ & $\approx0$ & $\approx0$ & Tests $S\approx\sum R\,E$. \\
ROI (top 5\%) rate & $\approx$constant & varies & Bright-region mean; energy dependence expected. \\
Intensity scale $K$ & $\approx1$ & $\approx1$ & Verifies intensity normalization ($G,C_e,\Omega$). \\
Intensity differences & $\to0$ & $\to0$ & Local intensity identity; only numerical noise. \\
Global intensity (map/expected) & constant & varies & Invariant to tessellation; follows spectral shape. \\
\hline
\end{tabular}
\label{tab:conservation_summary}
\end{table}

Overall, the diagnostics confirm that the pipeline conserves all physically meaningful quantities. Both analytical and numerical definitions of rate and intensity are self-consistent across tessellations and ESA steps. This demonstrates that the geometric, energetic, and exposure scaling relations are implemented correctly and that the mapping procedure is conservative.

\subsection{Resolution scaling - convergence testing}
\label{sec:convergence}

To verify that the pipeline converges numerically with increasing spatial resolution, we performed a \emph{resolution-scaling test}. Here, convergence is defined operationally: the map is considered converged when increasing $N_\mathrm{side}$ changes the reconstructed intensity field by less than the adopted RMSDiff threshold of 5\%, while preserving the Pearson correlation close to unity. In this test, the full pipeline was executed at multiple HEALPix tessellations ($N_\mathrm{side} = 16, 64, 128$), and the resulting intensity maps were compared pairwise in terms of their global structural similarity and normalized root mean-square difference (RMS). This analysis quantifies the degree to which the algorithm converges, i.e.,  gets progressively closer to a stable, resolution-independent limit.

\vspace{0.5em}
\noindent
\underline{Definitions } 

For each pair of maps $f_1$ and $f_2$ produced at resolutions $N_\mathrm{side} = Tess_1$ and $Tess_2$, we compute:

\begin{equation}
\mathrm{RMSDiff}(f_1, f_2) =
\frac{
\sqrt{\frac{1}{N_{\rm valid}}
\sum_{j\in\mathcal{V}}
\left[f_1(j)-\alpha f_2(j)\right]^2}
}{
\sqrt{\frac{1}{N_{\rm valid}}
\sum_{j\in\mathcal{V}} f_1(j)^2}
},
\label{eq:rmsdiff}
\end{equation}
where
\begin{equation}
\alpha =
\frac{\sum_{j\in\mathcal{V}} f_1(j)f_2(j)}
{\sum_{j\in\mathcal{V}} f_2(j)^2},
\end{equation}
and \(\mathcal{V}\) is the set of pixels for which both reprojected maps are valid. It is a dimensionless normalized residual after best-fit amplitude rescaling; it therefore measures the remaining map difference after removing a global multiplicative offset.

We also evaluate the linear Pearson correlation coefficient, as:
\begin{equation}
\rho(f_1, f_2) = 
\frac{\mathrm{Cov}(f_1, f_2)}
{\sigma(f_1)\,\sigma(f_2)},
\label{eq:corr}
\end{equation}
which quantifies morphological (structural) agreement. For the Pearson correlation coefficient, henceforth referred to just as ``the correlation'', a correlation of one means perfect structural alignment. In other words, it checks whether pixels that are bright in one map are also bright in the other, and whether pixels that are faint in one map are faint in the other. For two consecutive levels, we summarize the results in Figure \ref{fig:conv+res}.

\vspace{0.5em}
\noindent
\underline{Power-law convergence} 

In standard numerical schemes when sampling smooth functions on a uniform grid or a sphere, we often find the global error scaling of the form:
\[
\text{error} \;\propto\; h^{\,p},
\]
where \(h\) is the grid spacing and \(p\) is the order of the method \citep{Press2007, Fornberg1988}.  
Since for the HEALPix scheme the effective angular spacing scales as \(h \sim N_\mathrm{side}^{-1}\) (see eq.\,(24) in \citealt{gorski_etal:05a}),  we expect
\[
\mathrm{RMSDiff} \;\propto\; N_\mathrm{side}^{-\alpha}, \quad \alpha = p,
\]
with \(\alpha\) determined empirically by a log–log regression of \(\mathrm{RMSDiff}\) versus the effective resolution \(N_\mathrm{eff}\) (Eq. \ref{eq:neff}).

To characterize how the residual error decreases with increasing map resolution, we model the root-mean square difference between two successive maps as a power law in the effective HEALPix resolution. 
For a pair of tessellations $(N_1, N_2)$, the effective resolution is defined as the geometric mean
\begin{equation}
N_\mathrm{eff} = \sqrt{N_1 N_2},
\label{eq:neff}
\end{equation}
which represents the characteristic scale between the two levels of refinement. 
The empirical convergence relation is then expressed as
\begin{equation}
\mathrm{RMSDiff}(N_\mathrm{eff}) = C\,N_\mathrm{eff}^{\,\alpha},
\label{eq:powerlaw}
\end{equation}
or, in logarithmic form (for regression analysis),
\begin{equation}
\log(\mathrm{RMSDiff}) = A + \alpha\,\log(N_\mathrm{eff}),
\label{eq:fit}
\end{equation}
where $A = \log C$, and $\alpha$ is the convergence exponent or the order of spatial accuracy of the mapping algorithm.

\vspace{0.5em}
\noindent

\underline{Results}

The log–log regression of Equation~\ref{eq:fit} using the measured $\mathrm{RMSDiff(f_1, f_2)}$ values gives a two-point slope
\[
\alpha_\mathrm{fit} = -0.62,
\]
and intercept \[
A = -0.61.
\]
The two map pairs analyzed,
\[
(16,64): \ \mathrm{RMSDiff}=0.063,\ \rho=0.989; \qquad
(64,128): \ \mathrm{RMSDiff}=0.033,\ \rho=0.997,
\]
show a \textbf{decreasing} RMS difference and \textbf{increasing} correlation with the resolution, as expected. Both correlation values exceed $0.98$, indicating excellent structural consistency across resolutions. The 64--128 comparison satisfies the adopted 5\% RMSDiff threshold, while the coarser 16--64 comparison remains above it at \(6.32\%\). The diagnostics details are shown in Table in Figure \ref{fig:conv+res} (b).

Figure \ref{fig:conv+res} (a) shows the measured data (blue) in log–log space, overlaid with the best-fit trend (gray dashed line) and theoretical reference slopes for first $\alpha=-1$ and second-order $\alpha=-2$ convergence (red and green lines, respectively), all normalized at the mid-range $N_\mathrm{eff}$ for visual comparison. 

The observed two-point slope \(\alpha_\mathrm{fit}\approx -0.62\) indicates sub-linear convergence, meaning the residual discrepancies between successive resolutions diminish, but at a rate shallower than first order. As we mentioned before, in standard numerical discretizations of smooth functions on a uniform grid, the global error typically behaves as
\[
\text{error} \;\propto\; h^{\,p},
\]
where \(h\) is the grid spacing and \(p\) is the order of the method \citep[][see Ch.~5, §5.1]{Fornberg1988,Press2007}.  
For HEALPix, the effective angular spacing scales as \(h \sim N_\mathrm{side}^{-1}\) via the pixel-area relation 
\(\Omega_\mathrm{pix} = 4\pi / (12N_\mathrm{side}^2)\) \citep[eq.~(24)][]{gorski_etal:05a}; consequently, one expects a power-law trend of residuals with \(N_\mathrm{side}\), with the empirical exponent \(\alpha\) estimated from \(\log(\mathrm{RMSDiff})\) versus the effective resolution \(N_\mathrm{eff}=\sqrt{N_1N_2}\).

In practice, full-sky survey maps represent the true sky convolved with the instrument beam and pixel window, plus instrumental noise.  
The total observed angular power spectrum can be written as
\[
\langle \tilde C_\ell\rangle = 
\sum_{\ell'} M_{\ell\ell'}\,B_{\ell'}^{2}\,P_{\ell'}^{2}\,C_{\ell'} + \langle \tilde N_\ell\rangle,
\]
as shown in \citet[][eq.~(23)]{Hivon2002}.  
Here \(B_\ell\) and \(P_\ell\) describe the beam and pixel-window functions, which suppress power at high multipoles (small angular scales), while \(\langle \tilde N_\ell\rangle\) is the noise contribution.  
Once the pixel scale becomes smaller than the beam width, the map residuals are dominated by \(B_\ell\) and by the noise floor rather than by geometric discretisation.  
This behavior is explicitly noted in the \emph{Planck 2015 LFI mapmaking} paper \citep[see §6.3, “Smoothing and noise suppression,” and Fig.~11]{Planck2015LFIMapmaking}, where maps were convolved to $1^\circ$ FWHM to mitigate noise.  
Similarly, the \emph{Planck 2018 lensing} analysis \citep[§2.2]{Planck2018Lensing} emphasizes that small-scale modes are noise-dominated and beam-limited.  
Consequently, refining the HEALPix resolution beyond the beam scale produces a shallower-than-first-order convergence slope-consistent with our measured \(\alpha_\mathrm{fit}\approx -0.3\).

\vspace{0.5em}
\noindent

\underline{Figures and convergence criterion}

The analysis establishes that the algorithm achieves numerical stability between $N_\mathrm{side}=64$ and $128$, as RMSDiff of 3.31\% and correlation of 0.99, considering our convergence criterion of RMSDiff of 5\% and $\rho\approx1$.
Concluding, $N_\mathrm{side}=64$ is considered a sufficient working resolution for intensity-map analyses.

The first two diagnostic figures illustrate the complementary aspects of convergence.

\begin{figure}[h!]
  \centering
  \includegraphics[width=0.48\textwidth]{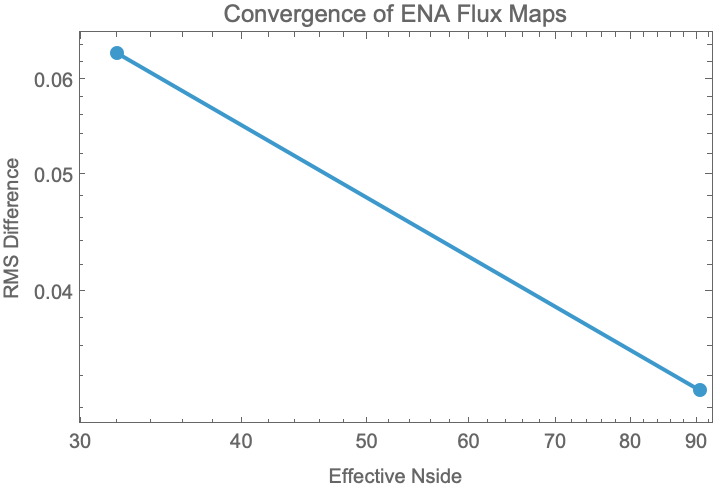}\includegraphics[width=0.48\textwidth]{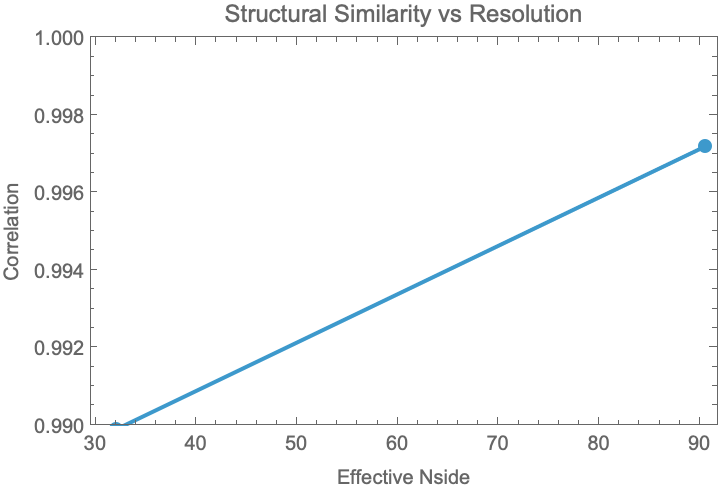}
  \caption{RMS difference between successive resolutions as a function of effective $N_\mathrm{side}$ (left) and Pearson correlation coefficient between intensity maps as a function of effective $N_\mathrm{side}$ (right).}
  \label{fig:rmsdiffANDcorrelation}
\end{figure}

\begin{figure}[h!]
  \centering

  \begin{subfigure}[t]{0.8\textwidth}
    \centering
    \includegraphics[width=\textwidth]{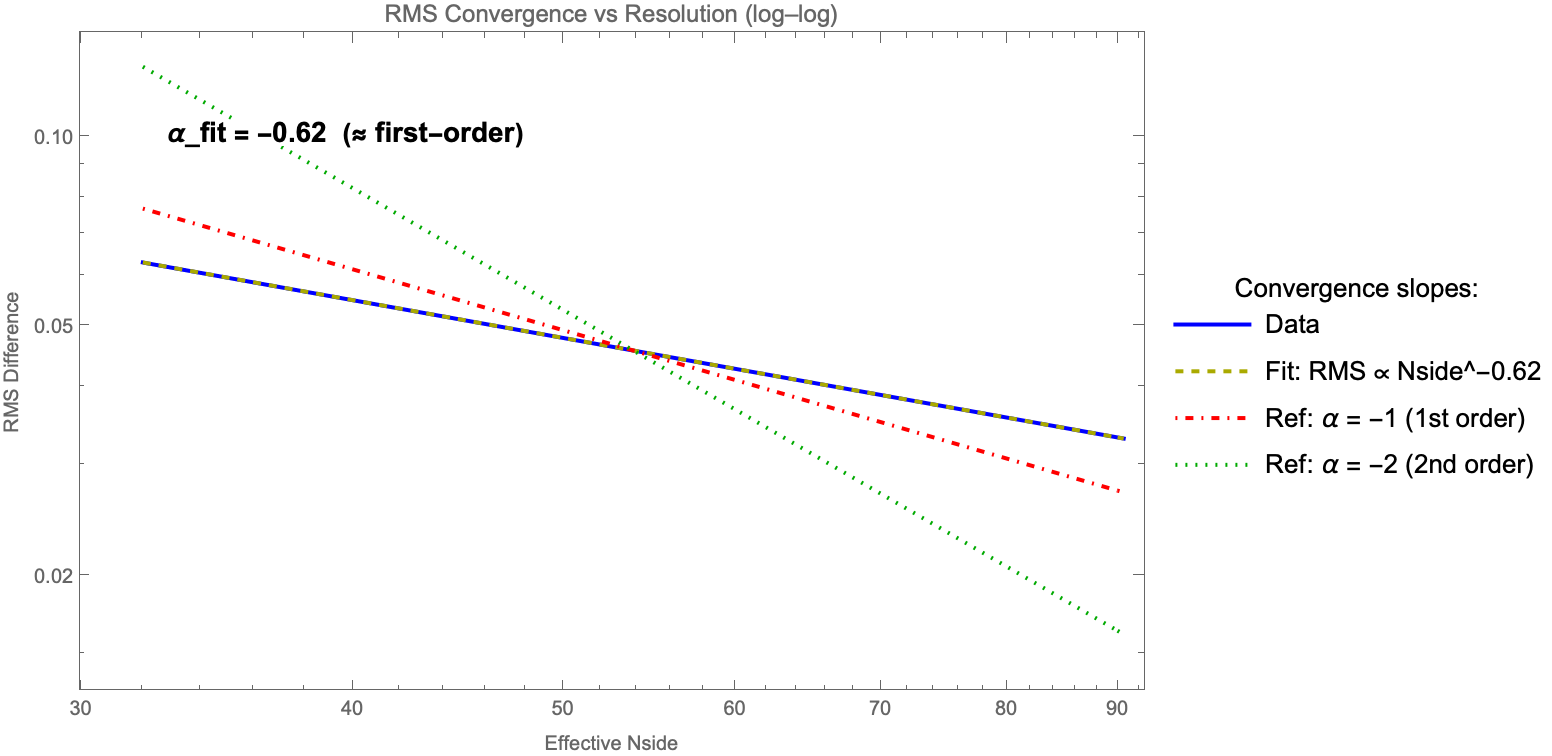}
    \caption{Log–log convergence test showing measured RMS differences (blue), two-point slope \(\alpha_\mathrm{fit}\simeq -0.62\) (gray dashed), and reference first- and second-order trends (red, green).}
    \label{fig:convergence}
  \end{subfigure}

  \vspace{0.5cm}

  \begin{subfigure}[t]{0.45\textwidth}
    \centering
    \includegraphics[width=\textwidth]{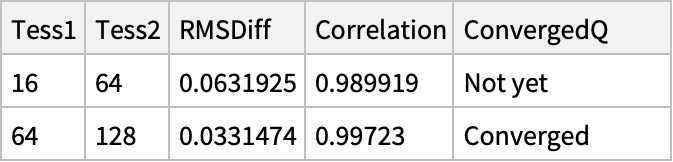}
    \caption{Resolution-scaling diagnostics.}
    \label{fig:resolution-scaling}
  \end{subfigure}

  \caption{Convergence and resolution-scaling diagnostics.  
  (a) Log–log convergence trends.  
  (b) Resolution comparison table.}
  \label{fig:conv+res}
\end{figure}

Figure \ref{fig:rmsdiffANDcorrelation}(left) shows the $\mathrm{RMSDiff}(N_\mathrm{eff})$, between intensity maps as a function of resolution. The monotonic decline of $\mathrm{RMSDiff}$ indicates that as the pixelization becomes finer, the absolute amplitude of residuals decreases - i.e., the total variance between consecutive maps becomes smaller. 
This reflects convergence in the \emph{L\textsuperscript{2}} sense (meaning that the squared residuals integrated over the sphere are going to zero as the grid is refined), quantifying how rapidly the numerical solution approaches an asymptotic form.

Figure~\ref{fig:rmsdiffANDcorrelation}(right) complements this by tracking 
the Pearson correlation coefficient $\rho(N_\mathrm{eff})$ between map pairs.
While the RMS difference is sensitive to overall normalization and amplitude offsets,  the correlation coefficient measures the \emph{structural consistency}  of the spatial pattern, independent of scale - it provides a normalized measure of shape similarity between two maps.  
In the context of spherical sky data, a correlation coefficient 
$\rho \gtrsim 0.99$ means that the large-scale angular distribution - 
the morphology of the (ENA) intensity pattern - remains stable across resolutions even if the overall variance slowly decreases.

This approach follows standard practices in image and map-comparison analysis, 
where the RMSDiff quantifies absolute convergence and the correlation captures morphological fidelity (e.g., \citep{gorski_etal:05a}). The simultaneous behavior observed here - decreasing RMS difference and increasing correlation toward unity - demonstrates that both the amplitude and structure of the (ENA) intensity maps stabilize with resolution, confirming that the pipeline converges in both intensity and morphology.

\vspace{0.5em}
\noindent
In summary, the intensity-map reconstruction demonstrates stable convergence across increasing HEALPix resolutions, reaching the asymptotic regime where additional refinement yields negligible morphological or statistical change.

\noindent
\underline{Relation to performance scaling}

The convergence analysis presented above establishes that the  pipeline achieves numerical stability by $N_\mathrm{side}=64$.  
This provides the necessary baseline for interpreting the computational performance tests in Section~\ref{sec:performance}.  
Since the total runtime and memory usage (Figure~\ref{fig:performance}) scale primarily with the number of processed pixels ($\propto 12\,N_\mathrm{side}^2$), verifying convergence ensures that higher resolutions do not improve scientific accuracy, only computational cost.

\subsection{Performance tests} \label{sec:performance}

To assess the parallel scalability of the HEALPix-based all-sky map generation pipeline, we conducted performance testing on two representative HEALPix tessellations, $N_\mathrm{side}=16$ and $N_\mathrm{side}=64$, using up to 16 parallel kernels. The choice of tessellations was dictated by the standard pixel angular resolution of astronomical instruments (about 1 $\arcdeg$). Each run measured total runtime and memory usage, and we calculated the speedup and parallel efficiency. The derived quantities were computed as follows:

\begin{itemize}
\item \textbf{Runtime:} $T(n)$ denotes the directly observed runtime for $n$ parallel kernels. 
\item \textbf{Speedup:}
\begin{equation}
S(n) = \frac{T(1)}{T(n)}
\end{equation}
quantifies how much faster the program runs using \textit{n} kernels compared to 1. An ideal linear speedup would satisfy $S(n) = n$.
\item \textbf{Parallel efficiency:}
\begin{equation}
E(n) = \frac{S(n)}{n} = \frac{T(1)}{n \, T(n)}
\end{equation}
measures how effectively the additional computational resources are being used.
\item \textbf{Peak memory usage:} 
\begin{equation}
M(n) = \max_t \, \mathrm{Memory}(t)
\end{equation}
represents the maximum physical memory footprint during a run, obtained using \textsc{MaxMemoryUsed[]} - it is directly measured from the profiling data.
\end{itemize}

\begin{figure}[h]
    \centering

    \begin{subfigure}[b]{\textwidth}
        \centering
        \includegraphics[width=\textwidth]{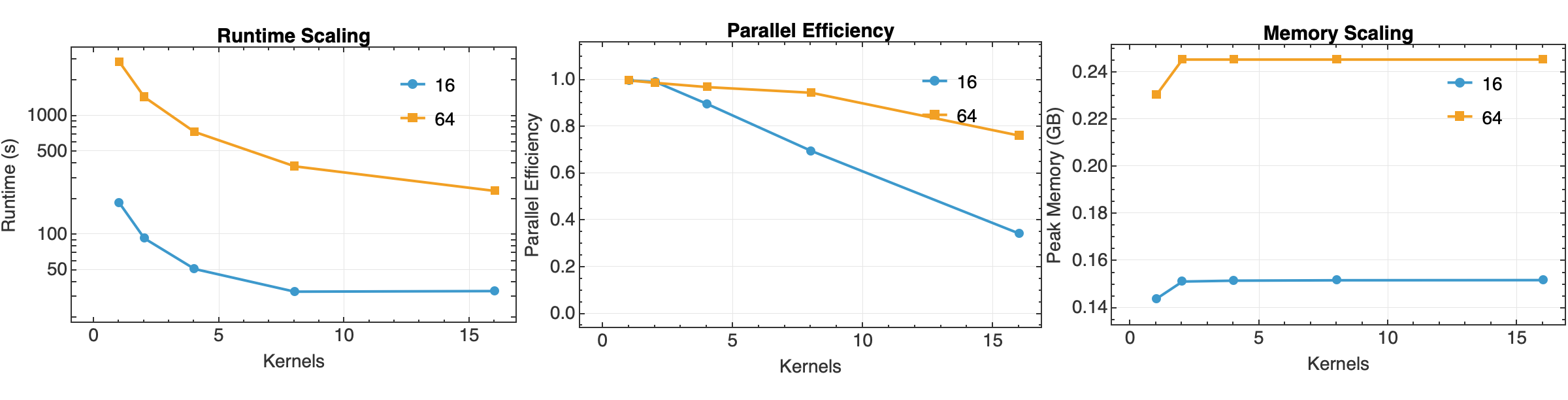}
    \end{subfigure}

    \begin{subfigure}[b]{0.35\textwidth}
        \centering
        \includegraphics[width=\textwidth]{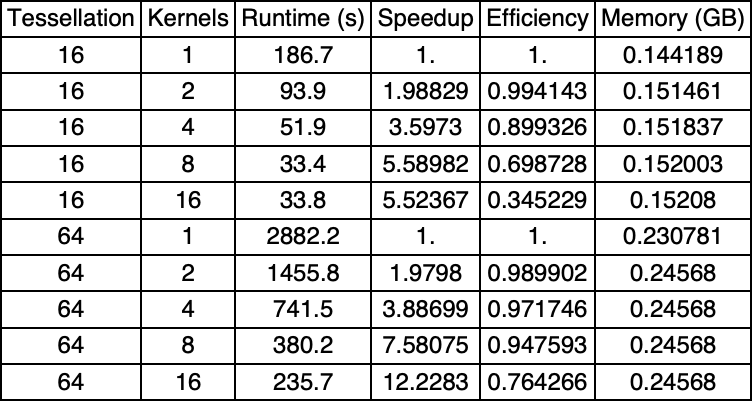}
    \end{subfigure}

    \caption{Performance scaling of the IBEX HEALPix pipeline. Top: 
 Left - Runtime scaling with kernel count for tessellations 16 and 64. Center - Parallel efficiency relative to single - 
  core baseline. Right - Peak memory usage as a function of kernel count, 
 showing consistent memory footprint across parallel runs. Bottom: Performance table summarizing all the measured and calculated quantities. Larger tessellation 64 maintains near-linear speedup up to 8 kernels with efficiency above 80$\%$.}
    \label{fig:performance}
\end{figure}

\vspace{0.5em}
\noindent
Figure~\ref{fig:performance} summarizes the results for both tessellations. The top panels show runtime scaling, parallel efficiency, and memory behavior, while the bottom table lists the quantitative values.

\noindent
\textbf{Runtime scaling}  
The runtime results show very efficient parallelization of the HEALPix map generation pipeline. 
For \(N_\mathrm{side}=16\), runtime decreased from \(T(1)=186.7\) s to \(T(8)=33.4\) s, corresponding to a speedup of \(S(8)=5.59\) and efficiency \(E(8)=69.9\%\). The 16-kernel run did not improve further, giving \(T(16)=33.8\) s and \(E(16)=34.5\%\), indicating saturation for the smaller workload.
For \(N_\mathrm{side}=64\), runtime decreased from \(T(1)=2882.2\) s to \(T(16)=235.7\) s, giving \(S(16)=12.23\) and \(E(16)=76.4\%\). The larger workload maintains near-linear scaling through 8 kernels, with \(S(8)=7.58\) and \(E(8)=94.8\%\).
Almost linear scaling up to 8 kernels (efficiency $>85\%$) confirms that the workload is dominated by independent per-pixel operations, that can be parallelized. 
At higher core counts ($n>8$), efficiency begins to drop probably due to coordination overheads between parallel kernels.

\noindent
\textbf{Scaling with tessellation}  
At a single kernel, the runtime ratio between the two tessellations is
\[
\frac{T_{64}(1)}{T_{16}(1)}
=
\frac{2882.2}{186.7}
\simeq 15.4,
\]
matching the expected $\sim16\times$ increase in computational cost given that the number of HEALPix pixels scales as $N_\mathrm{pix} \propto N_\mathrm{side}^2$. 
This confirms that the algorithm exhibits the intended $\mathcal{O}(N_\mathrm{pix})$ scaling with pixel count, validating both the geometric and computational aspects of the implementation.

\noindent
\textbf{Memory usage}  
Peak memory usage remains nearly constant across kernel counts, varying by about \(0.008\) GB for \(N_\mathrm{side}=16\) and about \(0.015\) GB for \(N_\mathrm{side}=64\). For \(N_\mathrm{side}=16\), \(M(n)\) stays near \(0.15\) GB, while for \(N_\mathrm{side}=64\) it stays near \(0.25\) GB. The increase from \(\sim0.15\) GB to \(\sim0.25\) GB is smaller than the \(16\times\) increase in pixel number from \(N_\mathrm{side}=16\) to 64, indicating that the measured peak memory includes a substantial fixed overhead and is not dominated solely by the final HEALPix arrays.
This mostly flat behavior shows efficient memory reuse and good workload distribution across parallel kernels. 
Minimal growth with $n$ indicates that no significant duplication of data occurs between kernels, and memory overhead per kernel is negligible.

\noindent
\textbf{Summary and interpretation}  
The overall performance results confirm that the pipeline is both computationally and memory efficient. 
Strong scaling remains close to ideal up to 8 kernels, with efficiency above 80\% for the higher-resolution case. 
Runtime scales almost linearly with total pixel count, verifying $\mathcal{O}(N_\mathrm{pix})$ complexity, while peak memory usage remains stable and well below 0.3~GB for all runs. 
The primary limitation observed is the efficiency drop beyond 8 kernels, pointing most likely to communication and coordination overhead.



\bibliography{bibliography, iplbib}{}

@article{Schwadron2009,
    author = {Schwadron, N. A. and Crew, G. and Vanderspek, R. and Allegrini, F. and Bzowski, M.},
    title = {The Interstellar Boundary Explorer Science Operations Center},
    journal = {Space Science Reviews},
    year = {2009},
    doi = {10.1007/s11214-009-9513-x}
}

@article{McComas_2020,
doi = {10.3847/1538-4365/ab8dc2},
url = {https://doi.org/10.3847/1538-4365/ab8dc2},
year = {2020},
month = {jun},
publisher = {The American Astronomical Society},
volume = {248},
number = {2},
pages = {26},
author = {McComas, D. J. and Bzowski, M. and Dayeh, M. A. and DeMajistre, R. and Funsten, H. O. and Janzen, P. H. and Kowalska-Leszczyńska, I. and Kubiak, M. A. and Schwadron, N. A. and Sokół, J. M. and Szalay, J. R. and Tokumaru, M. and Zirnstein, E. J.},
title = {Solar Cycle of Imaging the Global Heliosphere: Interstellar Boundary Explorer (IBEX) Observations from 2009–2019},
journal = {The Astrophysical Journal Supplement Series}
}

@article{Dayeh_2023,
   title={Investigating the IBEX Ribbon Structure a Solar Cycle Apart},
   volume={952},
   ISSN={1538-4357},
   url={http://dx.doi.org/10.3847/1538-4357/acda8b},
   DOI={10.3847/1538-4357/acda8b},
   number={1},
   journal={The Astrophysical Journal},
   publisher={American Astronomical Society},
   author={Dayeh, M. A. and Zirnstein, E. J. and Swaczyna, P. and McComas, D. J.},
   year={2023},
   month=jul, pages={19} }

@article{Hivon2002,
  author = {Hivon, E. and G{\'o}rski, K. M. and Netterfield, C. B. and Crill, B. P. and Prunet, S. and Hansen, F.},
  title = {MASTER of the CMB Anisotropy Power Spectrum: A Fast Method for Statistical Analysis of Large and Complex CMB Data Sets},
  journal = {The Astrophysical Journal},
  year = {2002},
  volume = {567},
  pages = {2--17},
  doi = {10.1086/338126}
}

@article{Planck2015LFIMapmaking,
  author = {Planck Collaboration},
  title = {Planck 2015 results. II. Low Frequency Instrument data processing},
  journal = {Astronomy \& Astrophysics},
  year = {2016},
  volume = {594},
  pages = {A2},
  doi = {10.1051/0004-6361/201525818}
}

@article{Planck2018Lensing,
  author = {Planck Collaboration},
  title = {Planck 2018 results. VIII. Gravitational lensing},
  journal = {Astronomy \& Astrophysics},
  year = {2020},
  volume = {641},
  pages = {A8},
  doi = {10.1051/0004-6361/201833886}
}

@article{Osthus02042024,
author = {Dave Osthus and Brian P. Weaver and Lauren J. Beesley and Kelly R. Moran and Madeline A. Stricklin and Eric J. Zirnstein and Paul H. Janzen and Daniel B. Reisenfeld},
title = {Towards Improved Heliosphere Sky Map Estimation with Theseus},
journal = {Technometrics},
volume = {66},
number = {2},
pages = {208--226},
year = {2024},
publisher = {ASA Website},
doi = {10.1080/00401706.2023.2271017}}

@ARTICLE{ribbon,
    author    = {P. C. Frisch and B.-G. Andersson and A. Berdyugin and V. Piirola and R. DeMajistre and H. O. Funsten and A. M. Magalhães and D. B. Seriacopi and D. J. McComas and N. A. Schwadron and J. D. Slavin and S. J. Wiktorowicz},
    title     = {The Interstellar Magnetic Field Close to the Sun II},
    journal   = {Astrophysical Journal},
    year      = {2012},
    volume    = {760},
    number    = {2},
    pages     = {106},
    doi       = {10.1088/0004-637x/760/2/106}
}

@book{Press2007,
  author = {Press, W. H. and Teukolsky, S. A. and Vetterling, W. T. and Flannery, B. P.},
  title = {Numerical Recipes: The Art of Scientific Computing},
  edition = {3rd},
  year = {2007},
  publisher = {Cambridge University Press}
}

@article{Fornberg1988,
  author = {Fornberg, Bengt},
  title = {Generation of Finite Difference Formulas on Arbitrarily Spaced Grids},
  journal = {Mathematics of Computation},
  year = {1988},
  volume = {51},
  pages = {699--706}
}

@ARTICLE{2009:funsten,
       author = {{Funsten}, H.~O. and {Allegrini}, F. and {Bochsler}, P. and {Dunn}, G. and {Ellis}, S. and {Everett}, D. and {Fagan}, M.~J. and {Fuselier}, S.~A. and {Granoff}, M. and {Gruntman}, M. and {Guthrie}, A.~A. and {Hanley}, J. and {Harper}, R.~W. and {Heirtzler}, D. and {Janzen}, P. and {Kihara}, K.~H. and {King}, B. and {Kucharek}, H. and {Manzo}, M.~P. and {Maple}, M. and {Mashburn}, K. and {McComas}, D.~J. and {Moebius}, E. and {Nolin}, J. and {Piazza}, D. and {Pope}, S. and {Reisenfeld}, D.~B. and {Rodriguez}, B. and {Roelof}, E.~C. and {Saul}, L. and {Turco}, S. and {Valek}, P. and {Weidner}, S. and {Wurz}, P. and {Zaffke}, S.},
        title = "{The Interstellar Boundary Explorer High Energy (IBEX-Hi) Neutral Atom Imager}",
      journal = {\ssr},
         year = 2009,
        month = aug,
       volume = {146},
       number = {1-4},
        pages = {75-103},
          doi = {10.1007/s11214-009-9504-y},
       adsurl = {https://ui.adsabs.harvard.edu/abs/2009SSRv..146...75F}
}

@misc{capra,
  author       = {Bukowiecka, Nikola},
  year         = {2025},
  title        = {HEALPix based system to create all sky maps of density distributions},
  howpublished = {\href{https://github.com/nikolabukowiecka/Capra}{GitHub repository}}
}

@misc{Bukowiecka_Capra,
  author       = {Nikola Bukowiecka},
  title        = {{nikolabukowiecka/Capra: First official release of
                   the mapping software Capra}},
  month        = jun,
  year         = 2026,
  publisher    = {Zenodo},
  version      = {v1},
  doi          = {10.5281/zenodo.20584912},
  url          = {https://doi.org/10.5281/zenodo.20584912},
  swhid        = {swh:1:dir:f9dc31d1aeca31082b4d4919d128d4c90026ca90
                   ;origin=https://doi.org/10.5281/zenodo.20584911;vi
                   sit=swh:1:snp:6146842b66483fae40adf0bb376d592b5df6
                   c53b;anchor=swh:1:rel:08c5de89b0cc3c0c3451506d1614
                   b172fddcf175;path=nikolabukowiecka-Capra-0744526
                  },
}

@ARTICLE{funsten_etal:09a,
        author     = {{Funsten}, H.~O. and {Allegrini}, F. and {Bochsler}, P. and {Dunn}, G. and {Ellis}, S. and {Everett}, D. and {Fagan}, M.~J. and {Fuselier}, S.~A. and {Granoff}, M. and {Gruntman}, M. and {Guthrie}, A.~A. and {Hanley}, J. and {Harper}, R.~W. and {Heirtzler}, D. and {Janzen}, P. and {Kihara}, K.~H. and {King}, B. and {Kucharek}, H. and {Manzo}, M.~P. and {Maple}, M. and {Mashburn}, K. and {McComas}, D.~J. and {Moebius}, E. and {Nolin}, J. and {Piazza}, D. and {Pope}, S. and {Reisenfeld}, D.~B. and {Rodriguez}, B. and {Roelof}, E.~C. and {Saul}, L. and {Turco}, S. and {Valek}, P. and {Weidner}, S. and {Wurz}, P. and {Zaffke}, S.},
        title      = "The {Interstellar Boundary Explorer} High Energy (IBEX-Hi) Neutral Atom Imager",
        journal    = {\ssr},
        year       = 2009,
        volume     = 146,
        pages      = {75--103},
        doi        = {10.1007/s11214-009-9504-y}}

@article{funsten_etal:26,
  author  = {Funsten, H. O. and Allegrini, F. and Reisenfeld, D. B. and Cardarelli, G. and Carpenter, B. C. and Christian, E. R. and Cortinas, S. and Craft, S. P. and Da Rocha, F. D. and Dauson, E. R. and De Los Santos, A. and Dunn, G. F. and Fernandes, P. A. and Fletcher, G. and Ford, K. A. and Ford, J. and Geros, E. G. and Gkioulidou, M. and Grubbs, G. and Guthrie, A. A. and Hanley, J. J. and Harvey, D. S. and Hemphill, R. L. and Hill, B. D. and Hoeper, P. J. and Hom-Crosier, R. D. and Hooks, D. E. and Janzen, P. H. and Kim, T. K. and Lanctot, S. I. and Liu, Y. and Maldonado, C. A. and Martinez, J. P. and McComas, D. J. and Merrill, A. S. and Moebius, E. and Moorhead-Rosenberg, Z. and Mosley, B. N. T. and Newell, R. T. and Noh, S. and Nunez, C. and Pacheco, T. R. and Pontoni, A. A. and Pope, S. E. and Rahmanifard, F. and Rodriguez, B. G. and Rodriguez, H. and Sandoval, B. F. and Schultz, T. B. and Schiferl, C. M. and Schwadron, N. A. and Skoug, R. M. and Storms, S. A. and Tapley, M. B. and Taylor, K. L. and Taylor, T. W. and Toczynski, W. and Trevino, J. A. and Tucker, C. J. and Venhaus, D. M. and Vigil, V. J. and Walia, N. K. and Wurz, P.},
  title   = {The Interstellar Mapping And Acceleration Probe High Energy ({IMAP-Hi}) Neutral Atom Imager},
  journal = {Space Science Reviews},
  year    = {2026},
  volume  = {222},
  number  = {4},
  pages   = {47},
  doi     = {10.1007/s11214-026-01298-3},
  url     = {https://doi.org/10.1007/s11214-026-01298-3},
  issn    = {1572-9672}
}

@ARTICLE{funsten_etal:09b,
        author     = {{Funsten}, H.~O. and {Allegrini}, F. and {Crew}, G.~B. and {DeMajistre}, R. and {Frisch}, P.~C. and {Fuselier}, S.~A. and {Gruntman}, M. and {Janzen}, P. and {McComas}, D.~J. and {M{\"o}bius}, E. and {Randol}, B. and {Reisenfeld}, D.~B. and {Roelof}, E.~C. and {Schwadron}, N.~A.},
        title      = "Structures and Spectral Variations of the Outer Heliosphere in {IBEX} Energetic Neutral Atom Maps",
        journal    = {Science},
        year       = 2009,
        volume     = 326,
        pages      = {964--966},
        doi        = {10.1126/science.1180927}}

@ARTICLE{fuselier_etal:09b,
        author     = {{Fuselier}, S.~A. and {Bochsler}, P. and {Chornay}, D. and {Clark}, G. and {Crew}, G.~B. and {Dunn}, G. and {Ellis}, S. and {Friedmann}, T. and {Funsten}, H.~O. and {Ghielmetti}, A.~G. and {Googins}, J. and {Granoff}, M.~S. and {Hamilton}, J.~W. and {Hanley}, J. and {Heirtzler}, D. and {Hertzberg}, E. and {Isaac}, D. and {King}, B. and {Knauss}, U. and {Kucharek}, H. and {Kudirka}, F. and {Livi}, S. and {Lobell}, J. and {Longworth}, S. and {Mashburn}, K. and {McComas}, D.~J. and {M{\"o}bius}, E. and {Moore}, A.~S. and {Moore}, T.~E. and {Nemanich}, R.~J. and {Nolin}, J. and {O'Neal}, M. and {Piazza}, D. and {Peterson}, L. and {Pope}, S.~E. and {Rosmarynowski}, P. and {Saul}, L.~A. and {Scherrer}, J.~R. and {Scheer}, J.~A. and {Schlemm}, C. and {Schwadron}, N.~A. and {Tillier}, C. and {Turco}, S. and {Tyler}, J. and {Vosbury}, M. and {Wieser}, M. and {Wurz}, P. and {Zaffke}, S.},
        title      = "The {IBEX-Lo} Sensor",
        journal    = {\ssr},
        year       = 2009,
        volume     = 146,
        pages      = {117--147},
        doi        = {10.1007/s11214-009-9495-8}}

@ARTICLE{gorski_etal:05a,
        author     = {{G{\'o}rski}, K.~M. and {Hivon}, E. and {Banday}, A.~J. and {Wandelt}, B.~D. and {Hansen}, F.~K. and {Reinecke}, M. and {Bartelmann}, M.},
        title      = "{HEALPix: A Framework for High-Resolution Discretization and Fast Analysis of Data Distributed on the Sphere}",
        journal    = {\apj},
        eprint     = {arXiv:astro-ph/0409513},
        year       = 2005,
        volume     = 622,
        pages      = {759--771},
        doi        = {10.1086/427976}}

@ARTICLE{mccomas:09a,
        author     = {{McComas}, D.~J.},
        title      = "{ENA} imaging of the inner heliosheath -- preparing for the {Interstellar Boundary Explorer (IBEX)}",
        journal    = {\ssr},
        year       = 2009,
        volume     = 143,
        pages      = {125--138},
        doi        = {10.1007/s11214-008-9410-8}}

@ARTICLE{mccomas_etal:09a,
        author     = {{McComas}, D.~J. and {Allegrini}, F. and {Bochsler}, P. and {Bzowski}, M. and {Collier}, M. and {Fahr}, H. and {Fichtner}, H. and {Frisch}, P. and {Funsten}, H.~O. and {Fuselier}, S.~A. and {Gloeckler}, G. and {Gruntman}, M. and {Izmodenov}, V. and {Knappenberger}, P. and {Lee}, M. and {Livi}, S. and {Mitchell}, D. and {M{\"o}bius}, E. and {Moore}, T. and {Pope}, S. and {Reisenfeld}, D. and {Roelof}, E. and {Scherrer}, J. and {Schwadron}, N. and {Tyler}, R. and {Wieser}, M. and {Witte}, M. and {Wurz}, P. and {Zank}, G.},
        title      = "{IBEX -- Interstellar Boundary Explorer}",
        journal    = {\ssr},
        year       = 2009,
        volume     = 146,
        pages      = {11--33},
        doi        = {10.1007/s11214-009-9499-4}}

@ARTICLE{mccomas_etal:09b,
        author     = {{McComas}, D.~J. and {Allegrini}, F. and {Bochsler}, P. and {Frisch}, P. and {Funsten}, H.~O. and {Gruntman}, M. and {Janzen}, P.~H. and {Kucharek}, H. and {M{\"o}bius}, E. and {Reisenfeld}, D.~B. and {Schwadron}, N.~A.},
        title      = "Lunar backscatter and neutralization of the solar wind: {First} observations of neutral atoms from the {Moon}",
        journal    = {\grl},
        year       = 2009,
        volume     = 36,
        pages      = {12104-+},
        doi        = {10.1029/2009GL038794}}

@ARTICLE{mccomas_etal:11a,
        author     = {{McComas}, D.~J. and {Carrico}, J.~P. and {Hautamaki}, B. and {Intelisano}, M. and {Lebois}, R. and {Loucks}, M. and {Policastri}, L. and {Reno}, M. and {Scherrer}, J. and {Schwadron}, N.~A. and {Tapley}, M. and {Tyler}, R.},
        title      = "A new class of long-term stable lunar resonance orbits: {Space} weather applications and the {Interstellar Boundary Explorer}",
        journal    = {Space Weather},
        year       = 2011,
        volume     = 9,
        eid        = {S11002},
        pages      = {S11002},
        doi        = {10.1029/2011SW000704}}

@ARTICLE{mccomas_etal:25a,
        author     = "D.J. McComas and E.R. Christian and N.A. Schwadron and M. Gkioulidou and F. Allegrini and D.N. Baker and M. Bzowski and G. Clark and C.M.S. Cohen and I. Cohen and C. Collura and M.J. Cully and S. Dalla and M.I. Desai and A. Driesman and D. Eng and H.O. Funsten and S.A Fuselier and A. Galli and J. Giacalone and J. Hahn and K.P. Hegarty and T. Horbury and M. Horanyi and L.M. Kistler and M.A. Kubiak and S. Kubota and S. Livi and N. Lugaz and C.O. Lee and J. Luhmann and W. Matthaeus and D.G. Mitchell and J.G. Mitchell and E. Moebius and S. Pope and E. Provornikova and J.S. Rankin and D.B. Reisenfeld and C. Reno and J.D. Richardson and C.T. Russell and M.M. Shaw-Lecerf and J. Scherrer and R.M. Skoug and M.M. Shen and H.E. Spence and Z. Sternovsky and M. Strumik and J.R. Szalay and M. Tapley and M. Tokumaru and D.L. Turner and S. Weidner and P. Wurz and G.P. Zank",
        title      = "{Interstellar Mapping and Acceleration Probe: The NASA IMAP} mission",
        year       = 2025,
        journal    = {\ssr},
        volume     = 221,
        pages      = 100,
        eid        = {100},
        doi        = {10.1007/s11214-025-01224-z}}
\bibliographystyle{aasjournal}



\end{document}